\documentclass[aps,prx,amsmath,floats,floatfix,twocolumn,
  superscriptaddress,nofootinbib,showpacs]{revtex4-1}

\pdfoutput=1 
\usepackage{diagbox}
\usepackage{multirow}
\usepackage{braket}
\usepackage{amsfonts}
\usepackage{amsmath}
\usepackage{amssymb}
\usepackage{amsthm}
\usepackage{bm}
\usepackage{graphicx}
\usepackage{graphics}
\usepackage[latin1]{inputenc}
\usepackage{latexsym}
\usepackage{rotating}
\usepackage[dvipsnames]{xcolor}
\usepackage{mathrsfs}
\usepackage{microtype}
\usepackage{verbatim}
\usepackage{url}

\usepackage[caption=false]{subfig}
\usepackage[normalem]{ulem}

\usepackage[breaklinks=true]{hyperref}
\hypersetup{
    colorlinks=true,
    linkcolor=NavyBlue,
    filecolor=Magenta,
    urlcolor=NavyBlue,
    citecolor=NavyBlue
}

\usepackage{yfonts}
\usepackage{float}
\usepackage{xspace} 
\usepackage{mathrsfs}
\usepackage{siunitx}
\usepackage{array}

\allowdisplaybreaks[4]

\newcommand{\be}{\begin{equation}}
\newcommand{\ee}{\end{equation}}

\newcommand{\h}{{\mbox{\tiny H}}}

\newcommand{\ba}{\begin{align}}
\newcommand{\ea}{\end{align}}

\makeatletter
\newcommand*{\rom}[1]{\expandafter\@slowromancap\romannumeral #1@}
\makeatother

\AtBeginDocument{%
    \newwrite\bibnotes
    \def\bibnotesext{Notes.bib}
    \immediate\openout\bibnotes=\jobname\bibnotesext
    \immediate\write\bibnotes{@CONTROL{REVTEX41Control}}
    \immediate\write\bibnotes{@CONTROL{%
    apsrev41Control,author="08",editor="1",pages="1",title="0",year="1"}}
    \if@filesw
    \immediate\write\@auxout{\string\citation{apsrev41Control}}%
    \fi
}

\allowdisplaybreaks

\begin{document}


\title{Spectral method for the gravitational perturbations of black holes: \\ Schwarzschild background case}

\author{Adrian Ka-Wai Chung}
\email{akwchung@illinois.edu}
\affiliation{Illinois Center for Advanced Studies of the Universe \& Department of Physics, University of Illinois at Urbana-Champaign, Urbana, Illinois 61801, USA}

\author{Pratik Wagle}
\email{wagle2@illinois.edu}
\affiliation{Illinois Center for Advanced Studies of the Universe \& Department of Physics, University of Illinois at Urbana-Champaign, Urbana, Illinois 61801, USA}

\author{Nicol\'as Yunes}
\affiliation{Illinois Center for Advanced Studies of the Universe \& Department of Physics, University of Illinois at Urbana-Champaign, Urbana, Illinois 61801, USA}

\date{\today}

\begin{abstract}

We develop a novel technique through spectral decompositions to study the gravitational perturbations of a black hole, without needing to decouple the linearized field equations into master equations and separate their radial and angular dependence.
We first spectrally decompose the metric perturbation in a Legendre and Chebyshev basis for the angular and radial sectors respectively, using input from the asymptotic behavior of the perturbation at spatial infinity and at the black hole event horizon. 
This spectral decomposition allows us to then transform the linearized Einstein equations (a coupled set of partial differential equations) into a linear matrix equation.
By solving the linear matrix equation for its generalized eigenvalues, we can estimate the complex quasinormal frequencies of the fundamental mode and various overtones of the gravitational perturbations simultaneously and to high accuracy. 
We apply this technique to perturbations of a nonspinning, Schwarzschild black hole in general relativity and find the complex quasinormal frequencies of two fundamental modes and their first two overtones. 
We demonstrate that the technique is robust and accurate, in the Schwarzschild case leading to relative fractional errors of $\leq 10^{-10} - 10^{-8}$ for the fundamental modes, $\leq 10^{-7} - 10^{-6}$ for their first overtones, $\leq 10^{-7} - 10^{-4}$ for their second overtones. 
This method can be applied to any black hole spacetime, irrespective of its Petrov type, making the numerical technique extremely powerful in the study of black hole ringdown in and outside general relativity.
\end{abstract}

\maketitle


\section{Introduction}
\label{sec:intro}

The LIGO-Virgo-KAGRA collaboration has successfully detected numerous gravitational-wave (GW) signals, most of which are emitted by binary black hole (BH) coalescence \cite{LIGO_01, LIGO_02, LIGO_03, LIGO_04, LIGO_05, LIGO_06, LIGO_07, LIGO_08, LIGO_09, LIGO_10, LIGO_11, LIGO_GW190412, LIGO_GW190814}. 
After the merger, the remnant eventually relaxes into a stationary and rotating BH by emitting GWs with a discrete set of quasinormal mode (QNM) frequencies, a coalescence stage known as ringdown.
These signals grant us pristine access to the properties of spacetime in the strong field, most dynamical and nonlinear regime, as these GWs travel mostly undisturbed, and thus, carry nondistorted information about their source. 
Thus far, all the GWs detected are consistent with general relativity (GR) \cite{LIGO_07, LIGO_11, LIGOScientific:2021sio, Gupta:2021vdj, Perkins:2021mhb, Cardenas-Avendano:2019zxd, Perkins:2018tir, Chamberlain:2017fjl}, indicating that Einstein's theory has now also passed the first GW tests. 
In the near future, the ongoing improvements in GW detector technology and the addition of new, next-generation detectors~\cite{Barausse:2020rsu, Perkins:2020tra} with improved sensitivity will allow us to listen to the Universe and decipher its physics better.

While GR has passed numerous astrophysical and solar system tests~\cite{Will2014,Stairs2003,Wex:2020ald,Yunes:2013dva,Will2014,Yagi:2016jml,Berti:2018cxi,Nair:2019iur,Berti:2018vdi} 
, several theoretical and observational issues remain. On the theoretical side, the existence of spacelike and timelike singularities and the hard-coded nature of locality in GR begs for a quantum completion of Einstein's classical theory that may resolve the BH information paradox \cite{PhysRevD.14.2460, Hawking:2015qqa} and allow for quantum entanglement even in the presence of horizons.
On the observational side, the matter-antimatter asymmetry of the Universe, its late-time acceleration~\cite{late_time_acceleration_01, late_time_acceleration_02} and galaxy rotation curves~\cite{rotation_curve_01, rotation_curve_02} require that GR be completed with additional parity-violating physics (that satisfy the Sakharov conditions~\cite{Sakharov:1967dj, Petraki:2013wwa, Gell-Mann:1991kdm, Alexander:2004us}), an ``unnaturally'' small cosmological constant~\cite{Nojiri:2006ri, Tsujikawa:2010zza} and a dark matter particle~\cite{peccei1977constraints, weinberg1978new, wilczek1978problem, Roszkowski:2017nbc} yet to be observed through direct detection particle experiments.   
These issues have inspired many modified gravity theories, such as Einstein-dilaton-Gauss-Bonnet gravity~\cite{EdGB_01, EdGB_02, EdGB_03, EdGB_04}, dynamical Chern-Simons gravity~\cite{dCS_01, dCS_02, dCS_03}, Einstein-aether theory~\cite{Eling:2004dk, Jacobson:2007fh, Campista:2018gfi, Haghani:2014ita}, Horndeski and beyond Horndeski gravity~\cite{Horndeski:1974wa, Kobayashi:2019hrl, Jana:2020vov}. 
In these modified theories, BHs still exist but they need not be described by their GR counterparts, instead acquiring certain modifications that may render them more generic (e.g.~of Petrov type I instead of D~\cite{Kanti:1995vq, Kanti:2004nr, Owen:2021eez}). As a result of the modified field equations and the non-GR corrections to BHs in these theories, their QNM spectra 
can be quite different than that predicted in GR~\cite{QNM_dCS_01, QNM_dCS_02, QNM_dCS_03, QNM_dCS_04, QNM_DHOST, QNM_EdGB_01, QNM_EdGB_02, QNM_EdGB_03,Molina:2010fb}, in principle allowing for new tests with GWs~\cite{NHT_Test_01, NHT_Test_02, NHT_Test_03, NHT_Test_04, Cheung_01, ringdown_test_01, ringdown_test_02, pyRing_01, pyRing_02, pyRing_03, pyRing_04, pyRing_05, pyRing_06, Chung_01, Chung_02, ringdown_test_03, Cardoso:2019rvt}. 

Ringdown GW tests of modified gravity, however, are hindered by the intrinsic difficulty in the computation of the gravitational QNM frequencies of rotating BHs in modified theories. 
In principle, the BH QNM frequencies can be computed by solving the linearized field equations in that theory, derived by expanding the field equations to first order in metric perturbations.
For a nonspinning BH background, the linearized field equations are a complicated set of coupled, partial differential equations, which one decouples to find \textit{master equations} for its propagating degrees of freedom through the use of special (Regee-Wheeler~\cite{Regge:PhysRev.108.1063} and Zerilli-Moncrief~\cite{Zerilli:1971wd, Moncrief:1974am}) master functions. 
For a rotating BH, the linearized field equations are an extremely complicated set of coupled, partial differential equations, which nobody has yet been able to decouple into master equations when working directly with metric perturbations \footnote{Nonetheless, the authors do note that numerical methods have been explored for solving the inhomogeneous coupled linearized equations with source for the perturbed Schwarzschild \cite{Barack:2005nr} and Kerr metric \cite{Dolan:2012jg}. }. 
Instead, for rotating BHs one can work with curvature perturbations through the Newman-Penrose (NP) formalism~\cite{Newman:1961qr} (in which the field equations are cast in terms of spinor coefficients, the Weyl scalars and differential operators) to derive a master function for these curvature perturbations. 
In this way, the NP formalism allows one to derive the Teukolsky master equation (i.e.~a separable wave equation for the NP scalars that represent propagating degrees of freedom), provided the rotating BH background is of Petrov-type D and the field equation is Einstein's~\cite{Teukolsky_01_PRL, Teukolsky:1973ha, Press:1973zz, Teukolsky:1974yv}. 
If the theory is not Einstein's, or if the BH is not of Petrov-type D, then there is no guarantee that one can decouple the field equations linearized in curvature perturbations through the NP formalism\footnote{We note that, in parallel with this work, recent progress has been made to extend the derivation of the Teukolsky equation to beyond-GR BHs\cite{Modified_TE_01, Modified_TE_02} by working to leading-order in GR deviations within an effective field theory treatment.}. 

This difficulty motivates us to explore new methods to compute the gravitational QNM frequencies of BH spacetimes. One necessary criterion that these new methods must satisfy is robustness and accuracy, which we can only assess by implementing them first within GR and comparing results to known gravitational QNM frequencies of Schwarzschild and Kerr BHs~\cite{Leaver:1985ax}. This is the main focus of this paper, focusing here on Schwarzschild BHs, a necessary step before tackling the Kerr case. One can attempt to construct many new methods that satisfy the above criteria, but one that has shown some promise in the past few decades is spectral methods. Spectral decomposition can be an effective method to handle complicated linearized field equations, as shown in \cite{Jansen:2017oag, Langlois:2021aji, Langlois:2021xzq, Spectral_03, Spectral_04, Spectral_05, Spectral_06, Spectral_07}. 
Using the completeness and orthonormal properties of certain special functions, like the Chebyshev polynomials and the Legendre polynomials, we can express any piecewise continuous function as a linear combination of these special functions.
The metric perturbations and the coefficient functions of the linearized field equations are at least $C^1$ outside the horizon, so we can accurately approximate them by using a finite number of spectral bases, which simplifies the calculation of QNM frequencies. 

Previous works have considered the use of spectral or pseudospectral collocation methods to study BH perturbations. 
These studies transformed various BH perturbation problems into (quadratic) eigenvalue problems via spectral decompositions in different ways and for different scenarios. 
One class of such studies focused on scalar and electromagnetic perturbations of BHs using spectral decompositions (e.g. \cite{Jansen:2017oag, Eperon:2019viw, Dias:2018ufh}). 
Another class of studies used the NP formalism to spectrally decompose the perturbed NP scalars and the NP equations (some with~\cite{Ripley:2022ypi, Cardoso:2013pza} and some without decoupling them~\cite{Ripley:2020xby, Dias:2015wqa, Dias:2021yju, Dias:2022oqm}).
A third class of studies used spectral or pseudospectral collocation methods to study the QNM frequencies of spherically symmetric BHs (e.g.~\cite{Jansen:2017oag, Langlois:2021aji, Langlois:2021xzq}). 
These studies solved the linearized field equations directly, through separation of variables with spherical harmonics (focusing on the zero magnetic number case) and a spectral decomposition of the radial sector.
A final class of studies explored spectral decompositions of metric perturbations and the linearized field equations without decoupling (e.g.~\cite{Dias:2014eua}). 
The spectral methods in such studies focused on the QNMs related to ultraspinning and bar-mode instabilities of higher-dimensional (Myers-Perry) BHs, and worked with scalar- and vector-mode perturbations separately~\cite{Dias:2014eua}\footnote{Spectral or pseudospectral collocation methods have also been used to study BH metric perturbations by transforming the linearized field equations into an eigenvalue problem (e.g.~\cite{Spectral_03, Spectral_04, Spectral_05, Santos:2015iua}). These studies, however, focused on BH thermodynamical properties and stability issues related to higher-dimensional BHs, which are not strictly relevant to the QNM frequencies of the ringdown phase of four-dimensional BHs.}.

Building on the work of~\cite{Jansen:2017oag, Langlois:2021aji, Langlois:2021xzq}, the goal of this paper is to develop a powerful, adaptable and extendable spectral method to study the QNMs that are likely to be measured by actual GW detectors in the near future.  
In particular, our spectral method works simultaneously with different sectors (scalar, vector and tensor) of the metric perturbations and with the linearized field equations of four-dimensional BHs, without decoupling the latter into master equations.
We begin by deriving the linearized Einstein equations that govern the metric perturbations of a Schwarzschild BH in the Regge-Wheeler gauge (Sec.~\ref{sec:EFEs}). 
We then use a product decomposition of the metric tensor into radial and angular functions, together with a spectral decomposition (of the angular sector in terms of associate Legendre polynomials) to turn the system of partial differential equations into a system of ordinary differential equations.  
By solving this system of ordinary differential equations asymptotically at spatial infinity and at the event horizon, we obtain the boundary conditions that the radial functions must satisfy (Sec.~\ref{sec:asymptotic_behaviour}).
The asymptotic behavior of the radial functions allows us to construct a radial \textit{Ansatz} that corrects the asymptotic behavior through a spectral sum of Chebyshev polynomials (Sec.~\ref{sec:separation}). 

The full spectral decomposition transforms the linearized Einstein equations into a system of linear \textit{algebraic} equations, whose generalized eigenvalues contain the QNM frequencies of the Schwarzschild BH. 
We compute these QNM frequencies numerically by solving for the generalized eigenvalues and we devise specific procedures to identify which generalized eigenvalues correspond to which QNM frequencies. 
We show that the reconstruction of the metric functions through this spectral decomposition is actually an asymptotic series by calculating its optimal truncation order (Sec.~\ref{sec:extraction}). 
We find that typically keeping 25 basis functions in the Chebyshev and the Legendre sectors suffices to identify six QNM frequencies, two of which correspond to fundamental modes, two to the first overtones and 2 to the second overtones. 
We also find that these QNM frequencies can be calculated fast and accurately, with relative fractional errors of $\leq 10^{-10} - 10^{-8}$ for the fundamental modes, $\leq 10^{-7} - 10^{-6}$ for their first overtones, and $\leq 10^{-7} - 10^{-4}$ for their second overtones. 

We conclude by analyzing the robustness of our spectral method (Sec.~\ref{sec:robustness}). 
We first check that our QNM frequency calculations are independent of the order ($m$) of the associated Legendre polynomial basis, an important feature of gravitational perturbation of spherically symmetric BHs. 
We then check that our QNM calculations are independent of the choice of radial scaling we choose in the \textit{Ansatz} for the radial function, further indicating the robustness of the spectral method. 
Finally, we check that the calculation of QNM frequencies is approximately insensitive to the set of 6 components of the linearized Einstein equations that we choose to solve for the six metric perturbation functions.  
This flexibility allows us to select the set of equations that is most convenient and to cross-check our results. Moreover, our approach allows us to better understand how different components of the metric perturbations oscillate, without having to rely on metric reconstruction or a specific set of components of the linearized equations. 

The work presented here is yet another avenue to calculate QNMs of perturbed BHs, but it is very promising and interesting for the following reasons. First, since we work with the metric perturbations directly, there is never a need to decouple the field equations and find master functions and equations. This is important because such a decoupling can be extremely complicated in modified theories of gravity, especially when the BH background is spinning and not of Petrov type D. Moreover, since we work with the metric perturbations directly, we automatically find solutions for all components of the metric itself without needing any further metric reconstruction. This could be useful when doing second-order BH perturbation theory \cite{Ripley:2020xby, Loutrel_Ripley_Giorgi_Pretorius_2020} and self-force calculations \cite{Barack:2009ux, Toomani:2021jlo}, which typically require metric reconstruction. Finally, the method presented here is fast, computationally efficient, accurate, robust and able to obtain QNM frequencies of not just the fundamental modes, but also of its overtones with similar speed, efficiency, accuracy, and robustness. 
This is important because, while some methods, such as \cite{Leaver:1985ax, Cook:2014cta}, can be used to estimate the QNM frequencies of higher overtones very precisely, the calculation of the higher-overtone frequencies can sometimes be noisy and not as accurate as that of the fundamental model using some other methods, such as direct numerical integration. Section~\ref{sec:conclusion} will further elaborate all of these features further and possible extensions of our work. 

Henceforth, we assume the following conventions: 
$x^{\mu} = (x^0, x^1, x^2, x^3) = (t, r, \chi, \phi)$, where $\chi = \cos \theta$ and $\theta$ is the azimuthal angle;
the signature of the metric tensor is $(-, +, +, +)$;
gravitational QNMs are labeled in the form of $n l m$ or $(n, l, m)$, where $n$ is the principal mode number, $l$ is the azimuthal mode number and $m$ is the magnetic mode number of the QNMs;
Greek letters in index lists stand for spacetime coordinates;
Greek letters in curly braces $\{ \mu \nu \}$ denote the collection of the $\mu \nu$ components of the perturbed Einstein equations, $G^{(1)}_{\mu \nu} = 0$. 
For example, $\{ tr, t \chi, t \phi, rr, r \chi, r \phi \} $ stands for $\{ G^{(1)}_{tr} = 0, G^{(1)}_{t \chi} = 0, ..., G^{(1)}_{r \phi}  = 0 \} $. 
For the convenience of the reader, we have presented a list of all definitions and symbols in Appendix~\ref{sec:Appendix_A}.

\section{Linearized Einstein field equations about a Schwarzschild black hole background}
\label{sec:EFEs}

In this section, we discuss our representation of the background Schwarzschild spacetime, present the linearized Einstein field equations for a perturbed Schwarzschild BH, and then conclude with a quick description of the spectral decomposition of the metric perturbations.

\subsection{Background spacetime, metric perturbation and the linearized Einstein equations}
\label{sec:metpert}

The solution to the vacuum Einstein equation $G_{\mu \nu} = 0$ that represents a stationary and spherically symmetric (nonspinning) BH is the Schwarzschild metrics $g_{\mu \nu}^{(0)}$. The line element associated with this metric can be written in Schwarzschild coordinates as 
\begin{align}\label{eq:metric}
ds^2_{(0)} &= g_{\mu \nu}^{(0)} dx^{\mu}  dx^{\nu} \nonumber \\
& =  -f(r) dt^2 +  \frac{dr^2}{f(r)} + \frac{r^2}{1 - \chi^2} d \chi^2 \nonumber \\
& \quad + r^2 \left( 1 - \chi^2 \right) d\phi^2 \,,
\end{align}
where $M$ is the BH mass, $\chi \equiv \cos \theta$ with $\theta$ the polar angle, $\phi$ is the azimuthal angle and 
\be\label{eq:Schwarzschild_fn}
f(r) = 1 - \frac{2M}{r} \,
\ee
is the so-called Schwarzschild factor. For a Schwarzschild BH in these coordinates, the event horizon is located at $r_{\h} = 2M$.

We now consider linear perturbations of the metric tensor, such that
\be \label{eq:metpert}
g_{\mu\nu} = g_{\mu\nu}^{(0)} + \epsilon \; h_{\mu\nu} \,,
\ee
where $g_{\mu\nu}^{(0)}$ is the background metric of Equation~\eqref{eq:metric}, $h_{\mu\nu}$ is the metric perturbation, and $\epsilon$ is a bookkeeping parameter for the perturbations. The metric perturbation is a function of spacetime coordinates and it can be decomposed into temporal, radial and angular components. Under a parity transformation (i.e., the simultaneous shifts $\theta \to \pi - \theta$ and $\phi \to \phi + \pi$) these components can be classified into odd (or ``axial'') and even (or ``polar'') sectors, depending on whether they pick up a factor of $(-1)^{\ell+1}$ or $(-1)^\ell$ respectively.
This allows us to decompose $h_{\mu\nu}$ as~\cite{Regge:PhysRev.108.1063, Berti_02, Zerilli:1971wd, Moncrief:1974am}
\be
h_{\mu\nu} (t,r,\chi,\phi) =  h^{\rm odd}_{\mu\nu} (t,r,\chi,\phi) + h^{\rm even}_{\mu\nu} (t,r,\chi,\phi) \,,
\ee
where\footnote{Our choice of signs for $h_3$ and $h_4$ is different from that in some of the literature, such as \cite{Zerilli_even}.}
\begin{widetext}
\begin{subequations} \label{eq:metpert}
\be \label{eq:odd}
	 h^{\rm odd}_{\mu\nu} = e^{im\phi-i\omega t}
\begin{pmatrix}
    0 & 0 & -im(1-\chi^2)^{-1} h_5(r,\chi) &  (1-\chi^2) \partial_\chi h_5(r,\chi) \\
    * & 0 & -im(1-\chi^2)^{-1} h_6(r,\chi) &  (1-\chi^2) \partial_\chi h_6(r,\chi) \\
	* & * & 0 & 0  \\
	* & * & * & 0
\end{pmatrix}
\,,
\ee
and
\be \label{eq:even}
h^{\rm even}_{\mu\nu} = - e^{im\phi-i\omega t}
\begin{pmatrix}
	f(r)h_1(r,\chi) & h_2(r,\chi) & 0 &  0 \\
	* & \frac{1}{f(r)} h_3(r,\chi) & 0 &  0 \\
	* & * & r^2 (1-\chi^2)^{-1} h_4(r,\chi) & 0  \\
	* & * & *  & r^2 (1-\chi^2) h_4(r,\chi)
\end{pmatrix}
\,,
\ee
\end{subequations}
\end{widetext}
and where we have made use of the Regge-Wheeler gauge~\cite{Regge:PhysRev.108.1063, Berti_02}.  
We have also assumed that both sectors depend on the same QNM frequency because both the axial and polar perturbations that are purely ingoing at the event horizon and outgoing at spatial infinity depend on the same complex QNM frequencies in GR, a manifestation of \textit{isospectrality}. If one were to generalize this method to beyond-GR theories that break isospectrality, then the above assumption may have to be relaxed.

With the \textit{Ansatz} defined, we can now find the system of equations that the metric perturbations $h_i(r,\chi)$ $\forall i \in (1,6)$ must satisfy. 
Unlike in the case of early studies in BH perturbations by Regge and Wheeler~\cite{Regge:PhysRev.108.1063}, Zerilli~\cite{Zerilli:1971wd} and Moncrief~\cite{Moncrief:1974am}, we do not treat the odd and even perturbations separately. Considering them simultaneously will allow us, in the future, to extend the spectral approach to QNMs of Kerr BHs, where these two parities cannot be separately studied easily\footnote{Nonetheless, the metric perturbations of the Kerr black hole of these two parities can be constructed using the procedures described in \cite{PhysRevD.11.2042} based on the Teukolsky equation. }.
Substituting Equation~\eqref{eq:metpert} into the vacuum Einstein equation, one finds a system of ten coupled, partial differential equations to solve for the six unknown functions $h_i(r,\chi)$. Only six of these equations, however, are independent of each other, so the remaining four can be eliminated by the use of perturbed Bianchi identities. In this paper, we will mainly focus on solving the $\{ tr, t \chi, t \phi, rr, r \chi, r \phi \} $ components, because we found empirically that this system is the most convenient to work with. In Sec.~\ref{sec:other_selections} and Appendix.~\ref{sec:Diagonalization_example}, we will show that using a different set of components of the linearized Einstein equations also allows us to find the Schwarzschild QNMs.

Let us now massage the linearized Einstein equations. First, note that the components of the background metric tensor $g_{\mu\nu}^{(0)}$ in Schwarzschild coordinates, whose line element is in Equation~\eqref{eq:metric}, are rational functions of $r$ and $\chi$. 
Therefore, the coefficient functions multiplying the metric perturbations $h_i$ in the linearized Einstein equations must also be rational functions of $r$ and $\chi$, since they can only depend on background quantities and their derivatives. 
With this understanding, 
we can always express the $i$th linearized field equation\footnote{Throughout this work, when multiplied by $m$ or $\omega$, $i$ stands for $\sqrt{-1}$. 
Otherwise, $i$ stands for one of the components of the linearized Einstein equations.}, after appropriate factorization and multiplying through the common denominator, as
\begin{align}\label{eq:pertFE-1}
& \sum_{j=1}^{6} \sum_{\alpha, \beta = 0}^{\alpha + \beta \leq 3} \sum_{\gamma=0}^{2} \sum_{\delta=0}^{d_{r}} \sum_{\sigma=0}^{d_{\chi}} \mathcal{G}_{i, \gamma, \delta, \sigma, \alpha, \beta, j} \omega^\gamma r^{\delta} \chi^{\sigma} \partial_{r}^{\alpha} \partial_{\chi}^{\beta} h_j = 0 \,,
\end{align}
where $\sum_{\alpha, \beta = 0}^{\alpha + \beta \leq 3}$ is a summation starting from $\alpha=0$ and $\beta=0$ up to $\alpha+\beta=3$ for all non-negative $\alpha$ and $\beta$, while $\mathcal{G}_{i, \gamma, \delta, \sigma, \alpha, \beta, j}$ is a complex function of $M$ and $m$ only.
The constants $d_r$ and $d_\chi$ are the degree of $r$ and $\chi$ of the coefficient of a given term in the equations respectively, which depend on the specific equation we are looking at and can thus be thought of to be dependent on the summation indices $\alpha,\, \beta, \, i, \, j$. 
When factorizing each of the linearized Einstein equations to obtain the common denominator, there can be prefactors, such as some powers of $1-\chi^2$, $r$ and $r-r_\h$, which contain no metric perturbation functions and are nonzero except at $r = r_\h$, $r = \infty$ and $\chi = \pm 1$. 
Since these common factors are never zero in the computational domain (except at the boundaries), we will divide by them to simplify the equations and improve the numerical stability of the linearized Einstein equations.
Equation~\eqref{eq:pertFE-1} represents a system of coupled, two-dimensional, third-order partial differential equations. Notice that the perturbed field equations for the even perturbations are at most second order, whereas for odd perturbations, due to $\partial_\chi h_i$ for $h_i \in \{h_5,h_6\}$, the system of equations is at most third order.

\subsection{Spectral decomposition of the metric perturbations}

In this subsection, we present the spectral decomposition along the radial and angular coordinates of our metric perturbations, introduced in the previous subsection. The metric perturbation functions $h_i(r,\chi)$ that enter the linearized Einstein equations are functions of $r$ and $\chi$. 
Using separation of variables, we can write these functions through the product decomposition
\be \label{eq:separation-variable}
h_i(r,\chi) = y_i(r) \Theta_i(\chi) \,, \quad i=1,...,6 \,,
\ee
with no summation over $i$ implied, where $y_i$ are new functions of $r$ only and $\Theta_i$ are functions of $\chi$ only.  

Let us now determine the angular dependence of the metric perturbation functions. 
We express the angular dependence as a linear combination of spectral function of $\chi$. 
To determine the explicit spectral basis, we note that in general, the angular dependence of metric perturbations can be expressed in terms of scalar, vector and tensor spherical harmonics \cite{Pani:2013pma, Berti_02, ColemanMiller:2021lky}, whose $\chi$-part is the associate Legendre polynomials of $\chi$. 
This is also the spectral function of $\chi$ used in the original Regge-Wheeler \cite{Regge:PhysRev.108.1063} and Zerilli-Moncrief calculations \cite{Zerilli:1971wd,Moncrief:1974am}. 
Taking all these into account, 
we represent the $\chi$ dependence using associated Legendre polynomials $P_\ell^m(\chi)$ of degree and order\footnote{Though $l$, the azimuthal number that labels QNMs, and $\ell$, the degree of the associated Legendre polynomials in the product decomposition of the metric perturbation functions, are the same for a Schwarzschild BH background, this is not necessarily the same in general, which is why we use different symbols for them here.} $(\ell,m)$, namely
\be \label{eq:Thetai-def}
\Theta_i(\chi) = \sum_{\ell = |m|}^\infty a_{i,\ell} \;  P_\ell^{|m|}(\chi) \,.
\ee

Absorbing the $a_{i,\ell}$ coefficients into the $y_i$ functions via $y_i^\ell(r) = a_{i,\ell} y_i(r)$, we then have
\be \label{eq:separation-variable-2}
h_i(r,\chi) = \sum_{\ell = |m|}^\infty y_i^\ell(r) P_\ell^{|m|}(\chi) \,, \quad i=1,..,6 \,.
\ee
In practice, only a finite number of associated Legendre polynomials need to be included in our approximations, so let $\mathcal{N}_{\chi}$ represent the maximum number of terms kept in these sums. In principle, different metric perturbation functions (i.e.~different $h_i$) could be represented by a different number of terms in the sum (i.e.~$\mathcal{N}_{\chi}$ could be different for different $h_i$ functions), but to maximize the symmetry of the spectral representation, we choose the same $\mathcal{N}_{\chi}$ for all $i$.

With the representation of the angular sector determined, let us now discuss the radial sector. Using the above product decomposition of Equation~\eqref{eq:separation-variable-2} in the left-hand side of Equation~\eqref{eq:pertFE-1}, we can rewrite any component of the linearized Einstein equation as
\begin{align}\label{eq:elliptic_eqn_v}
& \mathcal{G}_{i, \gamma, \delta, \sigma, \alpha, \beta, j} \omega^\gamma r^{\delta} \chi^{\sigma} \partial_{r}^{\alpha} \partial_{\chi}^{\beta} \left\{ \sum_{\ell = |m|}^{\mathcal{N}_{\chi}+|m|} y_j^{\ell}(r) P^{|m|}_{\ell}(\chi) \right\}  \nonumber \\
& = \sum_{\ell=|m|}^{\mathcal{N}_{\chi}+|m|} H_i^{\ell}(r) P^{|m|}_{\ell}(\chi)\, , 
\end{align}
where this equation defines the functions $H_i^\ell(r)$, and the repeated indices in the left-hand side of Equation~\eqref{eq:elliptic_eqn_v} implicitly represent the summations used in Equation~\eqref{eq:pertFE-1}. Since the linearized Einstein equations must be satisfied, Equation~\eqref{eq:elliptic_eqn_v} implies that 
\begin{equation} \label{eq:ODE_system}
H_i^{\ell}(r) = 0 
\end{equation}
for $\ell = |m|, |m|+1, ..., \mathcal{N}_{\chi} + |m|$ and $i=\{1,6\}$. 

Let us now derive an expression for the $H_i^\ell(r)$ expressions through the use of the orthogonality properties of the associated Legendre polynomials. Multiplying Equation~\eqref{eq:elliptic_eqn_v} by another associated Legendre polynomial of different degree and integrating over $\chi$, we find
\begin{equation}\label{eq:H_ell}
\begin{split}
& H_i^{\ell}(r) = \mathcal{G}_{i, \gamma, \delta, \sigma, \alpha, \beta, j} \omega^\gamma r^{\delta} \mathcal{I}^{\ell, \sigma, \alpha, \beta}_{j} \,, \\
& \mathcal{I}^{\ell, \sigma, \alpha, \beta}_{j} = \mathcal{N}_{\ell, m} \sum_{\ell'=|m|}^{\mathcal{N}_{\chi}+|m|} \partial_{r}^{\alpha} y_j^{\ell'}  \int_{-1}^{+1} d \chi P_{\ell}^{|m|} (\chi) \chi^{\sigma} \partial_{\chi}^{\beta} P_{\ell '}^{|m|} (\chi)\,, \\
\end{split} 
\end{equation}
where, again, the repeated indices represent the summations used in Equation~\eqref{eq:pertFE-1}, and
\begin{equation}
\mathcal{N}_{\ell, m} = 
(2 \ell+1) \frac{(\ell-m)!}{(\ell+m)!} \,. 
\end{equation}
Equation~\eqref{eq:ODE_system} then becomes 
\begin{equation}
\mathcal{G}_{i, \gamma, \delta, \sigma, \alpha, \beta, j} \omega^\gamma r^{\delta} \mathcal{I}^{\ell, \sigma, \alpha, \beta}_{j} = 0\,,
\end{equation}
which can be thought of as a coupled system of ordinary differential equations for the $y_i^\ell$ radial functions. 

Let us now convert this coupled system of ordinary differential equations into first-order form. First, we observe that the linearized Einstein equations can contain at most second-order radial derivatives of $y_i^{\ell}$; although the $\alpha$ sum in Equation~\eqref{eq:pertFE-1} ranges up to $\alpha + \beta \leq 3$, in practice when $\alpha + \beta = 3$ then $(\alpha, \beta) = (0, 3), (1, 2)$ or $(2, 1)$, so $\alpha =2$ at most.  
To convert this system of ordinary differential equations to first-order form, we now introduce the following auxiliary fields
\begin{equation}
Y_{i}^{\ell} \equiv \frac{d y_{i}^{\ell}}{dr} \,,
\end{equation}
where again $\ell = |m|, |m|+1, ...., \mathcal{N}_{\chi}+|m|$. Let us now promote these auxiliary fields to free fields and define the collection of all fields $\textbf{y}$ through the shortcut notation  $\textbf{y} = \{ y_i^\ell\} \cup \{Y_i^\ell\}$, or more explicitly,
\begin{equation}
\begin{split}
\textbf{y}=(& y_1^{|m|}, y_1^{|m|+1}, ..., y_1^{|m|+\mathcal{N}_{\chi}}, \\
& ..., \\
& y_6^{|m|}, y_6^{|m|+1}, ..., y_6^{|m|+\mathcal{N}_{\chi}}, \\
& Y_{1}^{|m|}, Y_{1}^{|m|+1}, ..., Y_{1}^{|m|+\mathcal{N}_{\chi}}, \\
& ..., \\
& Y_{6}^{|m|}, Y_{6}^{|m|+1}, ..., Y_{6}^{|m|+\mathcal{N}_{\chi}})^{\rm T} \,.
\end{split}
\end{equation}
Therefore, the resulting first-order system of ordinary differential equations  of equations can then be written as
\begin{equation} \label{eq:ODE_system_01}
\mathbb{Q}(r) \frac{d \textbf{y}}{d r} = \mathbb{R}(r) \textbf{y},
\end{equation}
where $\mathbb{Q}(r)$ and $\mathbb{R}(r)$ are square matrices of order $\mathcal{N}_{\chi} \cdot (6+6)$, whose elements are functions of the radial coordinate $r$ only. The procedure to solve for the QNMs now reduces to solving the above equation. Before doing so, however, we will simplify this system by peeling off the asymptotic behavior of the solution near the event horizon and spatial infinity in the next section, and then absorbing it into the radial \textit{Ansatz}. 

Equation~\eqref{eq:ODE_system_01} depends on $m$ only because of the metric-perturbation \textit{Ansatz} and the spectral basis of $\chi$ that we used. 
The original calculations of Regge and Wheeler~\cite{Regge:PhysRev.108.1063}, and of Zerilli and Moncrief~\cite{Zerilli:1971wd,Moncrief:1974am}, however, lead to master equations that do not explicitly depend on $m$; this constant does appear in their metric \textit{Ansatz} but it is eliminated when they decouple the perturbed field equations and derive their master equations. 
This implies that the QNM frequencies of a perturbed Schwarzschild BH should be $m$ independent, which is physically reasonable for gravitational perturbations of a spherically symmetric background spacetime. Our equations for the QNM frequencies [Equation~\eqref{eq:ODE_system_01}], however, do depend on $m$, and this is precisely because we are not decoupling the perturbed field equations to find master equations. Such $m$ dependence, nonetheless, can be put to good use: if our numerical calculations are correct, the QNM frequencies we calculate numerically should be invariant under shifts of $m$ in Equation~\eqref{eq:ODE_system_01}, i.e.~we should be able to compute QNM frequencies for any choice of $m$ in this equation and find the same numerical answer. We apply this cross-check in Sec.~\ref{sec:m_independence} and find that our results for the QNM frequencies we calculate are indeed $m$ independent. 

\section{Study of asymptotic behavior of linearized field equations}
\label{sec:asymptotic_behaviour}

To perform a spectral decomposition of the metric perturbations defined in Sec.~\ref{eq:metpert}, we need to construct an \textit{Ansatz} for the $y_i(r)$ functions that appear in Equation~\eqref{eq:separation-variable}. This \textit{Ansatz} must satisfy the appropriate boundary conditions at the BH event horizon and at spatial infinity. In order to simplify later analysis, we will construct a global \textit{Ansatz} for $y_i(r)$ by pulling out the asymptotic behavior of the solution at the two boundaries, similar to what was done in~\cite{Langlois:2021xzq, Jansen:2017oag}. 
In this section, we present this asymptotic analysis. Readers familiar with this topic may wish to skip to Sec.~\ref{sec:asymptotic_analysis_results}, where we summarize the results of this asymptotic analysis.

\subsection{Inversion of coefficient matrix}

\begin{figure}[tp!]
\includegraphics[width=\columnwidth]{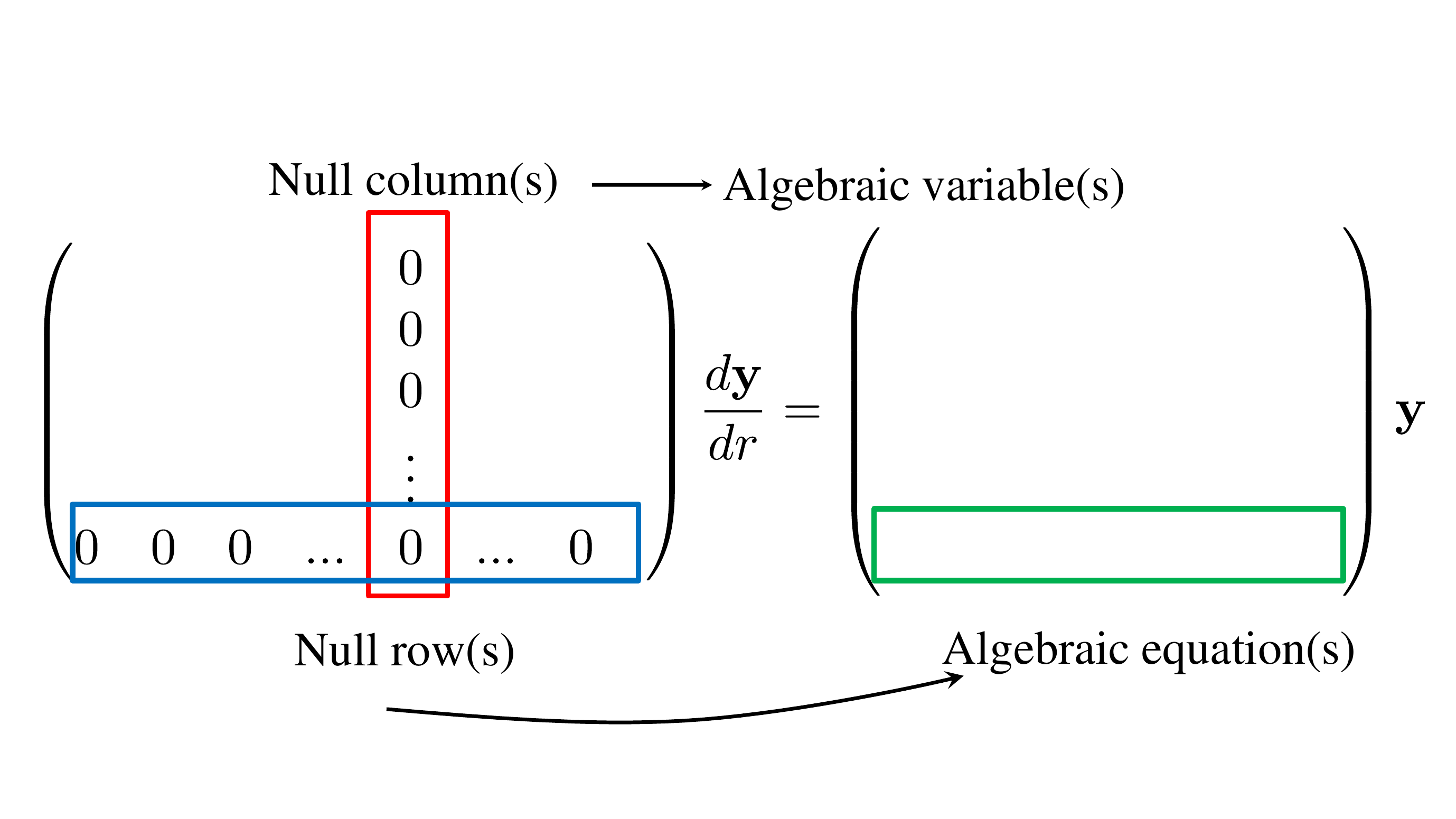}
\caption{
Schematic illustration of the structure of the system of ordinary differential equations obtained by spectral decomposition of the linearized Einstein field equations. 
The vector $\textbf{y}$ is related to the amplitude of metric perturbations at a given angular position. 
The matrix on the left-hand side is the coefficient matrix $\mathbb{Q}(r)$ of the $r$ derivatives of the system of ordinary differential equations. 
The null column (red rectangle) indicates the existence of algebraic variables, which are those whose $r$ derivatives are not contained in the differential equations. 
The null row (blue rectangle) indicates the existence of algebraic equations in different components of $\textbf{y}$ (green rectangle).
}
 \label{fig:Fig_0}  
\end{figure}

Let us begin by simplifying the first-order differential system of Equation~\eqref{eq:ODE_system_01}. Following~\cite{Langlois:2021xzq, Jansen:2017oag}, we multiply this equation by $\mathbb{Q}^{-1}(r)$ to recast it as
\begin{equation} 
\label{eq:ODE_system_02}
\frac{d \textbf{y}}{d r} = \tilde{\mathbb{M}}(r) \textbf{y} \,, 
\end{equation}
where $\tilde{\mathbb{M}}(r)$ is another square matrix of order $\mathcal{N}_{\chi} \cdot (6+6)$.

For a Schwarzschild or Kerr BH background, $\mathbb{Q}(r)$ is singular because some $y_i$ are \textit{algebraic variables}. Such variables are defined as those whose radial derivative is not present in the selected ordinary differential equations. If this is the case, then some columns and rows in $\mathbb{Q}(r)$ are null (see Fig.~\ref{fig:Fig_0} for a graphical illustration), which renders $\mathbb{Q}(r)$ noninvertible and singular. Algebraic variables can arise for two reasons. One reason is that the selected components of the linearized Einstein equations do not contain any explicit radial derivatives of some components of the metric perturbation functions. For example, the $\{tr, t \theta, t \phi, rr, r \theta, r \phi\}$ equations do not contain  $\partial_r h_{3}$ and $\partial_r^2 h_{3}$, and therefore $y_3$ and $Y_3$ are algebraic. Another reason is that, although the selected components of the linearized Einstein equations do contain radial derivatives of the $h_i$ functions, these can be eliminated by substituting in other components of the linearized Einstein equations. If the $\text{rank}(\mathbb{Q}) < \mathcal{N}_{\chi} \cdot (6+6)$, then the system of ordinary differential equations  contains $\mathcal{N}_{\chi} \cdot (6+6) - \text{rank}(\mathbb{Q})$ algebraic equations. All variables that are not algebraic (i.e.~those whose radial derivatives are present and cannot be eliminated from the system of ordinary differential equations) will be called \textit{differential variables}. 

Though $\mathbb{Q}(r)$ is singular, we can still write Equation~\eqref{eq:ODE_system_01} in the form of Equation~\eqref{eq:ODE_system_02} 
through the following procedure:
\begin{enumerate}
    \item We first identify $\mathcal{N}_{\chi} \cdot (6+6) - \text{rank}(\mathbb{Q})$ algebraic equations through elementary row operations.
    This step gives $\text{rank}(\mathbb{Q})$ differential equations and some zero rows of $\mathbb{Q}$. 
    
    \item We then identify the algebraic variable(s) of Equation~\eqref{eq:ODE_system_01} by reading the column(s) of $\mathbb{Q}(r)$ that is (are) null. 
    For, say, $N_{\rm alg}$ algebraic variables identified, we then select $N_{\rm alg}$ differential equations. This allows us to solve for the $N_{\rm alg}$ algebraic variables in terms of the differential variables and their first-order derivatives. These results can be verified to be independent of the choice of the differential equations made for these $N_{\rm alg}$ algebraic variables.
    
    Substituting these solved algebraic variables into the remaining unsolved equations leaves us with a system of $\mathcal{N}_{\chi} \cdot (6+6) - N_{\rm alg}$ differential variables. For convenience, we represent these $\mathcal{N}_{\chi} \cdot (6+6) - N_{\rm alg}$ unsolved differential variables by $\tilde{\textbf{y}}$. Therefore, the remaining unsolved equations can then be written as
    \begin{equation}\label{eq:ODE_system_AV_removed}
    \tilde{\mathbb{Q}}(r) \frac{d \tilde{\textbf{y}}}{d r} = \tilde{\mathbb{R}}(r) \tilde{\textbf{y}}, 
    \end{equation}
    where $ \tilde{\mathbb{Q}}(r) $ and $\tilde{\mathbb{R}}(r)$ are two square matrices of order $[\mathcal{N}_{\chi} \cdot (6+6) - N_{\rm alg}]$. 
    Since $N_{\rm alg}$ differential equations are eliminated,  $\text{rank}(\tilde{\mathbb{Q}}) = \text{rank}(\mathbb{Q}) - N_{\rm alg}$. 
    
    \item Some of the algebraic variables may contain $r$ derivatives, which upon substitution may convert some of the algebraic equations into differential equations. 
    Using elementary row operations, we can then identify $\mathcal{N}_{\chi} \cdot (6+6) - N_{\rm alg} - \text{rank}(\tilde{\mathbb{Q}})$ algebraic equations.
    These algebraic equations allow us to express $\mathcal{N}_{\chi} \cdot (6+6) - N_{\rm alg} - \text{rank}(\tilde{\mathbb{Q}})$ differential variables in terms of the remaining $\text{rank}(\tilde{\mathbb{Q}})$ differential variables and possibly the algebraic variables. 
    We can then eliminate another $\mathcal{N}_{\chi} \cdot (6+6) - N_{\rm alg} - \text{rank}(\tilde{\mathbb{Q}})$ equations from the system by differentiating $\mathcal{N}_{\chi} \cdot (6+6) - N_{\rm alg} - \text{rank}(\tilde{\mathbb{Q}})$ differential variables and expressing the first-order radial derivatives of $\mathcal{N}_{\chi} \cdot (6+6) - N_{\rm alg} - \text{rank}(\tilde{\mathbb{Q}})$ differential variables with the remaining differential variables and their first-order radial derivatives. 
    This leaves us with a system of $\text{rank}(\tilde{\mathbb{Q}})$ ordinary differential equations of $\text{rank}(\tilde{\mathbb{Q}})$ differential variables.
    We then denote the $\text{rank}(\tilde{\mathbb{Q}})$ differential variables with a $\text{rank}(\tilde{\mathbb{Q}})$-vector $\textbf{z}$ and the resulting system can then be expressed as
    \begin{equation}\label{eq:ODE_system_03}
    \frac{d \textbf{z}}{d r} = \mathbb{M}(r) \textbf{z}, 
    \end{equation}
    where $\mathbb{M}(r)$ is a $ \text{rank}(\tilde{\mathbb{Q}}) \times \text{rank}(\tilde{\mathbb{Q}}) $ square matrix, such that $\text{rank}(\mathbb{M}) = \text{rank}(\tilde{\mathbb{Q}}) = \text{rank}(\mathbb{Q}) - N_{\rm alg}$.
\end{enumerate} 

The procedure presented above allows us to construct a differential system without singular matrices, but in order to calculate the asymptotic behavior of the solution we must diagonalize it. We will do so through the algorithm presented in~\cite{Langlois:2021xzq}, whose essence involves asymptotically expanding $\mathbb{M}(r)$ as a matrix-valued series in (positive or negative) powers of $r$ at spatial infinity and $r-r_\h$ at the event horizon, both of which are irregular singular points. 
Explicitly, at spatial infinity, we asymptotically expand $\mathbb{M}(r)$ as
\begin{equation}\label{eq:M_r}
\mathbb{M}(r) = \sum_{k=-1}^{p_{\infty}} \mathbb{M}_{k} r^{k} + \mathcal{O} \left(\frac{1}{r^2} \right) \,.
\end{equation}
Here $p_{\infty}$ is the Poincar\'e rank of $\mathbb{M}(r)$ at spatial infinity, and $\mathbb{M}_{k}$ are matrices independent of $r$. We have also discarded terms that decay faster than $r^{-1}$ at $r = \infty$, as they have negligible effects at spatial infinity.
The asymptotic behavior at the horizon can be studied similarly by a change of variable. Defining $\epsilon = (r-r_\h)^{-1} $, where recall that $r_\h$ is the radial location of the event horizon, the differential system of Equation~\eqref{eq:ODE_system_03} is correspondingly transformed to
\begin{equation}\label{eq:ODE_system_H}
\frac{d \textbf{z}}{d \epsilon} = - \frac{1}{\epsilon^2}\mathbb{M}(\epsilon) \textbf{z}, 
\end{equation}
where $\mathbb{M}(\epsilon) = \mathbb{M}(r(\epsilon))$ is the asymptotic expansion of $\mathbb{M}(r)$ near the event horizon. 
Since the leading-order term in an $\epsilon \ll 1$ expansion of $\mathbb{M}_{\epsilon}$ may be nilpotent, we discard the terms that decay faster than $\epsilon^{-2} $ \cite{Langlois:2021xzq},  
\begin{equation}
    - \frac{1}{\epsilon^2}\mathbb{M}(\epsilon) = \sum_{k=-2}^{p_{H}} \mathbb{M}_{k} \epsilon^{k} + \mathcal{O} \left(\frac{1}{\epsilon^3} \right), 
\end{equation}
where $p_{H}$ is the Poincar\'e rank of $\- \epsilon^{-2} {M}(\epsilon)$ at the horizon. 
The algorithm in \cite{Langlois:2021xzq} can reduce the Poincar\'e rank and consecutively diagonalize every $\mathbb{M}_k$ through successive transformations. 
Once every $\mathbb{M}_r$ is diagonalized, we can immediately integrate the system of ordinary differential equations to give the asymptotic behavior of $\textbf{z}$. In Appendix~\ref{sec:Diagonalization_example}, we provide an explicit and concrete example of the implementation of the above procedure. 

Although our spectral analysis formalism essentially requires the algorithm presented in~\cite{Langlois:2021xzq, Jansen:2017oag}, unlike the previous works, our formalism does not require the decoupling between the $r$ and $\chi$ dependence of $h_i(r, \chi)$, thereby enabling us to estimate the asymptotic behavior of the metric perturbations without explicitly separating $r$ and $\chi$, and rendering the spectral method more easily applicable to non-GR BH spacetimes.
%

\subsection{Summary of asymptotic behavior}
\label{sec:asymptotic_analysis_results}
Let us now summarize the results of applying the above procedure to determine the asymptotic behavior of the metric perturbation functions.  Since we aim to study GW QNMs, we require purely ingoing boundary conditions at the horizon $r_\h$ and purely outgoing boundary conditions at spatial infinity, such that
\begin{align} \label{eq:bdd1}
h_i \propto \bigg\{ ~
\begin{matrix}
e^{-i \omega r_*}\,, &  r \to r_\h \,, \\
e^{i \omega r_*} \,, & r \to \infty \,, \\
\end{matrix}
\end{align}
where $r_*$ is the tortoise coordinate, and for a Schwarzschild BH in Schwarzschild coordinates is given by 
\be \label{eq:tortoise_coordinate}
r_* = r + 2 M \log \left(\frac{r}{2 M}-1\right)\,.
\ee
Applying the above procedure (see Appendix~\ref{sec:Diagonalization_example} for a concrete example), the asymptotic behavior of $y_i^{\ell}(r)$ that is consistent with these boundary conditions is 
\begin{align}
\label{eq:asymptotic_limits}
& \lim \limits_{r \rightarrow \infty} y^{\ell}_i (r) \sim e^{i \omega r} r^{i \omega r_\h + \rho_{\infty}^{(i)}} \sum^{\infty}_{k=0} \frac{a_{\ell k}}{r^k}, \\
\label{eq:asymptotic_limits2}
& \lim \limits_{r \rightarrow r_\h} y^{\ell}_i (r) \sim \left( r-r_\h \right)^{-i \omega r_\h - \rho_H^{(i)}} \sum_{k=0}^{\infty} b_{\ell k} (r-r_\h)^{k}, 
\end{align}
where $a_{\ell k}$ and $b_{\ell k}$ are constants and 
\begin{equation}\label{eq:rhos}
\begin{split}
\rho_H^{(i)} & = 
\begin{cases}
1, ~~ \text{for $i \neq 4\text{  and  } 5$, }\\
0, ~~ \text{otherwise,}
\end{cases} \\
\rho_{\infty}^{(i)} & = 
\begin{cases}
1, ~~ \text{for $i \neq 4$, }\\
0, ~~ \text{for $i = 4$.}
\end{cases}
\end{split} 
\end{equation}
Note that the controlling factors, the factors multiplying the series, do not depend on $\ell$.
Appendix~\ref{sec:Diagonalization_example} shows that this asymptotic behavior is consistent with that in the literature.

Let us conclude this section by stressing that Equation~\eqref{eq:asymptotic_limits} is the asymptotic expansion of the metric perturbations at spatial infinity and the event horizon \cite{Langlois:2021xzq, wasow2018asymptotic}, as we mentioned before. This is because these expansions are obtained by solving Equation~\eqref{eq:ODE_system_03} with $\mathbb{M}(r)$ replaced by its asymptotic expansion at $r = \infty$ and $r = r_\h$. Both of these expansion points are irregular and singular. One can therefore show that the approximate solutions satisfy the criteria of an asymptotic series~\cite{bender1999advanced}. 

\section{Separation of the linearized Einstein equations through a spectral decomposition}
\label{sec:separation}

In this section, we present a spectral decomposition of the linearized Einstein equations in Equation~\eqref{eq:pertFE-1} through the use of the product decomposition presented in Equation~\eqref{eq:separation-variable}, or equivalently Equation~\eqref{eq:separation-variable-2}. We begin with a refinement of the radial \textit{Ansatz}, which we then apply to the linearized Einstein equations to turn the differential system into a linear algebra problem. 

\subsection{Refinements of the radial functions}
\label{sec:specdec-r}

Since the radial functions $y_i^\ell(r)$ must satisfy the appropriate boundary conditions at the event horizon and at spatial infinity, it is convenient to pull out this asymptotic behavior in the radial \textit{Ansatz}. Let us then write
\be \label{eq:radspec}
y_i^{\ell}(r) = A^{\ell}_i(r)  u^{\ell}_i(r) \,,
\ee
where $A^{\ell}_i(r)$ is the asymptotic controlling factor of the radial function $y_i^\ell(r)$ and $u^{\ell}_i(r)$ is a correction factor that is both bounded and has trivial boundary conditions.
Using Equation~\eqref{eq:asymptotic_limits}, we are motivated to construct $A_i^{\ell}(r)$ as 
\begin{equation}\label{eq:asym_prefactor}
   A^{\ell}_i(r) = e^{i \omega r} r^{i \omega r_\h + \rho_{\infty}^{(i)}} \left( \frac{r-r_{H}}{r}\right)^{-i \omega r_\h - \rho_H^{(i)}}\,,
\end{equation}
because then $u^{\ell}_i(r) $ approaches to a constant both at the event horizon and spatial infinity. 

Since the computational domain is finite, let us introduce one more refinement of our \textit{Ansatz} through compactification. More specifically, the radial coordinate $r$ is semi-infinite, and thus, it is computationally inconvenient to perform spectral decompositions along this coordinate because the decomposition involves the evaluation of improper integrals. 
Let us then reduce the computational complexity by defining the compactified variable, $z$ \cite{Langlois:2021xzq, Jansen:2017oag}, via
\begin{equation}\label{eq:z}
z = \frac{2r_\h}{r} - 1, 
\end{equation}
so that $u_i$ is a bounded function in the finite domain $z \in [-1, +1] $. 

Finally, since $u^{\ell}_i(z)$ is finite for $z \in [-1, +1]$, we can express $u^{\ell}_i(z)$ as a linear combination of a spectral function of $z$. 
In this work, we choose to represent $u^{\ell}_i(z)$ through a Chebyshev polynomials $T_n(z)$ basis, which is uniformly convergent~\cite{CHUGUNOVA2009794}. These functions are commonly used in numerical studies of gravitational physics~\cite{Langlois:2021xzq,Jansen:2017oag,Monteiro:2009ke,Dias:2009iu,Dias:2010eu,Cardoso:2013pza,Ferrari:2007rc,Ripley:2020xby,Kidder:2000yq,Grandclement:2007sb,Boyd-1981,Ripley:2022ypi} for their computational advantages and accuracy when approximating certain functions. 

Combining all of these refinements,  Equation~\eqref{eq:separation-variable-2} with  Eqs.~\eqref{eq:radspec},~\eqref{eq:asym_prefactor} and a Chebyshev polynomial expansion takes the form 
\begin{equation}\label{eq:spectral_decoposition_factorized}
h_i (z, \chi) = A_i(z) \sum_{n=0}^{\infty} \sum_{\ell=|m|}^{\infty} v_i^{n \ell}  T_{n}(z) P^{|m|}_{\ell}(\chi). 
\end{equation}
where $ v_i^{n \ell}$ are \textit{constant} coefficients, which one can think of as the component of $u_i (r)$ along the basis of $T_{n}(z)$ and $P^{|m|}_{\ell}(\chi)$. 
Note that we have dropped the superscript $\ell$ from $A_i(r)$, as this quantity is the same for all $\ell$, and we have factorized it out of the summation. 
Equation~\eqref{eq:spectral_decoposition_factorized} gives us the full spectral decomposition of the metric perturbation along the angular coordinate $\chi$ and the compactified spatial coordinate $z$.

In practice, however, we will only include a \textit{finite} number of spectral bases in our representation of the metric perturbation functions. 
More precisely, henceforth we will set 
\begin{equation}\label{eq:spectral_decoposition_factorized_finite}
h_i (r, \chi) = A_i(r) \sum_{n=0}^{\mathcal{N}_z} \sum_{\ell=|m|}^{\mathcal{N}_{\chi}+|m|} v_i^{n \ell}  T_{n}(z) P^{|m|}_{\ell}(\chi)\,,
\end{equation}
where $\mathcal{N}_z$ and $\mathcal{N}_{\chi}$ are respectively the number of Chebyshev polynomials and associated Legendre polynomials included. In the rest of this paper, we will investigate how our calculation of the QNM frequencies is affected by choice of $\mathcal{N}_z$ and $\mathcal{N}_{\chi}$.

Before we substitute Equation~\eqref{eq:spectral_decoposition_factorized_finite} into the linearized Einstein equations, let us consider what type of series solution Equation~\eqref{eq:spectral_decoposition_factorized_finite} is.
Let us first consider this series expansion near spatial infinity. Since $z = 2 r_\h / r - 1$, the Chebyshev polynomials of $z$ are actually power series in $r^{-1}$. Thus, as $r \rightarrow \infty$, Equation~\eqref{eq:spectral_decoposition_factorized_finite} is asymptotic to 
\begin{equation}\label{eq:metric_asymptotic_expand_infty}
\begin{split}
h_i (z, \chi) \sim & e^{i \omega r} r^{i \omega r_\h + \rho_{\infty}^{(i)}} \\
& \times \sum_{\ell} \left( \tilde{a}_{0 \ell} + \frac{\tilde{a}_{1 \ell}}{r} + \frac{\tilde{a}_{2 \ell}}{r^2} + ... \right)P^{|m|}_{\ell}(\chi), 
\end{split}
\end{equation}
where $\tilde{a}_{k \ell}$ ($k = 0, 1, 2, ...$) are constants. 
If Equation~\eqref{eq:metric_asymptotic_expand_infty} is to agree with Equation~\eqref{eq:asymptotic_limits}, $\tilde{a}_{i \ell} = a_{i \ell} $. Moreover, since Equation~\eqref{eq:asymptotic_limits} is an asymptotic expansion of the metric perturbation at spatial infinity, by the uniqueness of asymptotic expansions \cite{bender1999advanced}, Equation~\eqref{eq:spectral_decoposition_factorized_finite} is also an asymptotic expansion of the metric perturbations as $r \rightarrow \infty$. 

Let us now study the behavior of the function near the horizon. 
As $r \rightarrow r_\h$, $z = {2 r_\h}/{r} - 1 \sim (r - r_\h)/{r_\h}$, the Chebyshev polynomials of $z$ are asymptotic to power series of $r-r_\h$ as $r \rightarrow r_\h$. Thus, near the event horizon, Equation~\eqref{eq:spectral_decoposition_factorized_finite} is asymptotic to  
\begin{equation}\label{eq:metric_asymptotic_expand_hor}
\begin{split}
h_i (z, \chi) \sim & e^{i \omega r} r^{i \omega r_\h + \rho_{\infty}^{(i)}} \\
& \times \sum_{\ell} \left[ \tilde{b}_{0 \ell} + \tilde{b}_{1 \ell} (r-r_\h) + ... \right] P^{|m|}_{\ell}(\chi), 
\end{split}
\end{equation}
where $\tilde{b}_{i \ell}$ are constants. If Equation~\eqref{eq:metric_asymptotic_expand_hor} is to agree with Equation~\eqref{eq:asymptotic_limits2}, then $\tilde{b}_{i \ell} = {b}_{i \ell}$. Therefore, applying the same uniqueness argument presented above, Equation~\eqref{eq:spectral_decoposition_factorized_finite} is also an asymptotic expansion of the metric perturbations as $r \rightarrow r_\h$. 
In other words, even though, by itself, the series$ \sum_{n} v_i^{n \ell} T_{n}(z) $ represents a continuous function that can be approximated by the Chebyshev polynomials with polynomial convergence \cite{CHUGUNOVA2009794}, as written in Equation~\eqref{eq:spectral_decoposition_factorized_finite}, the entire series behaves like an asymptotic one near the irregular singular points of the domain, due to the asymptotic nature of the controlling factor $A_i(r)$.

\subsection{The linearized Einstein equations as a linear algebraic eigenvalue problem}
\label{sec:evp}

Let us now use the spectral decomposition of the metric perturbation functions of Equation~\eqref{eq:spectral_decoposition_factorized} in the linearized Einstein equations to transform the latter into a system of linear algebraic equations. First, we note that the first or second radial derivatives of the asymptotic controlling factor are proportional to the product of a rational function of $r$ and the controlling factor itself.
Therefore, on substituting Equation~\eqref{eq:spectral_decoposition_factorized_finite} into the linearized Einstein equations, we can factorize the partial differential equations as
\begin{equation}\label{eq:system_3}
\begin{split}
& \sum_{j=1}^{6} \sum_{\alpha, \beta = 0}^{\alpha + \beta \leq 3} \sum_{\gamma=0}^{2} \sum_{\delta=0}^{d_{z}} \sum_{\sigma=0}^{d_{\chi}} \mathcal{K}_{i, \gamma, \delta, \sigma, \alpha, \beta, j} \omega^\gamma z^{\delta} \chi^{\sigma} \\
& \quad \quad \quad \times \partial_{z}^{\alpha} \partial_{\chi}^{\beta} \Bigg\{\sum_{n=0}^{\mathcal{N}_z} \sum_{\ell=|m|}^{\mathcal{N}_{\chi}+|m|} v_j^{n \ell} T_{n}(z) P^{|m|}_{\ell}(\chi) \Bigg\} = 0 \,.
\end{split}
\end{equation}
Here $d_z$ and $d_{\chi}$ are the degree of $z$ and $\chi$ of the coefficient of the partial derivative $\partial_{z}^{\alpha} \partial_{\chi}^{\beta} \{...\} $ in the equations respectively, while $\mathcal{K}_{i, \alpha, \beta, \gamma, \delta, \sigma, j}$ is a complex number that depends on $M$ and $m$.
As Equation~\eqref{eq:system_3} now involves only ordinary derivatives of the spectral functions with respect to the respective coordinates, we make use of their defining equations to factor and simplify Equation~\eqref{eq:system_3}, namely
\begin{equation}
\begin{split}
\frac{d^2 T_n}{d z^2} & = \frac{1}{1-z^2} \left( z \frac{d T_n}{dz} - n^2 T_n \right), \\
\frac{d^2 P_{\ell}^{|m|}}{d \chi^2} & = \frac{1}{1-\chi^2} \Big( 2 \chi \frac{d P_{\ell}^{|m|}}{d \chi} - \ell(\ell+1) P_{\ell}^{|m|} \\
& \quad \quad \quad \quad \quad - \frac{m^2}{1-\chi^2} P_{\ell}^{|m|} \Big). 
\end{split}
\end{equation}
These equations allow us to pull out more factors of $1-\chi^2, 1-z$ or $1+z$, further simplifying Equation~\eqref{eq:system_3}.  

To simplify our notation, we now rewrite the left-hand side of Equation~\eqref{eq:system_3} in terms of the spectral functions as 
\begin{equation}\label{eq:elliptic_eqn_v2}
\sum_{n=0}^{\mathcal{N}_z} \sum_{\ell=|m|}^{\mathcal{N}_{\chi}+|m|} w_i^{n \ell} T_{n}(z) P^{|m|}_{\ell}(\chi) = 0\,,
\end{equation}
where $w_i^{n \ell}$ is hiding much of the complexity of Equation~\eqref{eq:system_3}.
The orthogonality of $T_{n}(z) P^{|m|}_{\ell}(\chi) $ implies that $ w_i^{n \ell} = 0 $ for every $i, n$ and $\ell$.
Comparing Equation~\eqref{eq:system_3} and Equation~\eqref{eq:elliptic_eqn_v}, we can relate $w_i^{n \ell}$ to $v_i^{n \ell}$ by a linear combination, 
\begin{equation}
\label{eq:pertFE-2}
\begin{split}
w_i^{n \ell} = \sum_{j=1}^{6} \sum_{n'=0}^{\mathcal{N}_z} \sum_{\ell'=|m|}^{\mathcal{N}_{\chi}+|m|} \left[ \mathbb{D}_{n \ell, n' \ell'}(\omega) \right]_{ij} v_j^{n' \ell'} = 0, 
\end{split}
\end{equation}
where $\mathbb{D}_{n \ell, n' \ell'}(\omega)$ are quadratic matrix polynomials of $\omega$,
\begin{equation}\label{eq:quadratic_matrix_polynomials}
\mathbb{D}_{n \ell, n' \ell'}(\omega) = \sum_{\gamma = 0}^{2} \mathbb{D}_{n \ell, n' \ell', \gamma} \omega^{\gamma}, 
\end{equation}
and $ \mathbb{D}_{n \ell, n' \ell', \gamma} $ are constant $6 \times 6$ matrices, whose $ij$ th element is given by 
\begin{equation}
\begin{split}
& \left[ \mathbb{D}_{n \ell, n' \ell', 0} \right]_{ij} \\
& = \mathcal{N} \int_{-1}^{+1} dz \int_{-1}^{+1} d \chi (1-z^2)^{-\frac{1}{2}} T_{n}(z) P^{|m|}_{\ell}(\chi) \\
& \quad \quad \quad \quad \times \mathcal{K}_{i, 0, \delta, \sigma, \alpha, \beta, j} z^{\delta} \chi^{\sigma} \partial_{z}^{\alpha} \partial_{\chi}^{\beta} \left[ T_{n'}(z) P^{|m|}_{\ell'}(\chi) \right] \,, \\
& \left[ \mathbb{D}_{n \ell, n' \ell', 1} \right]_{ij} \\
& = \mathcal{N} \int_{-1}^{+1} dz \int_{-1}^{+1} d \chi (1-z^2)^{-\frac{1}{2}} T_{n}(z) P^{|m|}_{\ell}(\chi) \\
& \quad \quad \quad \quad \times \mathcal{K}_{i, 1, \delta, \sigma, \alpha, \beta, j} z^{\delta} \chi^{\sigma} \partial_{z}^{\alpha} \partial_{\chi}^{\beta} \left[ T_{n'}(z) P^{|m|}_{\ell'}(\chi) \right] \,, \\
& \left[ \mathbb{D}_{n \ell, n' \ell', 2} \right]_{ij} \\
& = \mathcal{N} \int_{-1}^{+1} dz \int_{-1}^{+1} d \chi (1-z^2)^{-\frac{1}{2}} T_{n}(z) P^{|m|}_{\ell}(\chi) \\
& \quad \quad \quad \quad \times \mathcal{K}_{i, 2, \delta, \sigma, \alpha, \beta, j} z^{\delta} \chi^{\sigma} \partial_{z}^{\alpha} \partial_{\chi}^{\beta} \left[ T_{n'}(z) P^{|m|}_{\ell'}(\chi) \right] \,. 
\end{split}
\end{equation}
Here the repeated indices implicitly represent the summation defined in Equation~\eqref{eq:pertFE-2} (except for $\gamma$), and the prefactor $\mathcal{N}$ is
\begin{equation}\label{eq:normalization_const}
\mathcal{N} = 
\begin{cases}
\frac{2 \ell+1}{\pi} \frac{(\ell-m)!}{(\ell+m)!}~~~\text{if}~~~ n \neq 0\\
\frac{2 \ell+1}{2 \pi} \frac{(\ell-m)!}{(\ell+m)!}~~~\text{if}~~~ n = 0 \,.
\end{cases}
\end{equation}

Equation~\eqref{eq:pertFE-2} can be cast into a quadratic eigenvalue problem with the QNM frequencies of the perturbed Schwarzschild BH being its generalized eigenvalues. 
To see this, we first introduce the following vector notation: 
\begin{equation}
\begin{split}
& \textbf{v}_{n \ell} = \left( v_1^{n \ell}, v_2^{n \ell}, v_3^{n \ell}, v_4^{n \ell}, v_5^{n \ell}, v_6^{n \ell} \right)^{\rm T} \,, \\
& \textbf{w}_{n \ell} = \left( w_1^{n \ell}, w_2^{n \ell}, w_3^{n \ell}, w_4^{n \ell}, w_5^{n \ell}, w_6^{n \ell} \right)^{\rm T} \,.
\end{split}
\end{equation}
Then Equation~\eqref{eq:pertFE-2} can be written as 
\begin{equation}\label{eq:vector_equations}
\textbf{w}_{n \ell} = \sum_{n'=0}^{\mathcal{N}_z} \sum_{\ell'=|m|}^{\mathcal{N}_{\chi}+|m|} \mathbb{D}_{n \ell, n' \ell'}(\omega) \textbf{v}_{n' \ell'} = 0\,,
\end{equation}
where the $\mathbb{D}_{n \ell, n' \ell'}$ matrix is now dotted into our new vector $\textbf{v}_{n' \ell'}$. 
Furthermore, let us define a vector $\textbf{v}$ and $\textbf{w}$, which respectively stores all $\textbf{v}_{n \ell}$ and $\textbf{w}_{n \ell}$, 
\begin{equation}
\begin{split}
\textbf{v} & = \left\{ \textbf{v}_{00}^{\rm T}, \textbf{v}_{01}^{\rm T}, ..., \textbf{v}_{0 \mathcal{N}_{\chi}}^{\rm T}, ..., \textbf{v}_{1 \mathcal{N}_{\chi}}^{\rm T}, ...,  \textbf{v}_{\mathcal{N}_{z} \mathcal{N}_{\chi}}^{\rm T} \right\}^{\rm T}, \\
\textbf{v}_{n \ell} &= \left( v_1^{n \ell}, v_2^{n \ell}, v_3^{n \ell}, v_4^{n \ell}, v_5^{n \ell}, v_6^{n \ell} \right)^{\rm T}. 
\end{split}
\end{equation}
and the following block matrix, 
\begin{widetext}
\begin{equation}\label{eq:augmented_matrix_02}
\begin{split}
& \tilde{\mathbb{D}} (\omega) = \\
& 
\begin{pmatrix}
\mathbb{D}_{0|m|, 0|m|} & \mathbb{D}_{0|m|, 0(1+|m|)} & ... & \mathbb{D}_{0|m|, 0  \mathcal{N}_{\chi}} & ... & \mathbb{D}_{0|m|,  1  \ell_{\rm max}} & ... &  \mathbb{D}_{0|m|, \mathcal{N}_{z}  \ell_{\rm max}}\\
\mathbb{D}_{0(1+|m|), 0|m|} & \mathbb{D}_{0(1+|m|), 0(1+|m|)} & ... & \mathbb{D}_{0(1+|m|), 0  \mathcal{N}_{\chi}} & ... & \mathbb{D}_{0(1+|m|) , 1  \ell_{\rm max}} & ... &  \mathbb{D}_{0(1+|m|), \mathcal{N}_{z}  \ell_{\rm max}}\\
... & ... & ... & ... & ... & ... & ... &  ... \\
\mathbb{D}_{0(\mathcal{N}_{\chi}+|m|), 0|m|} & \mathbb{D}_{0(\mathcal{N}_{\chi}+|m|), 0(1+|m|)} & ... & \mathbb{D}_{0(\mathcal{N}_{\chi}+|m|), 0  \mathcal{N}_{\chi}} & ... & \mathbb{D}_{0(\mathcal{N}_{\chi}+|m|), 1  \ell_{\rm max}} & ... &  \mathbb{D}_{0(\mathcal{N}_{\chi}+|m|), \mathcal{N}_{z}  \ell_{\rm max}}\\
... & ... & ... & ... & ... & ... & ... &  ... \\
\mathbb{D}_{1\ell_{\rm max}, 0|m|} & \mathbb{D}_{1\ell_{\rm max}, 0(1+|m|)} & ... & \mathbb{D}_{1\ell_{\rm max}, 0  \mathcal{N}_{\chi}} & ... & \mathbb{D}_{1\ell_{\rm max}, 1  \ell_{\rm max}} & ... &  \mathbb{D}_{1 \mathcal{N}_{\chi}, \mathcal{N}_{z}  \ell_{\rm max}}\\
... & ... & ... & ... & ... & ... & ... &  ... \\
\mathbb{D}_{\mathcal{N}_{z}  \ell_{\rm max}, 0|m|} & \mathbb{D}_{\mathcal{N}_{z}  \ell_{\rm max}, 0(1+|m|)} & ... & \mathbb{D}_{\mathcal{N}_{z}  \ell_{\rm max}, 0  \mathcal{N}_{\chi}} & ... & \mathbb{D}_{\mathcal{N}_{z}  \ell_{\rm max},  1  \ell_{\rm max}} & ... &  \mathbb{D}_{\mathcal{N}_{z}  \ell_{\rm max}, \mathcal{N}_{z}  \ell_{\rm max}}\\
\end{pmatrix}\,.
\end{split}
\end{equation}
\end{widetext}
Then, the system of linear vector equations (Equation~\eqref{eq:pertFE-2}) can be more compactly written as 
\begin{equation}\label{eq:augmented_matrix_01}
\begin{split}
\tilde{\mathbb{D}} (\omega) \textbf{v} = \left[ \tilde{\mathbb{D}}_0 + \tilde{\mathbb{D}}_1 \omega + \tilde{\mathbb{D}}_2 \omega^2 \right] \textbf{v} = \textbf{0} \,, 
\end{split}
\end{equation}
which is a quadratic eigenvalue problem. 
Since $\textbf{v} \neq \textbf{0}$ in the ringdown, $\text{det} [\tilde{\mathbb{D}}(\omega)] = 0 $ for QNM frequencies. 

Numerically solving this quadratic eigenvalue equation, however, is computationally demanding. We can improve the numerical efficiency if we define 
\begin{equation}
\textbf{x} = 
\begin{pmatrix}
& \textbf{v} \\
& \omega \textbf{v}
\end{pmatrix}, 
\end{equation}
so that the quadratic eigenvalue problem is transformed\footnote{In numerical linear algebra, such a transformation is more commonly known as ``linearization" \cite{https://doi.org/10.1002/nme.2076, 10.2307/3649752}. However, through this paper, the name ``linearization" has been reserved solely for the linearization of the Einstein equation. To avoid confusion, we call the process that casts a quadratic eigenvalue problem into a generalized eigenvalue problem a ``transformation". } into a generalized eigenvalues problem that is linear in $\omega$ \cite{Langlois:2021xzq, Jansen:2017oag, https://doi.org/10.1002/nme.2076, 10.2307/3649752}, namely 
\begin{equation}\label{eq:matrix_equation}
\begin{split}
& M_0 \textbf{x} = - \omega M_1 \textbf{x} \,,  
\;\;\; M_0 = \begin{pmatrix}
\tilde{\mathbb{D}}_0 & \tilde{\mathbb{D}}_1 \\
0 & I \\
\end{pmatrix}, 
\; M_1 = \begin{pmatrix}
0 & \tilde{\mathbb{D}}_2 \\
-I & 0 \\
\end{pmatrix}. 
\end{split}
\end{equation}
The QNM frequencies of the Schwarzschild BH are then the generalized eigenvalues of Equation~\eqref{eq:matrix_equation}. The converse, however, is not true: not every generalized eigenvalue of Equation~\eqref{eq:matrix_equation} is a QNM frequency. 
As we will see in the next section, many surplus eigenvalues, which are not physically meaningful, will emerge, but we will develop a systematic method to identify the meaningful ones.

To explicitly illustrate how one can derive Equation~\eqref{eq:augmented_matrix_01} from Equation~\eqref{eq:vector_equations}, let us consider an example with $N_z = 1$ and $N_{\chi} = 0$. In this example, the only components of $\textbf{w}_{n \ell}$ are 
\begin{equation}
\begin{split}
\textbf{w}_{0 2} & = \mathbb{D}_{0 2, 0 2}(\omega) \textbf{v}_{0 2} + \mathbb{D}_{0 2, 1 2}(\omega) \textbf{v}_{1 2} = \textbf{0}, \\
\textbf{w}_{1 2} & = \mathbb{D}_{1 2, 0 2}(\omega) \textbf{v}_{0 2} + \mathbb{D}_{1 2, 1 2}(\omega) \textbf{v}_{1 2} = \textbf{0}. 
\end{split}
\end{equation}
Hence, as a block matrix, $\tilde{\mathbb{D}} (\omega)$ can be written as 
\begin{equation}
\tilde{\mathbb{D}} (\omega) =
\begin{pmatrix}
\mathbb{D}_{0 2, 0 2}(\omega) &  \mathbb{D}_{0 2, 1 2}(\omega) \\
\mathbb{D}_{1 2, 0 2}(\omega) &  \mathbb{D}_{1 2, 1 2}(\omega). 
\end{pmatrix}
\end{equation}
Explicitly, the nonzero elements of $\tilde{\mathbb{D}} (\omega)$ are  
\begin{widetext}
\begin{align*}
\tilde{\mathbb{D}}_{1,4}(\omega)&=\frac{18}{35} \pi  (7-2 i \omega ), \quad \tilde{\mathbb{D}}_{1,6}(\omega)=-\frac{9}{35} \pi  (8 \omega +3 i), \quad \tilde{\mathbb{D}}_{1,10}(\omega)=\frac{72}{35} (\pi +i \pi  \omega ), \quad \tilde{\mathbb{D}}_{1,12}(\omega)=\frac{18}{35} \pi  (8 \omega -i),\\
\tilde{\mathbb{D}}_{2,1}(\omega)&=\frac{216 \pi }{35}, \quad \tilde{\mathbb{D}}_{2,3}(\omega)=-\frac{18}{35} \pi  (4 \omega -i), \quad \tilde{\mathbb{D}}_{2,4}(\omega)=-\frac{288}{35} i \pi  \omega, \quad 
 \tilde{\mathbb{D}}_{2,6}(\omega)=-\frac{288 \pi  \omega }{35} \quad \tilde{\mathbb{D}}_{2,7}(\omega)=\frac{108 \pi }{35}, \\
 \tilde{\mathbb{D}}_{2,9}(\omega)& =\frac{144 \pi  \omega }{35},\\
\tilde{\mathbb{D}}_{3,1}(\omega)&=\frac{288 i \pi  \omega }{35},\quad\tilde{\mathbb{D}}_{3,3}(\omega)=\frac{288 \pi  \omega }{35},,\quad\tilde{\mathbb{D}}_{3,4}(\omega)=\frac{36 \pi }{35},\quad\tilde{\mathbb{D}}_{3,6}(\omega)=-\frac{9 i \pi }{7},\quad\tilde{\mathbb{D}}_{3,10}(\omega)=\frac{18 \pi }{35},\\
\tilde{\mathbb{D}}_{3,12}(\omega) & =-\frac{18 i \pi }{35},\\
\tilde{\mathbb{D}}_{4,1}(\omega)&=\frac{9}{35} \pi  \omega  (8 \omega +i),\quad\tilde{\mathbb{D}}_{4,3}(\omega)=\frac{18 \pi  \omega }{35},\quad\tilde{\mathbb{D}}_{4,4}(\omega)=-\frac{9}{560} \pi  \left(256 \omega ^2-5 i \omega -8\right),\quad\tilde{\mathbb{D}}_{4,6}(\omega)=-\frac{27 \pi  \omega }{140},\\
\tilde{\mathbb{D}}_{4,7}(\omega)& =-\frac{18}{35} \pi  \omega  (8 \omega -i),\quad\tilde{\mathbb{D}}_{4,10}(\omega)=\frac{9}{280} (\pi +8 i \pi  \omega ),\quad\tilde{\mathbb{D}}_{4,12}(\omega)=\frac{9 \pi  \omega }{70},\\
\tilde{\mathbb{D}}_{5,2}(\omega)&=-\frac{9}{280} \pi  \left(152 \omega ^2-54 i \omega +19\right),\quad\tilde{\mathbb{D}}_{5,5}(\omega)=\frac{18}{35} \pi  \omega  (4 \omega +i),\quad\tilde{\mathbb{D}}_{5,8}(\omega)=\frac{9}{70} \pi  \left(8 \omega ^2+6 i \omega -1\right),\\
\tilde{\mathbb{D}}_{5,11}(\omega)&=-\frac{144 \pi  \omega ^2}{35},\\
\tilde{\mathbb{D}}_{6,2}(\omega)&=-\frac{72 \pi  \omega ^2}{35},\quad\tilde{\mathbb{D}}_{6,5}(\omega)=\frac{18}{35} \pi  \left(16 \omega ^2-1\right),\quad\tilde{\mathbb{D}}_{6,8}(\omega)=\frac{36}{35} \pi  \omega  (4 \omega -i),\quad\tilde{\mathbb{D}}_{6,11}(\omega)=-\frac{9 \pi }{70},\\
\tilde{\mathbb{D}}_{7,4}(\omega)&=\frac{9}{35} \pi  (9+8 i \omega ),\quad\tilde{\mathbb{D}}_{7,6}(\omega)=\frac{36}{35} \pi  (4 \omega -i),\quad\tilde{\mathbb{D}}_{7,10}(\omega)=\frac{9}{140} \pi  (31-4 i \omega ),\\
\tilde{\mathbb{D}}_{7,12}(\omega)&=-\frac{9}{140} \pi  (8 \omega +9 i),\\
\tilde{\mathbb{D}}_{8,1}(\omega)&=\frac{108 \pi }{35},\quad\tilde{\mathbb{D}}_{8,3}(\omega)=\frac{18}{35} \pi  (8 \omega -i),\quad\tilde{\mathbb{D}}_{8,7}(\omega)=\frac{108 \pi }{35},\quad\tilde{\mathbb{D}}_{8,9}(\omega)=-\frac{18 \pi  \omega }{35},\quad\tilde{\mathbb{D}}_{8,10}(\omega)=-\frac{144}{35} i \pi  \omega,\\
\tilde{\mathbb{D}}_{8,12}(\omega) &=-\frac{144 \pi  \omega }{35},\\
\tilde{\mathbb{D}}_{9,4}(\omega)&=\frac{18 \pi }{35},\quad\tilde{\mathbb{D}}_{9,6}(\omega)=-\frac{18 i \pi }{35},\quad\tilde{\mathbb{D}}_{9,7}(\omega)=\frac{144 i \pi  \omega }{35},\quad\tilde{\mathbb{D}}_{9,9}(\omega)=\frac{144 \pi  \omega }{35},\quad\tilde{\mathbb{D}}_{9,10}(\omega)=\frac{18 \pi }{35},\\
\tilde{\mathbb{D}}_{9,12}(\omega)&=-\frac{81 i \pi }{140},\\
\tilde{\mathbb{D}}_{10,1}(\omega)&=-\frac{36}{35} \pi  \omega  (4 \omega -i),\quad\tilde{\mathbb{D}}_{10,4}(\omega)=\frac{9 \pi  (21+64 i \omega )}{2240},\quad\tilde{\mathbb{D}}_{10,6}(\omega)=\frac{9}{560} \pi  (8 \omega -3 i),\\
\tilde{\mathbb{D}}_{10,7}(\omega)&=\frac{9}{140} \pi  \omega  (8 \omega +7 i),\tilde{\mathbb{D}}_{10,9}(\omega)=\frac{9 \pi  \omega }{70},\quad\tilde{\mathbb{D}}_{10,10}(\omega)=-\frac{9 \pi  \left(1024 \omega ^2-76 i \omega -25\right)}{4480},\\
\tilde{\mathbb{D}}_{10,12}(\omega)&=-\frac{9 \pi  (4 \omega +i)}{1120},\\
\tilde{\mathbb{D}}_{11,2}(\omega)&=\frac{9}{70} \pi  \left(8 \omega ^2+5 i \omega -1\right),\quad\tilde{\mathbb{D}}_{11,5}(\omega)=-\frac{18}{35} \pi  \omega  (8 \omega -i),\\
\tilde{\mathbb{D}}_{11,8}(\omega)&=-\frac{9}{560} \pi  \left(200 \omega ^2-70 i \omega +9\right), \quad \tilde{\mathbb{D}}_{11,11}(\omega)=\frac{9}{35} \pi  \omega  (2 \omega +i),\\
\tilde{\mathbb{D}}_{12,2}(\omega)&=\frac{18}{35} \pi  \omega  (8 \omega -3 i),\quad\tilde{\mathbb{D}}_{12,5}(\omega)=-\frac{9 \pi }{70},\quad\tilde{\mathbb{D}}_{12,8}(\omega)=-\frac{9}{70} \pi  \omega  (4 \omega +5 i),\quad\tilde{\mathbb{D}}_{12,11}(\omega)=\frac{9}{70} \pi  \left(32 \omega ^2-1\right).
\end{align*}
\end{widetext}
By reading the coefficient of different terms, we can read $\tilde{\mathbb{D}}_0$, $\tilde{\mathbb{D}}_1$ and $\tilde{\mathbb{D}}_2$, and we find 
\begin{widetext}
\begin{align*}
& \tilde{\mathbb{D}}_0 = 
\begin{pmatrix}
 0 & 0 & 0 & \frac{18 \pi }{5} & 0 & -\frac{27 i \pi }{35} & 0 & 0 & 0 & \frac{72 \pi
   }{35} & 0 & -\frac{18 i \pi }{35} \\
 \frac{216 \pi }{35} & 0 & \frac{18 i \pi }{35} & 0 & 0 & 0 & \frac{108 \pi }{35} & 0 & 0
   & 0 & 0 & 0 \\
 0 & 0 & 0 & \frac{36 \pi }{35} & 0 & -\frac{9 i \pi }{7} & 0 & 0 & 0 & \frac{18 \pi }{35}
   & 0 & -\frac{18 i \pi }{35} \\
 0 & 0 & 0 & \frac{9 \pi }{70} & 0 & 0 & 0 & 0 & 0 & \frac{9 \pi }{280} & 0 & 0 \\
 0 & -\frac{171 \pi }{280} & 0 & 0 & 0 & 0 & 0 & -\frac{9 \pi }{70} & 0 & 0 & 0 & 0 \\
 0 & 0 & 0 & 0 & -\frac{18 \pi }{35} & 0 & 0 & 0 & 0 & 0 & -\frac{9 \pi }{70} & 0 \\
 0 & 0 & 0 & \frac{81 \pi }{35} & 0 & -\frac{36 i \pi }{35} & 0 & 0 & 0 & \frac{279 \pi
   }{140} & 0 & -\frac{81 i \pi }{140} \\
 \frac{108 \pi }{35} & 0 & -\frac{18 i \pi }{35} & 0 & 0 & 0 & \frac{108 \pi }{35} & 0 & 0
   & 0 & 0 & 0 \\
 0 & 0 & 0 & \frac{18 \pi }{35} & 0 & -\frac{18 i \pi }{35} & 0 & 0 & 0 & \frac{18 \pi
   }{35} & 0 & -\frac{81 i \pi }{140} \\
 0 & 0 & 0 & \frac{27 \pi }{320} & 0 & -\frac{27 i \pi }{560} & 0 & 0 & 0 & \frac{45 \pi
   }{896} & 0 & -\frac{9 i \pi }{1120} \\
 0 & -\frac{9 \pi }{70} & 0 & 0 & 0 & 0 & 0 & -\frac{81 \pi }{560} & 0 & 0 & 0 & 0 \\
 0 & 0 & 0 & 0 & -\frac{9 \pi }{70} & 0 & 0 & 0 & 0 & 0 & -\frac{9 \pi }{70} & 0 \\
\end{pmatrix}, \\
& \tilde{\mathbb{D}}_1 = 
\begin{pmatrix}
0 & 0 & 0 & -\frac{36 i \pi }{35} & 0 & -\frac{72 \pi }{35} & 0 & 0 & 0 & \frac{72 i \pi
   }{35} & 0 & \frac{144 \pi }{35} \\
 0 & 0 & -\frac{72 \pi }{35} & -\frac{288 i \pi }{35} & 0 & -\frac{288 \pi }{35} & 0 & 0 &
   \frac{144 \pi }{35} & 0 & 0 & 0 \\
 \frac{288 i \pi }{35} & 0 & \frac{288 \pi }{35} & 0 & 0 & 0 & 0 & 0 & 0 & 0 & 0 & 0 \\
 \frac{9 i \pi }{35} & 0 & \frac{18 \pi }{35} & \frac{9 i \pi }{112} & 0 & -\frac{27 \pi
   }{140} & \frac{18 i \pi }{35} & 0 & 0 & \frac{9 i \pi }{35} & 0 & \frac{9 \pi }{70} \\
 0 & \frac{243 i \pi }{140} & 0 & 0 & \frac{18 i \pi }{35} & 0 & 0 & \frac{27 i \pi }{35}
   & 0 & 0 & 0 & 0 \\
 0 & 0 & 0 & 0 & 0 & 0 & 0 & -\frac{36 i \pi }{35} & 0 & 0 & 0 & 0 \\
 0 & 0 & 0 & \frac{72 i \pi }{35} & 0 & \frac{144 \pi }{35} & 0 & 0 & 0 & -\frac{9 i \pi
   }{35} & 0 & -\frac{18 \pi }{35} \\
 0 & 0 & \frac{144 \pi }{35} & 0 & 0 & 0 & 0 & 0 & -\frac{18 \pi }{35} & -\frac{144 i \pi
   }{35} & 0 & -\frac{144 \pi }{35} \\
 0 & 0 & 0 & 0 & 0 & 0 & \frac{144 i \pi }{35} & 0 & \frac{144 \pi }{35} & 0 & 0 & 0 \\
 \frac{36 i \pi }{35} & 0 & 0 & \frac{9 i \pi }{35} & 0 & \frac{9 \pi }{70} & \frac{9 i
   \pi }{20} & 0 & \frac{9 \pi }{70} & \frac{171 i \pi }{1120} & 0 & -\frac{9 \pi }{280}
   \\
 0 & \frac{9 i \pi }{14} & 0 & 0 & \frac{18 i \pi }{35} & 0 & 0 & \frac{9 i \pi }{8} & 0 &
   0 & \frac{9 i \pi }{35} & 0 \\
 0 & -\frac{54 i \pi }{35} & 0 & 0 & 0 & 0 & 0 & -\frac{9 i \pi }{14} & 0 & 0 & 0 & 0 \\
\end{pmatrix}, \\
& \tilde{\mathbb{D}}_2 = 
\begin{pmatrix}
 0 & 0 & 0 & 0 & 0 & 0 & 0 & 0 & 0 & 0 & 0 & 0 \\
 0 & 0 & 0 & 0 & 0 & 0 & 0 & 0 & 0 & 0 & 0 & 0 \\
 0 & 0 & 0 & 0 & 0 & 0 & 0 & 0 & 0 & 0 & 0 & 0 \\
 \frac{72 \pi }{35} & 0 & 0 & -\frac{144 \pi }{35} & 0 & 0 & -\frac{144 \pi }{35} & 0 & 0
   & 0 & 0 & 0 \\
 0 & -\frac{171 \pi }{35} & 0 & 0 & \frac{72 \pi }{35} & 0 & 0 & \frac{36 \pi }{35} & 0 &
   0 & -\frac{144 \pi }{35} & 0 \\
 0 & -\frac{72 \pi }{35} & 0 & 0 & \frac{288 \pi }{35} & 0 & 0 & \frac{144 \pi }{35} & 0 &
   0 & 0 & 0 \\
 0 & 0 & 0 & 0 & 0 & 0 & 0 & 0 & 0 & 0 & 0 & 0 \\
 0 & 0 & 0 & 0 & 0 & 0 & 0 & 0 & 0 & 0 & 0 & 0 \\
 0 & 0 & 0 & 0 & 0 & 0 & 0 & 0 & 0 & 0 & 0 & 0 \\
 -\frac{144 \pi }{35} & 0 & 0 & 0 & 0 & 0 & \frac{18 \pi }{35} & 0 & 0 & -\frac{72 \pi
   }{35} & 0 & 0 \\
 0 & \frac{36 \pi }{35} & 0 & 0 & -\frac{144 \pi }{35} & 0 & 0 & -\frac{45 \pi }{14} & 0 &
   0 & \frac{18 \pi }{35} & 0 \\
 0 & \frac{144 \pi }{35} & 0 & 0 & 0 & 0 & 0 & -\frac{18 \pi }{35} & 0 & 0 & \frac{144 \pi
   }{35} & 0 \\
\end{pmatrix}. 
\end{align*}
\end{widetext}
From this example, we see that $\tilde{\mathbb{D}}_0$, $\tilde{\mathbb{D}}_1$ and $\tilde{\mathbb{D}}_2$ are sparse, singular and nonsymmetric. With these matrices in hand, one can now straightforwardly calculate the generalized eigenvalues of Equation~\eqref{eq:matrix_equation}, a subset of which will represent the QNMs of a Schwarzschild BH. 

\section{Extraction of the Quasinormal Frequencies}
\label{sec:extraction}
In this section, we present our numerical analysis of the solutions to Equation~\eqref{eq:matrix_equation} for the QNM frequencies of a Schwarzschild BH. We begin with a description of the numerical setup, followed by the distribution of eigenvalues and the presentation of a method to identify the modes obtained. 

\subsection{Numerical setup}
\label{sec:setup}

To simplify our discussion, from here on we assume $\mathcal{N}_z = \mathcal{N}_{\chi} = N $ and denote the eigenvalues computed using $N\times N$ spectral functions by $\lambda(N)$. 
Therefore, $\tilde{\mathbb{D}}(\omega)$ becomes a $6 (N+1)^2$ square matrix and $M_0$ and $M_1$ are $12(N+1)^2$ square matrices. 
For a given $(m,\rho_{\infty}^{(i)},\rho_H^{(i)})$, we solve Equation~\eqref{eq:matrix_equation} for its generalized eigenvalues (from now just ``eigenvalues") using the function \texttt{Eigenvalues} of \texttt{Mathematica} with double precision; this algorithm is sufficient for our purposes because the background spacetime is spherically symmetric and the modulus of the coefficients ($\mathcal{K}_{i, \gamma, \delta, \sigma, \alpha, \beta, j}$ of Equation~\eqref{eq:system_3}) are roughly of the same order of magnitude.
We have checked that our results are not significantly affected by increasing the working precision in \texttt{Mathematica} beyond double. 
Since Schwarzschild BHs are stable, the imaginary part of their QNM frequencies is negative, so we only study the eigenvalues of the negative imaginary part and the positive real part. 

Since we are working in spherical symmetry, the QNM frequencies should be independent of the $m$ index of spherical harmonics. For concreteness, we hereafter set $m=2$ (except in Sec.~\ref{sec:m_independence}, in which we check whether our results are truly independent of $m$), with the understanding that the QNM frequencies of the Schwarzschild black hole do not depend on $m$ (e.g.~$\omega_{040}=\omega_{042}=\omega_{044}$). 

\subsection{Possible sources of inaccuracies}
\label{sec:error_sources}

Although the error in approximating a continuous function by a spectral function decreases with $N$, one should not expect that the accuracy of the QNM frequencies computed using the spectral basis will always increase with $N$. We have identified three possible sources of inaccuracies, which we list below:
\begin{enumerate}
    \item \textit{Asymptotic nature}. 
    As mentioned earlier, Equation~\eqref{eq:spectral_decoposition_factorized} is an asymptotic expansion with an asymptotic basis constructed from spectral functions. 
    Typically, asymptotic expansions diverge if a large number of terms are included in the expansion \cite{bender1999advanced}. 
    Thus, the accuracy of the QNM frequencies estimated using Equation~\eqref{eq:spectral_decoposition_factorized_finite} cannot be improved indefinitely as $N$ is increased.
    \item \textit{Numerical precision}. Any numerical calculation is always an approximation to the exact answer that is limited by the precision with which we perform the calculation. 
    Within a given precision, the accuracy of the eigenvalues computed using a spectral method can deteriorate with unsuitably many spectral functions included. 
    Nonetheless, as mentioned before, we have checked that the results of our calculations are not affected by precision error (i.e.~there are other sources of inaccuracies that dominate). 
    \item \textit{Transformation inaccuracies}. This is the error induced by transforming the quadratic eigenvalue problem (Equation~\eqref{eq:augmented_matrix_01}) into a generalized eigenvalue problem (Equation~\eqref{eq:matrix_equation}). 
    In fact, given a quadratic eigenvalue problem, there exist infinite transformations that cast the problem into a generalized eigenvalue problem. Each transformation has its own numerical sensitivity and stability issues \cite{https://doi.org/10.1002/nme.2076, 10.2307/3649752}. 
    The specific transformation used in this work is chosen following \cite{Langlois:2021xzq, Jansen:2017oag}, where it was found to be accurate for computing BH QNM frequencies. 
    But to improve the numerical condition of the matrices we work with, through this work, we scale $\tilde{\mathbb{D}_0}$ and $\tilde{\mathbb{D}_2}$ such that their two-norm is one, as proposed and used in \cite{doi:10.1137/S0895479803434914, 2022arXiv220407424K}, before calculating the generalized eigenvalues. We refer the reader to Appendix.~\ref{sec:Normwise scaling} for the details of the scaling. 
\end{enumerate}
With all these three types of possible errors taken into account, 
one should expect the estimated QNM frequencies to be the most accurate at an \textit{optimal} $N$, with the accuracy deteriorating as $N$ is increased further. 
In the subsequent sections, we will show that this deterioration of accuracy 
indeed emerges in our calculations, but, through the scheme we prescribe below, we can still accurately extract the QNM frequencies with a surprisingly high relative fractional precision. 

\subsection{Distribution of the generalized eigenvalues}
\label{sec:evalue_distribution}

\begin{figure*}[tp!]
\centering  
\subfloat{\includegraphics[width=0.47\linewidth]{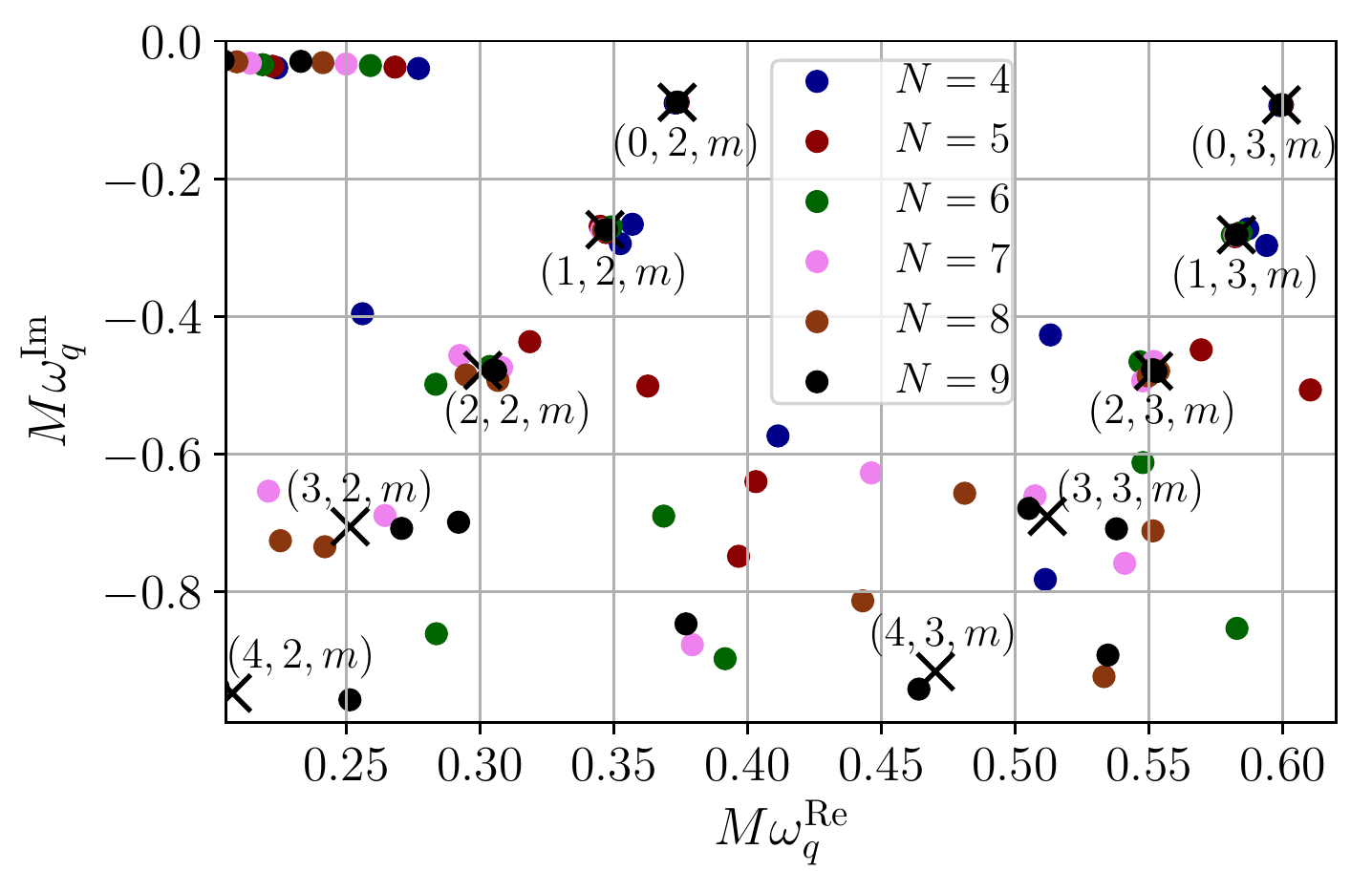}}
\subfloat{\includegraphics[width=0.47\linewidth]{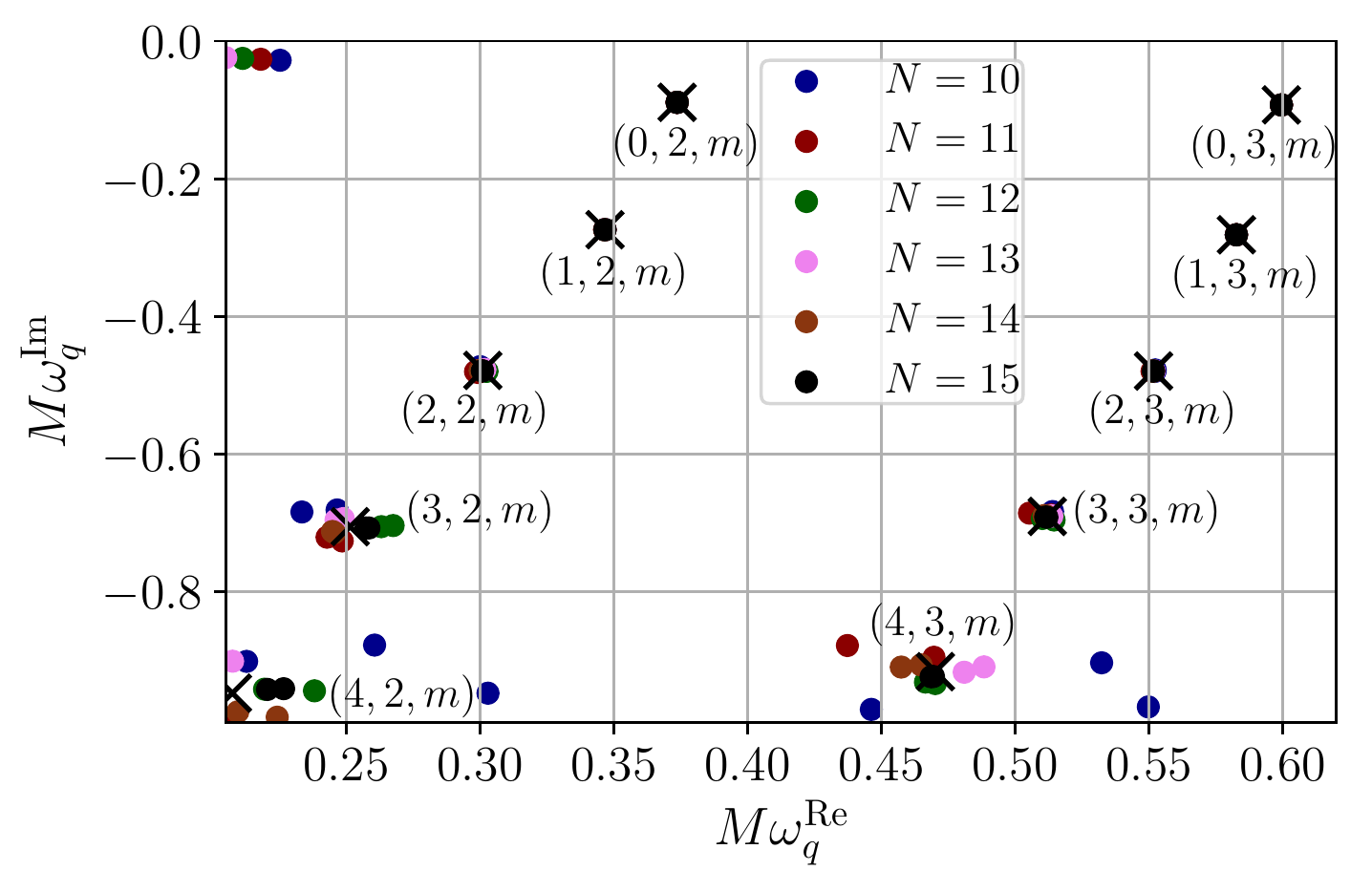}}
\qquad
\subfloat{\includegraphics[width=0.47\linewidth]{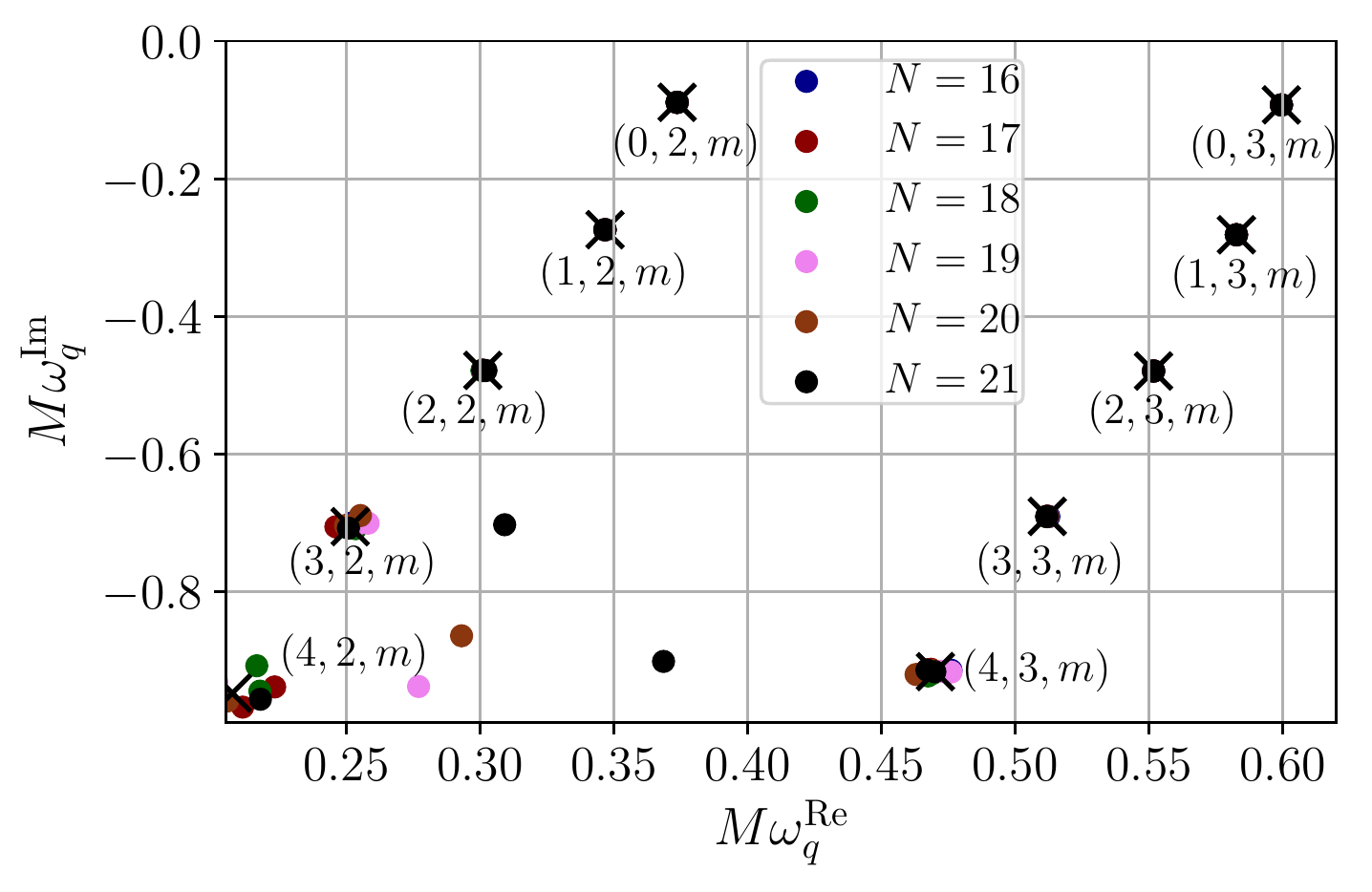}}
\subfloat{\includegraphics[width=0.47\linewidth]{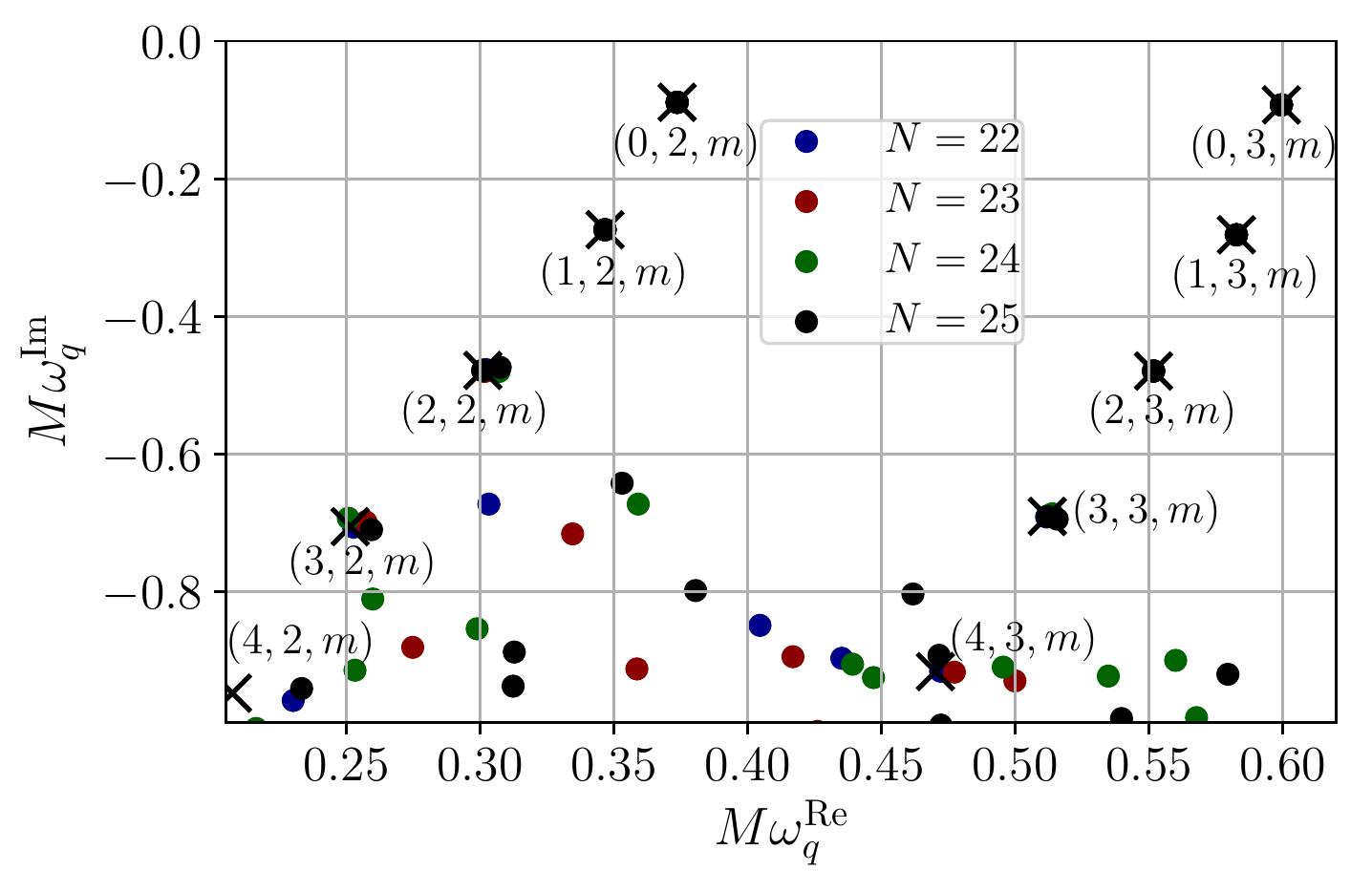}}
\caption{The distribution of the eigenvalues in the complex plane for $4 \leq N \leq 25$, where $N$ is the number of spectral functions used in the spectral decomposition (see Sec.~\ref{sec:evp}). Observe that the eigenvalues of the system of equations start to group together at certain points in the complex plane as N is increased. For comparison, we have also shown the corresponding QNM frequency calculated with Leaver's method~\cite{Leaver:1985ax}, using black crosses. 
The labels near the crosses follow a $(n, l, m)$ notation, where $n$ is the principal mode number, $l$ is the azimuthal mode number and $m$ is the magnetic mode number.
Since the QNM frequencies of the Schwarzschild BH do not depend on $m$, we have left this quantity unspecified in the labels. 
An animated version of these plots is available in the Supplemental Material.
}
\label{fig:Fig_1}
\end{figure*}

Let us now solve the $\{ tr, t \chi, t \phi, rr, r \chi, r \phi \} $ linearized Einstein equations and show how the eigenvalues emerge as we increase $N$. 
Figure~\ref{fig:Fig_1} shows the distribution of the eigenvalues in the complex plane from $N = 4$ to $ N = 25$ in four panels.
In general, the modulus of the eigenvalues ranges from $\sim 0$ to $10^{9}$. 
For QNM studies, we focus on eigenvalues in the range $0.2 \lesssim \text{Re} \lambda \lesssim 0.6$ and $ -1	\lesssim \text{Im} \lambda \lesssim 0$, which is also the range of the complex plane covered by Fig.~\ref{fig:Fig_1}.

Figure~\ref{fig:Fig_1} allows us to make several observations. 
As we begin to increase $N$ starting at $N=4$, groups of eigenvalues begin to cluster around certain areas in the complex plane. As $N$ is increased further to $\sim 20$, these clusters shrink to tiny areas, indicating that the eigenvalues are beginning to approach to certain values. Each tiny clustering area contains several slightly different eigenvalues, with relative differences in the real and imaginary parts of $\sim 10^{-7}$. The distances between these slightly different eigenvalues are much smaller than the typical distances between the clustering areas. Once the eigenvalues begin to cluster inside some small areas, any surplus eigenvalue begins to disappear as $N$ is increased, indicating that these surplus eigenvalues have no physical meaning.

As we further increase $N$ above $\sim 20$, surplus eigenvalues emerge again, indicating that the aforementioned sources of inaccuracies begin to affect the calculations.
There is therefore an optimal $N$ at which the eigenvalues have gotten as close as possible to the exact answer. 
These optimal eigenvalues coincide almost exactly with the Schwarzschild QNM frequencies computed by solving the Teukolsky equation, which we marked with crosses in Fig.~\ref{fig:Fig_1}. 
We will discuss later, in Sec.~\ref{sec:robustness}, what the relative fractional accuracy of the QNM frequencies computed with the spectral method is relative to other numerical solutions. 

The above observations suggest a method for the identification of the QNM frequencies. 
In essence, the QNM frequencies can be identified by searching for repeatedly emerging eigenvalues of the matrix equation before the accuracy deteriorates. 
In the next section, we will explain this method in more detail and explain how it can be used to accurately identify different QNMs. 

\subsection{Mode search}
\label{sec:mode_search}

As shown in the previous subsection, not all eigenvalues represent actual QNM frequencies. For a Schwarzschild or Kerr background, we could determine which eigenvalues are correct by comparing them to known solutions found through other methods, such as Leaver's method~\cite{Leaver:1985ax}. In modified gravity, theories, however, such other solutions may not be known, and thus, it would be ideal to find a self-contained method to identify which eigenvalues correspond to physical QNM frequencies. In essence, this method must answer the following question: 
What complex number is a given cluster of eigenvalues approaching and does it correspond to an actual QNM frequency? 

The answer to this question can be deduced from Fig.~\ref{fig:Fig_1}, which suggests that QNM frequencies can be identified by studying the cluster of eigenvalues that appear repeatedly in a small area in the complex plane for various choices of $N$.
More explicitly, we propose the following \textit{search method}: 
\begin{enumerate}
    \item Since not every eigenvalue is physical, keep only the eigenvalues in a region in the complex plane where QNM frequencies are expected to reside. 
    In this work, we keep eigenvalues whose real part is  $0.2 \leq \text{Re} \lambda \leq 0.6$ and imaginary part $ -1	\leq \text{Im} \leq 0$. 
    In general, this region can be adjusted based on the BH spacetime that needs to be studied.
    \item Compute the distance of the $i$ th eigenvalue obtained using $N\times N$ spectral functions, $\lambda_{i}(N)$, and the $j$ th eigenvalue using $(N+1)\times(N+1)$ functions, $\lambda_{j}(N+1)$. 
    If $\lambda_{i}(N)$ and $\lambda_{j}(N+1)$ are approaching a QNM frequency, their distance in the complex plane should be small. 
    Thus, store all eigenvalues that satisfy 
    \begin{equation}\label{eq:evalue_convergence}
    |\lambda_{i}(N) - \lambda_{j}(N+1)| \leq \text{threshold}, 
    \end{equation}
    where the threshold is a small number, which we choose here to be $10^{-3}$. This number corresponds to an error much smaller than the current relative uncertainty in the QNM frequency measurement of the detected ringdown signals~\cite{LIGO_07, LIGO_10, LIGO_11, pyRing_01, pyRing_02, pyRing_03, pyRing_04, pyRing_05, pyRing_06, NHT_Test_04, ringdown_test_01, ringdown_test_02, ringdown_test_03}. 
    \item As pointed out in Sec.~\ref{sec:evalue_distribution}, the stored eigenvalues may be slightly different from each other, and yet approach the same QNM frequency, because the separation between them in the complex plane is much smaller than the separation between different $n l m$ QNM frequencies.
    We thus select the average of these slightly different eigenvalues as the QNM frequency of mode $q= n l m$ and denote it $\omega_{q} (N)$. 
    \item Finally, just before the accuracy deteriorates, the difference of a mode-frequency between successive basis numbers, $|\omega_{\rm q}(N+1) - \omega_{\rm q} (N)|$, should reach its minimum. 
    Thus, we select the optimally\footnote{The optimal $N$ discussed here concerns the calculations of $\omega$, not the asymptotic expansion of the metric perturbations.} truncated QNM frequencies as 
    \begin{equation}\label{eq:w_optimally_truncated}
    \begin{split}
    \omega_{q}^{\rm opt} & = \omega_{q} (N_{\rm opt}) \,, \\
    N_{\rm opt} & = \text{arg}\min\limits_{N} |\omega_{q}(N+1) - \omega_{q}(N)| \,,
    \end{split}
    \end{equation}
    where we note that $N_{\rm opt}$ depends on the mode $q$. 
\end{enumerate}

Let us give an example of this search method in action by focusing on the $q=0 2$ mode. For any given $N>4$, we find various eigenvalues clustered around $M \omega_q \sim 0.37 - 0.1 i$. For example, at $N=4$ we find a cluster with the following eigenvalues
\begin{equation}
\begin{split}
\lambda(4) = \{ & 0.3737202242 - 0.0886296139i, \\
&0.3729295055 -0.0899641035i\}, 
\end{split}
\end{equation}
whose average is 
\begin{equation}
\omega_{0 2}^{N=4} = 0.37332486485 - 0.0892968587i. 
\end{equation}
Similarly, at $N=5$ we find a cluster with the eigenvalues
\begin{equation}
\begin{split}
\lambda(5) = \{& 0.3740968407 - 0.0888069404i, \\
& 0.3737492335 - 0.088943824i\}
\end{split}
\end{equation}
whose average is
\begin{equation}
\omega_{0 2}^{N=5} = 0.3739230371 - 0.0888753822i. 
\end{equation}
As we increase $N$, we find that the difference between the values of $\omega_{02}$ for adjacent values of $N$ first decreases, until $N \sim 20$, after which point the difference between adjacent averaged eigenvalues begins to increase. More concretely, we find that 
\begin{equation}
\begin{split}
& ..., \\
|\omega_{02}^{21}-\omega_{02}^{20}| = & 3.19 \times 10^{-8}, \\
|\omega_{02}^{22}-\omega_{02}^{21}| = & 9.04 \times 10^{-9}, \\
|\omega_{02}^{23}-\omega_{02}^{22}| = & 1.67 \times 10^{-8}, \\
|\omega_{02}^{24}-\omega_{02}^{23}| = & 3.47 \times 10^{-8}, \\
.... 
\end{split}
\end{equation}
From this sequence, we see that the optimal truncation is at $N=21$, and the optimal eigenvalue is 
\begin{equation}
\omega_{02}^{\rm opt} = 0.3736716790-0.0889623151i, 
\end{equation}
which demonstrates concretely how our search method works.  

\subsection{Mode identification}
\label{sec:mode_identification}

Once the QNM frequencies have been found through the search method of the previous subsection, we must now figure out which $n l m$ mode has been found. Again, for QNMs of a Schwarzschild or Kerr BH, this identification is easy, since we can compute the QNM frequencies through other robust methods. In modified gravity, however, such methods are typically not available, so one must create a robust procedure that answers the following question: 
Which QNMs (i.e. which $n l m$?) do the optimally truncated frequencies correspond to? 

Before we can establish an identification procedure, we need to first understand some general properties of the QNMs we are studying. 
To determine $n$ and $l$, we notice the following. For a fixed $n$, the real part of the QNM frequencies is much more sensitive to $l$ than the imaginary part. Similarly, for a fixed $l$, the imaginary part of the QNM frequencies is much more sensitive to $n$ than the real part~\cite{Leaver_01}. Although these trends hold strictly in GR, we expect them to also hold in effective-field theory-like modified theories in which BH solutions can be treated as small deformations of Schwarzschild and Kerr BHs with a continuous GR limit~\cite{QNM_dCS_01, QNM_dCS_02, QNM_dCS_03, QNM_dCS_04, QNM_EdGB_01, QNM_EdGB_02, QNM_EdGB_03}.  

We can understand this dependence from the eikonal approximation~\cite{2012PhRvD..86j4006Y, 2015APS..APR.Y7001M, Ferrari:1984zz, ColemanMiller:2021lky, Cardoso:2008bp} (valid when $l \gg 1$) and the geodesic analogy. 
In this approximation, the real part of the QNM frequency is roughly proportional to $l \Omega_{ph}$, where $\Omega_{ph}$ is the orbital frequency of the photon ring around the BH. Similarly, the imaginary part of the QNM frequency is roughly proportional to the Lyapunov exponent of photon ring, which does not sensitively depend on $l$~\cite{ColemanMiller:2021lky}.

With this understanding, let us now answer the question above by proposing the following \textit{identification procedure}: 
\begin{enumerate}
    \item We divide the optimally truncated frequencies into groups of similar imaginary parts. 
    \item The group with the least negative imaginary parts takes $n = 0$, and the group with the second least negative imaginary parts takes $n = 1$. We repeat this assignment of $n$ until we exhaust all the groups. 
    \item In a given group, the frequency with the smallest real part takes $l = 2$, and the frequency with the second-smallest real part takes $l = 3$. We repeat this assignment of $l$ until we exhaust all the frequencies in the same group. 
\end{enumerate}

Let us provide a concrete example of this procedure. 
When $N=24$, we have the following frequencies 
\begin{equation}
\begin{split}
\omega^{\rm{opt}} = \{ & 0.3736716813-0.0889623387i, \\
& 0.5994432887-0.0927030486i, \\
& 0.3467101908-0.2739044520i, \\
& 0.5826436957-0.2812978402i, \\
& 0.3010607141-0.4783191864i, \\
& 0.5517068087-0.4790929296i \}. 
\end{split}
\end{equation}
We immediately see that this list of optimally truncated frequencies can be divided into three groups of similar imaginary parts, namely, the first group consisting of the first and second frequencies, the second group of the third and fourth and the third of the fifth and sixth. 
Since the first group has the least negative imaginary parts, it takes $n=0$, corresponding to the fundamental modes. 
Amongst the first group, the frequency with the smallest real part takes the smallest azimuthal mode number, i.e. $l=2$, hence 
\begin{equation}
\omega_{02} = 0.3736716813-0.0889623387i, 
\end{equation}
and the frequency with a larger real part takes the next azimuthal mode number, i.e. $l=3$, 
\begin{equation}
\omega_{03} = 0.5994432887-0.0927030486i. 
\end{equation}
Then, we move on to the second group with more negative imaginary parts, which takes the next principal mode number, i.e. $n=1$, and the last group takes $n=2$. 
The azimuthal number of the frequencies in these groups can be labeled as we did for the first group. 
Explicitly, the frequencies of the second and third groups are labeled as 
\begin{equation}
\begin{split}
& \omega_{12} = 0.3467101908-0.2739044520i, \\
& \omega_{13} = 0.5826436957-0.2812978402i, \\
& \omega_{22} = 0.3010607141-0.4783191864i, \\
& \omega_{23} = 0.5517068087-0.4790929296i. 
\end{split}
\end{equation}

Following this procedure, we can confidently identify six QNMs ($q = \{02,03,12,13,22,23\})$, which is a smaller number than what was shown in Fig.~\ref{fig:Fig_1}. The reason that we cannot confidently identify the remaining modes (although they seem to clearly correspond to $q = \{32,33,43,42\}$) is that the absolute difference in any one of these clusters of eigenvalues is not yet smaller than the threshold defined in Equation~\eqref{eq:evalue_convergence}. If we had gone to higher $N$, then this difference would continue to decrease and we would have been able to confidently make the remaining identifications.

\subsection{Accuracy quantification}
\label{sec:error_quantification}

\begin{figure*}[htp!]
\centering  
\subfloat{\includegraphics[width=0.47\linewidth]{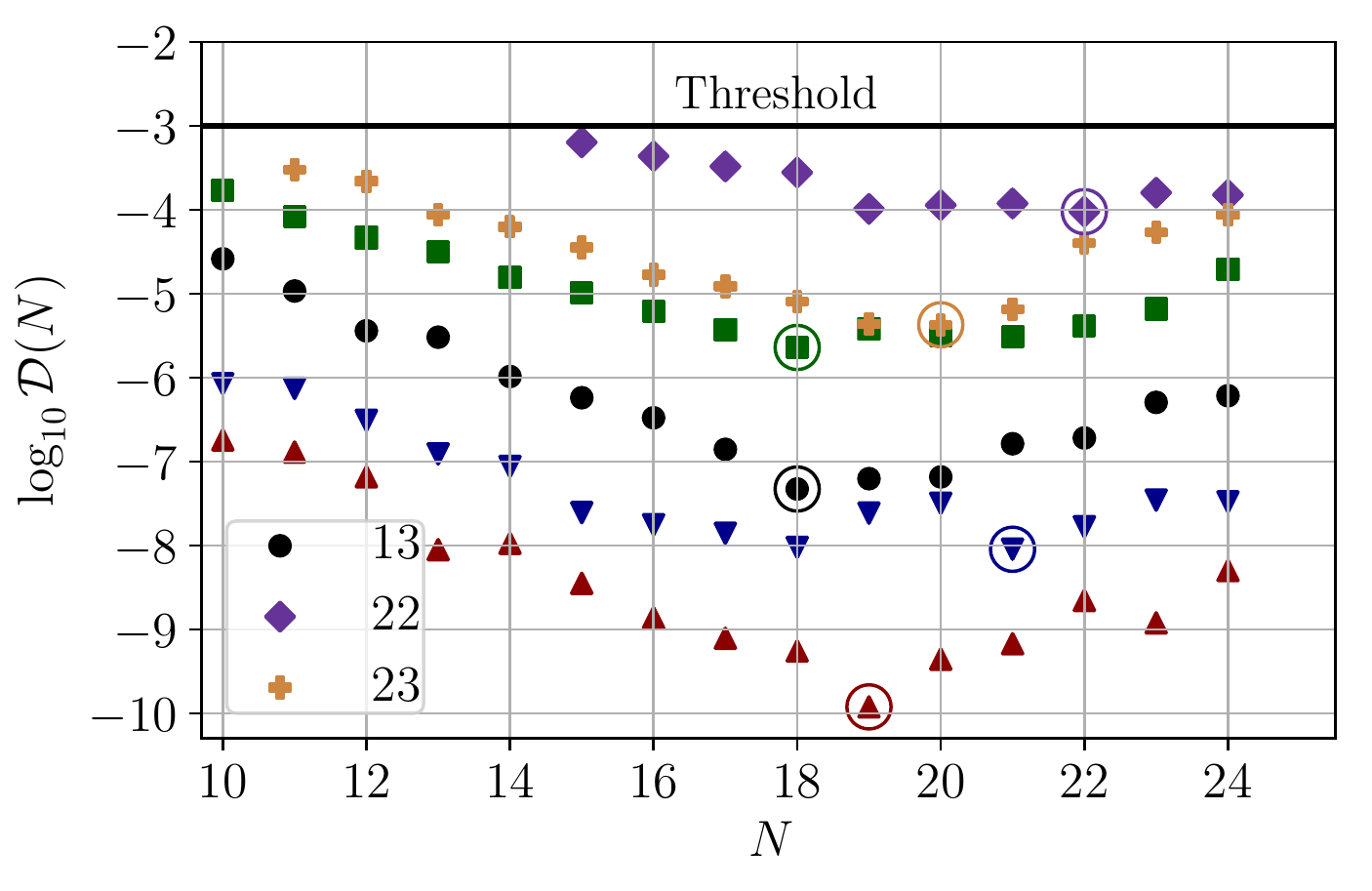}}
\subfloat{\includegraphics[width=0.47\linewidth]{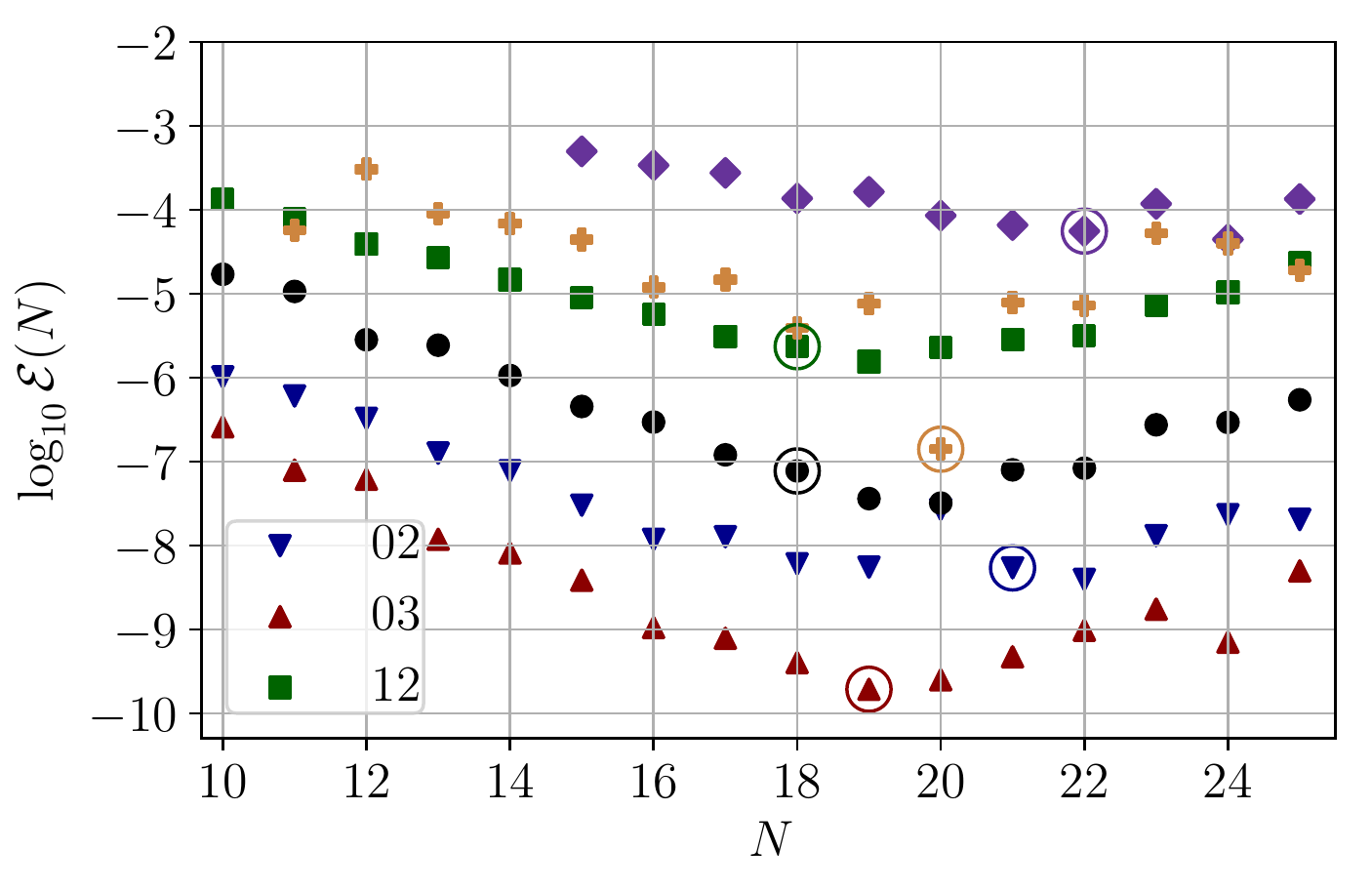}}
\qquad
\subfloat{\includegraphics[width=0.47\linewidth]{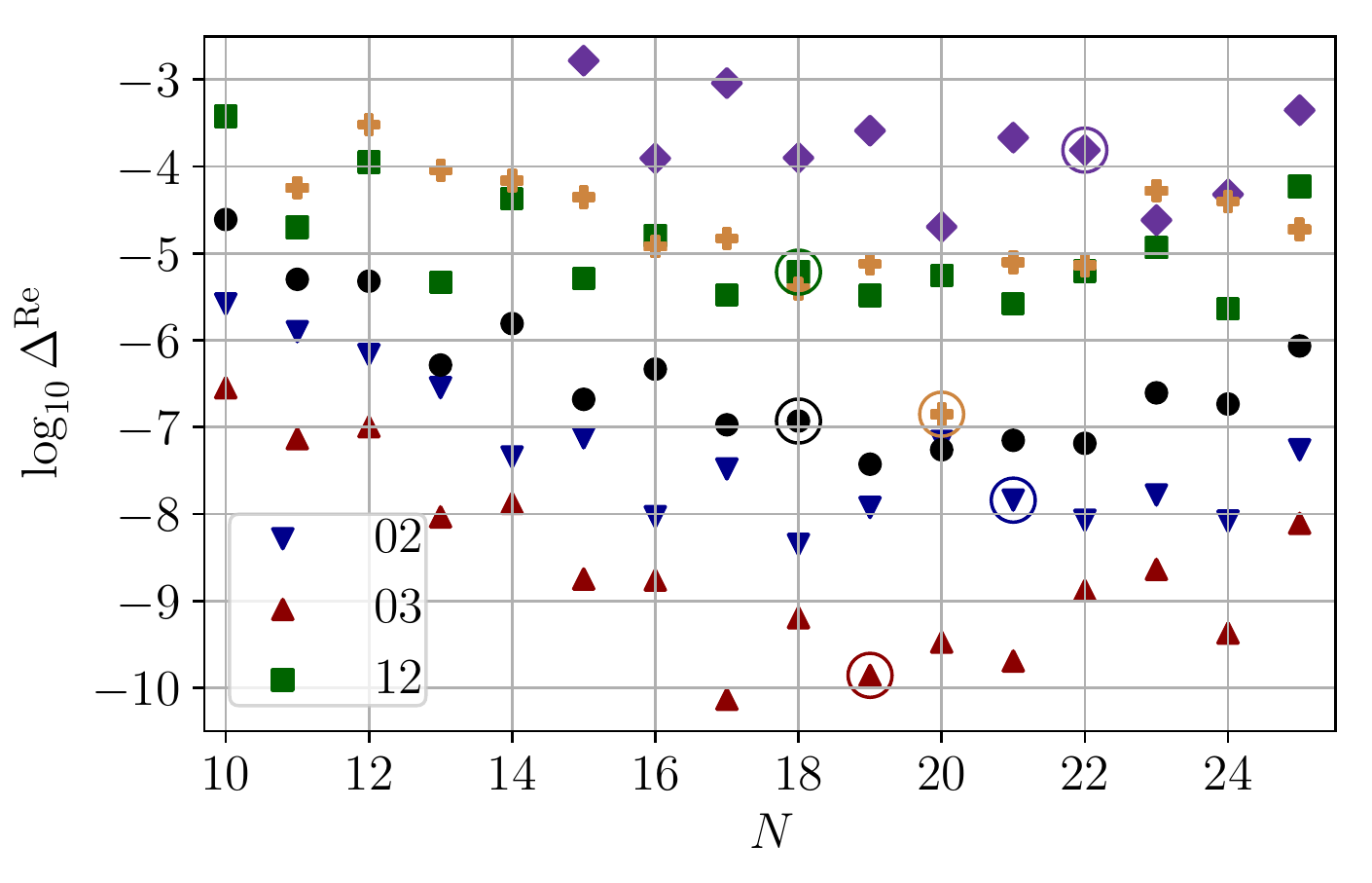}}
\subfloat{\includegraphics[width=0.47\linewidth]{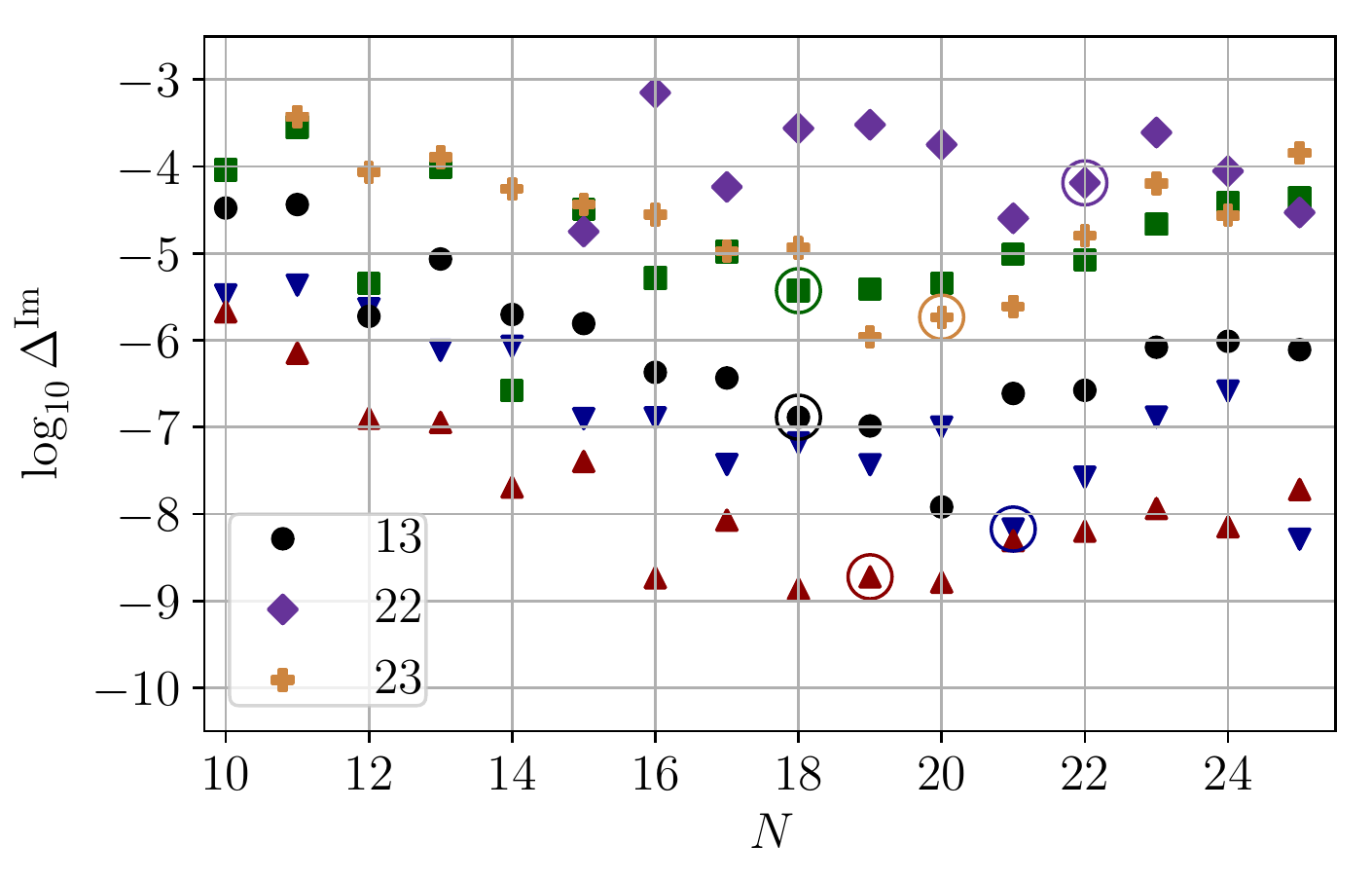}}
\caption{
The top, left panel shows the absolute difference between the QNM frequencies computed with adjacent $N$s, $\mathcal{D}(N) = |\omega(N+1) - \omega(N) |$, with the threshold $10^{-4}$ denoted by the horizontal solid black line. 
The $N$ that minimizes $\mathcal{D}(N)$, $N_{\rm opt}$ corresponds to the optimal truncation order and selects the optimal approximation $\omega(N_{\rm opt})$ to the QNM frequency (circled symbol). 
To gauge the accuracy of the spectral method, we compare the QNM frequencies computed using this spectral method at various $N$ [$\omega (\rm spectral)$] to those computed through Leaver's method~\cite{Leaver:1985ax} [$\omega (\rm L)$]. 
The top, right panel shows the absolute error 
$\mathcal{E}(N) = |\omega (\rm spectral) - \omega (\rm L)|$ as a 
function of $N$, while the bottom panels show the relative fractional 
error in the real [$\Delta^{\rm Re} = \left|1-{\omega^{\rm Re} (\text{spectral})}/{\omega^{\rm Re} (\text{L})}\right|$] and 
imaginary [$\Delta^{\rm Im} = \left|1-{\omega^{\rm Im} (\text{spectral})}/{\omega^{\rm Im} (\text{L})}\right|$ right panel] 
parts.
Observe that the QNM frequencies calculated with the spectral method are highly accurate for the fundamental mode and its overtones.
}
\label{fig:Fig_3}
\end{figure*}

Let us now assess the accuracy of the QNMs we have just calculated. To do so, let us define the following four accuracy measures:
\begin{enumerate}
    \item Difference over successive calculations, 
    \begin{equation}\label{eq:diff_over_iterations}
    \mathcal{D}(N) = |\omega^{\rm opt}_{q}(N+1) - \omega^{\rm opt}_{q}(N)|, 
    \end{equation}
    which characterizes how the QNMs approach a given answer as $N$ is increased, until a given optimal truncation order is achieved, after which point the estimates deteriorate. 
    \item The absolute error between the QNM frequencies computed using the spectral method, $\omega(\text{spectral})$, and Leaver's method to solve for the QNM modes $\omega (\text{L})$, 
    \begin{equation}\label{eq:absolute_error}
    \mathcal{E}(N) =  |\omega(\text{spectral}) - \omega (\text{L})|\,. 
    \end{equation}
    \item The relative fractional error in the real and imaginary parts of the QNM frequencies computed using the spectral method and Leaver's method ~\cite{Leaver:1985ax}, 
    \begin{equation} \label{eq:relative_error}
    \begin{split}
    & \Delta^{\rm Re / Im} = \left|1-\frac{\omega^{\rm Re / Im} (\text{spectral})}{\omega^{\rm Re / Im} (\text{L})}\right|\,,
    \end{split}
    \end{equation}
    where $\omega^{\rm Re / Im} (\text{spectral})$ and $\omega^{\rm Re / Im} (\text{L})$ stand for the real and imaginary parts of $\omega(\text{spectral})$ and $\omega (\text{L})$ respectively.
    \item Numerical uncertainty due to the deterioration of the accuracy with increasing $N$, $\delta$, defined as
    \begin{widetext}
    \begin{equation}\label{eq:computational_uncertainty}
    \begin{split}
    \delta^{\rm Re / Im} = 
    & \begin{cases}
    & \dfrac{\max \left( |\omega (N_{\rm opt}+1) - \omega (N_{\rm opt})|, |\omega (N_{\rm opt}) - \omega (N_{\rm opt} -1)| \right)}{|\omega^{\rm Re/Im}(N_{\rm opt})|}, \qquad \text{if $N_{\rm opt} < N_{\rm max} $~} \\ \\
    & \dfrac{|\omega (N_{\rm opt}) - \omega (N_{\rm opt}-1)|}{|\omega^{\rm Re/Im}(N_{\rm opt})|} , \qquad \text{if $N_{\rm opt} = N_{\rm max} $~}.
    \end{cases}
    \end{split}
    \end{equation}
    \end{widetext}
    This quantity gauges how the accuracy of the spectral method is limited by the possible sources of inaccuracies mentioned in Sec.~\ref{sec:error_sources}. 
    This measure will be useful to estimate the performance of the spectral method when applied to different systems of equations, as we do in Sec.~\ref{sec:robustness}. 
\end{enumerate}

To compute the above measures, we solve the Teukolsky equation in the zero-spin limit using Leaver's method of continued fractions~\cite{Leaver:1985ax} to find $\omega(\rm L)$.
Specifically, $\omega(\rm L)$ is computed using Leaver's method with 1000 terms in the continued fractions. We find that $\sim 200$ terms are already enough to converge to 14 digits of accuracy for the fundamental mode frequencies. Using 1000 terms, the first 16 digits of the real and imaginary parts of the QNM frequencies also converge for all modes studied here.
For the convenience of the reader, we list the QNM frequencies obtained through this method below:
\allowdisplaybreaks[4]
\begin{equation}
\begin{split}\label{eq:Leaver_QNMFs}
\omega_{02}(\text{L}) & = 0.37367168441804 - 0.08896231568894 \, i,\\
\omega_{03}(\text{L}) & = 0.59944328843749 - 0.09270304794495 \, i,\\
\omega_{12}(\text{L}) & = 0.34671099687916 - 0.27391487529123 \, i,\\
\omega_{13}(\text{L}) & = 0.58264380303330 - 0.28129811343504 \, i,\\
\omega_{22}(\text{L}) & = 0.30105345461237 - 0.47827698322307 \, i,\\
\omega_{23}(\text{L}) & = 0.55168490077845 - 0.47909275096696 \, i.
\end{split}
\end{equation}
We note that the above frequencies are identical to the frequencies published in \cite{Cook:2014cta}, except for differences in rounding off of the last digits.

The first three measures defined above are presented in Fig.~\ref{fig:Fig_3} as a function of $N$. 
The top left, right and bottom panels respectively show the base-10 logarithms of $\mathcal{D}(N)$, $\mathcal{E}(N)$, $\Delta^{\rm Re}$ (bottom left) and $\Delta^{\rm Im}$ (bottom right) of the QNM frequencies as a function of $N$. 
In general, all three measures first decrease as $N$ increases from $N = 10 $ to a QNM-dependent $N$. This indicates that our QNM frequency calculations become increasingly accurate as $N$ increases. 
Beyond the QNM-dependent $N$, all three measures begin to increase, indicating the emergence of effects due to possible sources of numerical inaccuracies, consistent with our observations of Fig.~\ref{fig:Fig_1}. The optimal truncation order,
$N_{\rm opt}$, minimizes $\mathcal{D}(N)$ and also approximately minimizes $\mathcal{E}(N)$ and $\delta^{\rm Re / Im}$, as we show with a circle in the figure.
Observe that the relative fractional error of the optimal truncation is very small for all six QNM frequencies computed. Observe also that 
the higher the mode number, the fewer the errors we can present and the less accurate the QNM frequencies are. This is because the higher the mode number, the more the number of basis terms that are required for the eigenvalues to be within the threshold tolerance we selected. 

\section{Robustness of Quasinormal Frequency Extraction}
\label{sec:robustness}

In this section, we study the robustness of the calculations presented in the previous section. 
In particular, we first focus on the $m$ independence of the QNM frequencies, which ought to hold for perturbations of a Schwarzschild background. We then study the effects of our choice of boundary conditions for the $\rho$ function on the QNM calculation. Finally, we consider the use of other combinations of linearized Einstein equations. 

\subsection{$m$ independence of the quasinormal spectrum}
\label{sec:m_independence}

\begin{figure*}[htp!]
\centering  
\subfloat{\includegraphics[width=0.47\linewidth]{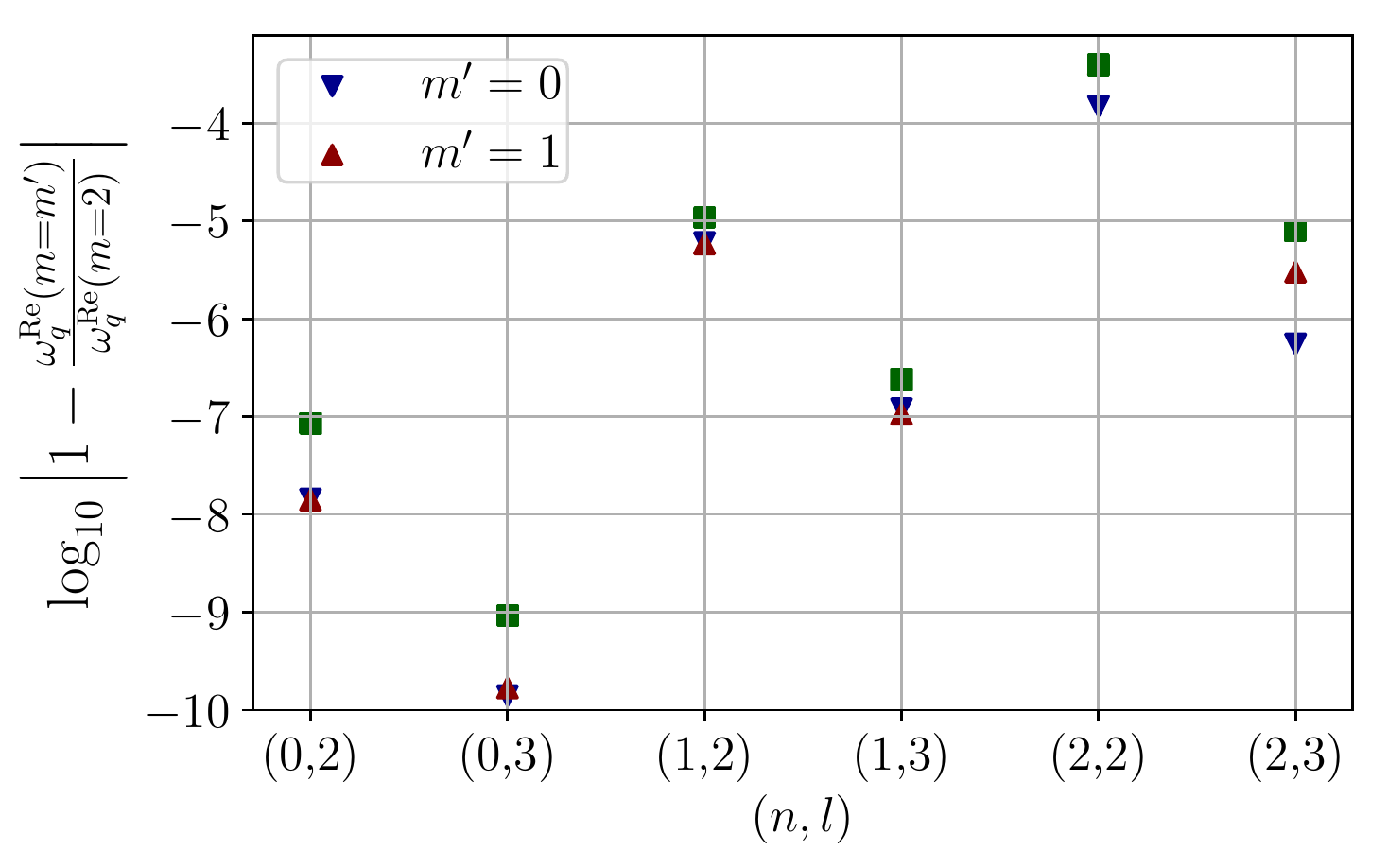}}
\subfloat{\includegraphics[width=0.47\linewidth]{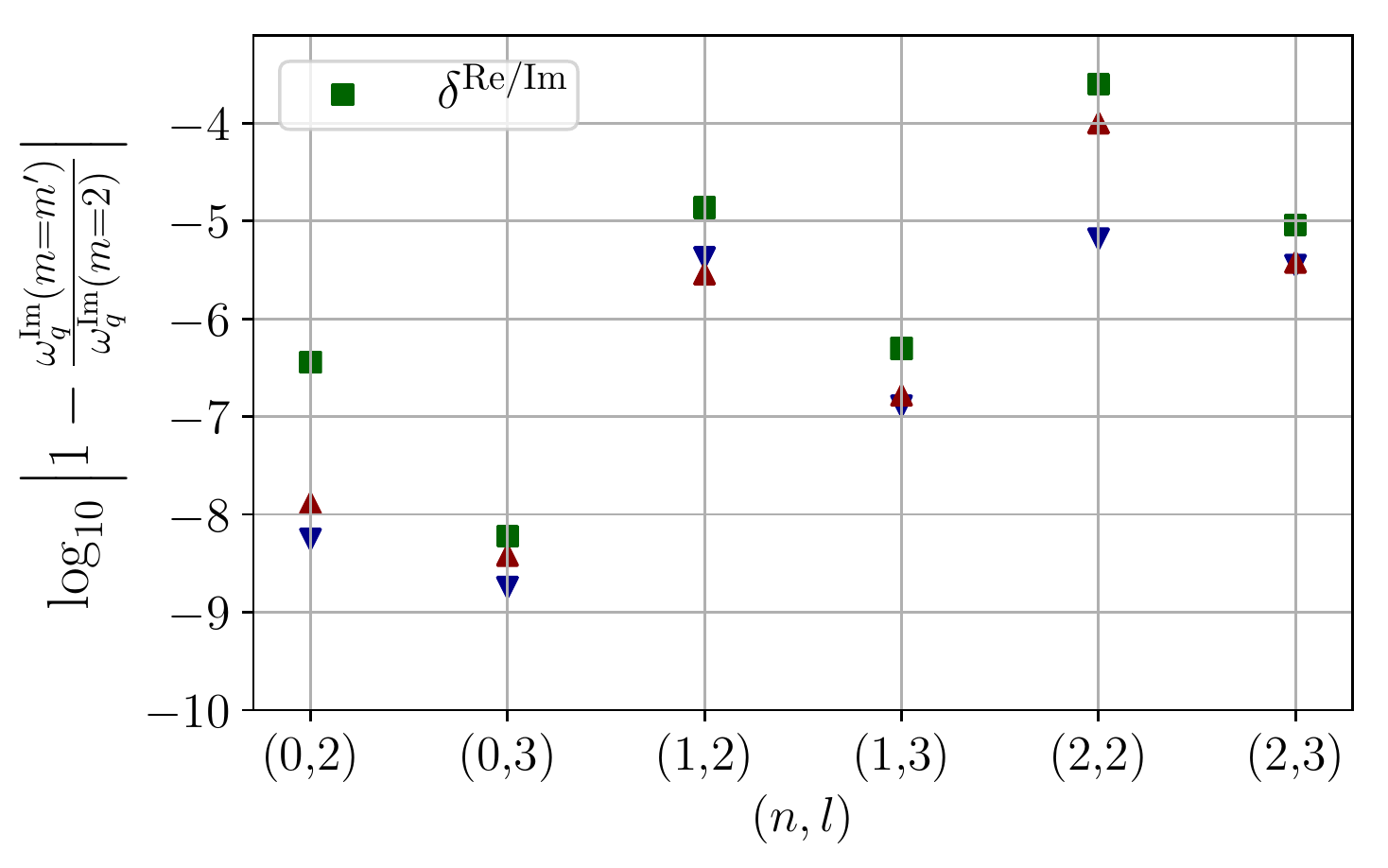}}
\caption{Base-10 logarithm of the relative fractional differences between the real (left) and imaginary parts (right) of the QNM frequencies when setting $m = 0$ and $m=2$, both computed using our spectral method of at most $25 \times 25$ spectral functions. 
The relative fractional difference for different QNMs is between $10^{-10}$ and $10^{-4}$, which is smaller than, or at worst approximately equal, to the numerical uncertainty of the $m=2$ frequencies (green squares). 
Thus, effectively, the QNM frequencies computed by setting $m$ to different values in our spectral method are the same. Such $m$ independence of our results is a nontrivial verification of the correctness and robustness of the spectral method. 
}
 \label{fig:abs_error_diff_m}
\end{figure*}

One important feature of gravitational perturbations of spherically symmetric BHs is the independence of the QNM spectra on $m$. Our matrix equations, however, explicitly depend on $m$ because we have not decoupled the linearized Einstein equations to find master equations. Therefore, validating the $m$ independence of our QNM frequency calculations constitutes a nontrivial test of the robustness of our spectral method.

Before comparing the QNMs computed by setting $m$ to different values, let us comment on the structure of the linearized Einstein equations when $m=0$. We have derived the linearized Einstein equations for general $m$, so when we take the $m=0$ limit, we find that each linearized EFE can be factorized with an additional term that is a power of $(1-\chi^2)$. Following Sec.~\ref{sec:EFEs}, it is usually desirable to divide such prefactors out (since they are never zero for a BH) to simplify the equations and potentially improve the accuracy and stability of the numerical calculations. Doing so then yields a somewhat simpler $\tilde{\mathbb{D}} (\omega)$ matrix, whose generalized eigenvalues contain the QNM frequencies of a Schwarzschild BH. 

With that in hand, let us now compute the QNM frequencies by solving the linearized Einstein equations setting $m=0$ and $m=1$ and compare them to the results we obtained above when we set $m=2$. We find that these two sets of QNM frequencies are very close to each other. 
Figure~\ref{fig:abs_error_diff_m} shows the relative fractional difference between the real (left) and imaginary (right) parts of the $m=2$ frequencies and the $m=0$ (blue inverse triangles) and $m=1$ frequencies (red triangles) for different $(n,l)$. Observe that this relative fractional difference ranges from $10^{-10}$ to $10^{-4}$. 
Comparing the relative differences with the numerical uncertainty of the $m=2$ frequencies (green squares), we see that the relative fractional differences are smaller or approximately equal to the numerical uncertainty, which suggests that the differences between the $m=0$ or the $m=1$ frequencies and the $m=2$ frequencies are due to numerical uncertainty. 
Thus, effectively, the spectral method obtains the same QNM frequency for a given $n$ and $l$ regardless of the value of $m$ we choose in our calculations. 

\subsection{Effects of $\rho_H$ and $\rho_{\infty}$}
\label{sec:rho_effects}

\begin{figure*}[htp!]
\centering  
\subfloat{\includegraphics[width=0.47\linewidth]{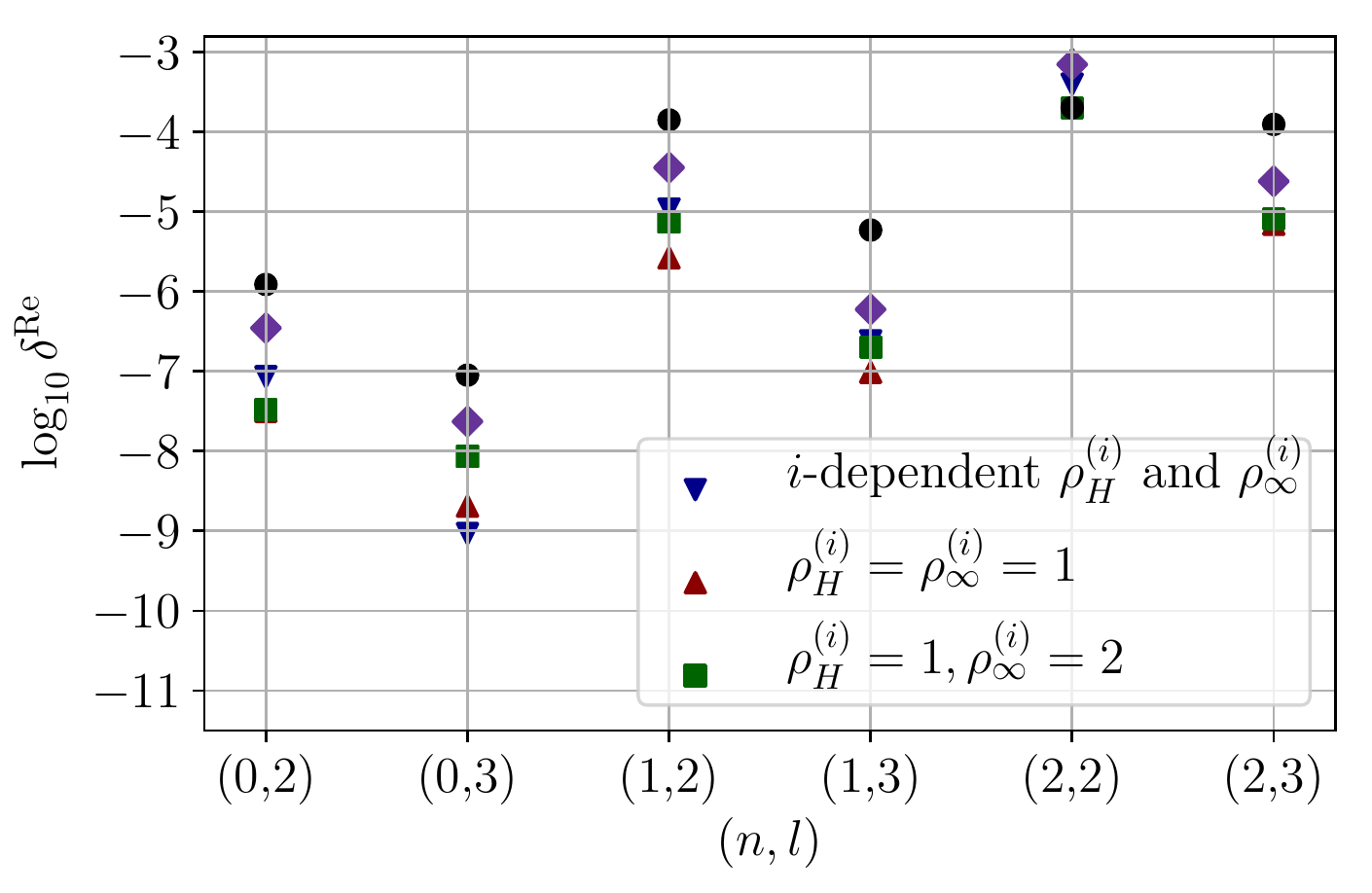}}
\subfloat{\includegraphics[width=0.47\linewidth]{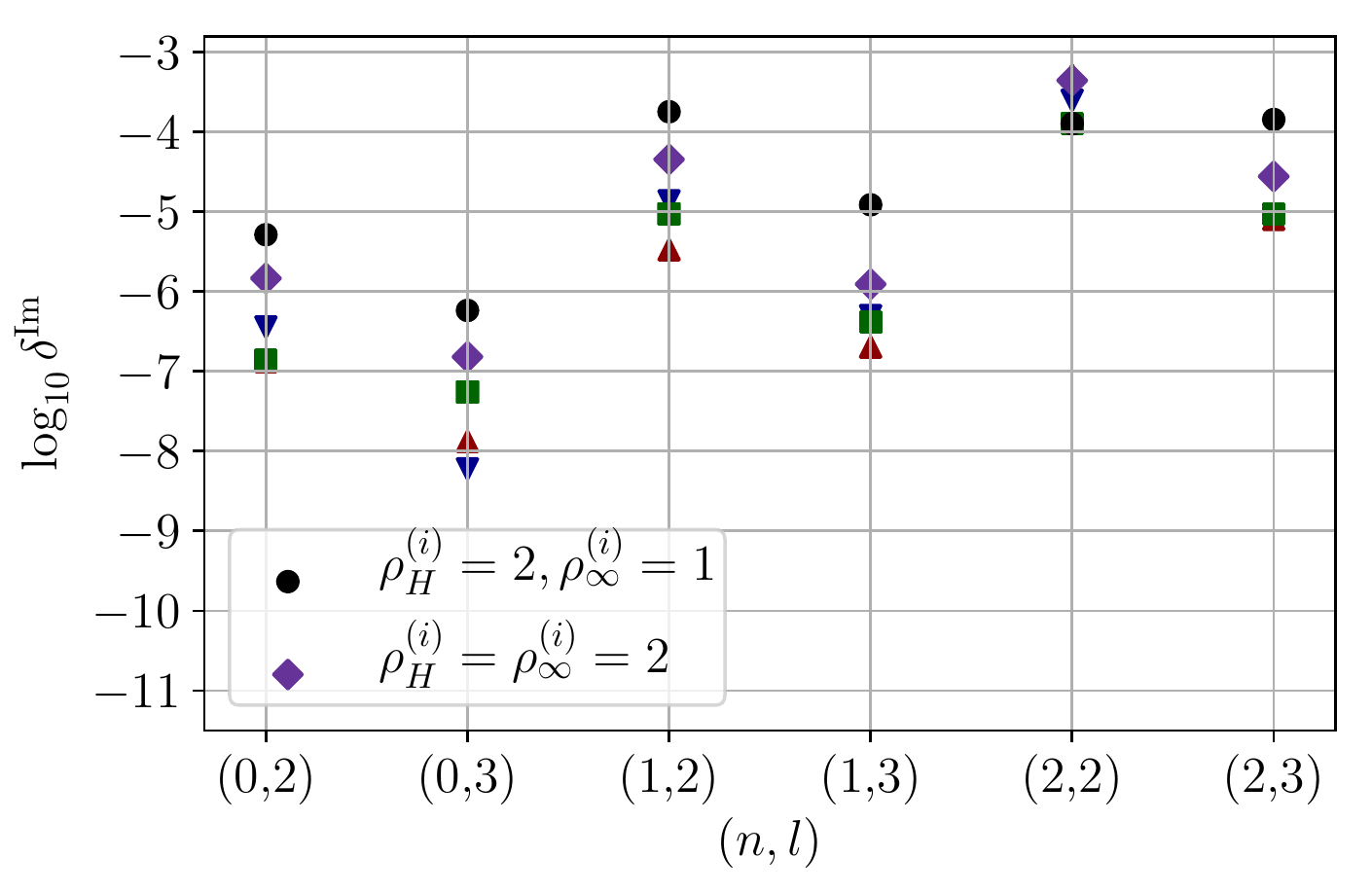}}
\caption{Numerical uncertainty of the real (left) and imaginary parts (right) of different QNM frequencies ($\delta^{\rm Re/Im}$, see Equation~\eqref{eq:computational_uncertainty}) computed using the spectral method with at most $25\times25$ spectral functions and assuming different $\rho_H^{(i)}$ and $\rho_{\infty}^{(i)}$ boundary conditions. Observe that the accuracy of our series solution, computed with different boundary conditions, is approximately the same, indicating the robustness of the spectral method.}
 \label{fig:numerical_uncertainty_diff_rhos}
\end{figure*}

The asymptotic behavior of the metric perturbation functions obtained in Sec.~\ref{sec:asymptotic_analysis_results} depends on the component of metric perturbations. 
We find that the extracted QNMs are not affected if we assume $\rho_{H}^{(i)}$ and $\rho_{\infty}^{(i)}$ to be the same number for all $i$, provided that the assumed $\rho_{H}^{(i)} \geq \max\limits_{1 \leq i \leq 6} \rho_{H}^{(i)} $ and $\rho_{\infty}^{(i)} \geq \max\limits_{1 \leq i \leq 6} \rho_{\infty}^{(i)}$. 
To illustrate this property, Fig.~\ref{fig:numerical_uncertainty_diff_rhos} shows $\delta^{\rm Re/Im}$ of the six previously identified QNMs, obtained by numerically solving the linearized Einstein equations, using $25\times25$ spectral functions and assuming that for all $i$ 
\begin{itemize}
\item $\rho_H^{(i)} = \rho_{\infty}^{(i)} = 1$ (inverted blue triangles), 
\item $\rho_H^{(i)} = 1, \rho_{\infty}^{(i)} = 2 $ (red triangles), 
\item $\rho_H^{(i)} = 2, \rho_{\infty}^{(i)} = 1 $ (green squares), and 
\item $\rho_H^{(i)} = 2, \rho_{\infty}^{(i)} = 2 $ (black circles). 
\end{itemize}
Figure~\ref{fig:numerical_uncertainty_diff_rhos} shows that if we assume a $\rho_{H}^{(i)}$ and $\rho_{\infty}^{(i)}$ that is larger than the exponents obtained by our asymptotic analysis in Sec.~\ref{sec:asymptotic_behaviour}, we can still accurately extract the QNMs of the Schwarzschild BH. 
As the figure shows the minimal $\delta^{\rm Re/Im}$ of different QNM frequencies depends on $\rho_H^{(i)}$ and $\rho_{\infty}^{(i)}$, but we leave further analysis of this relation to future work. 

The extremely mild dependence of the QNM frequencies on $\rho_H^{(i)}$ and $\rho_{\infty}^{(i)}$ is actually reasonable and can be understood as follows. 
Let us focus first on the $\rho_H^{(i)} = \rho_{\infty}^{(i)} = 1$ case.
Even if we assume these boundary conditions, the boundary conditions obtained in Sec.~\ref{sec:asymptotic_behaviour} for all $h_i$ are still satisfied, except when $i = 4$. 
When $i=4$, Sec.~\ref{sec:asymptotic_analysis_results} tells us that the ``correct" \textit{Ansatz} for $y_4$ is 
\begin{equation}
y_4 (r) = e^{i \omega r} r^{i \omega r_\h} \left( \frac{r-r_{H}}{r}\right)^{-i \omega r_\h} u_4^{\rm (corr)}(r), 
\end{equation}
where $u_4^{\rm (corr)}(r)$ is the finite part of $y_4$ that we must calculate numerically. 
If we assume $\rho_H^{(i)} = \rho_{\infty}^{(i)} = 1$ instead, we are actually imposing the \textit{Ansatz}
\begin{equation}
y_4(r) = e^{i \omega r} r^{i \omega r_\h+1} \left( \frac{r-r_{H}}{r}\right)^{-i \omega r_\h-1} u_4^{\rm (asum)}(r), 
\end{equation}
where $u_4^{\rm (asum)}(r)$ is now the finite part of $y_4$.
If these two \textit{Ans\"atze} are to agree, we must have that
\begin{equation}
u_4^{\rm (asum)}(r) = \frac{r-r_\h}{r} \frac{1}{r} u_4^{\rm (corr)}(r). 
\end{equation}
Hence, $ r \in [r_\h, \infty)$, $u_4^{\rm (asum)}(r)$ is bounded if $u_4^{\rm (corr)}(r)$ is also bounded because $(r-r_{H})/{r^2}$ is finite. 
Thus, the spectral decomposition can still be applied regardless of our assumptions on the boundary conditions for $\rho$.
This argument also applies for an even larger $\rho_H$ and $\rho_{\infty}$. 

This independence of our calculations on the choice of  $\rho_H^{(i)}$ and $ \rho_{\infty}^{(i)}$ has three advantages. 
First, it can simplify the prescriptions of the boundary conditions for numerical computations because we can simply use the same $\rho_{H}^{(i)}$ and $\rho_{\infty}^{(i)}$. 
Second, we can cross-check our results by repeating our calculations for different values of $\rho_{H}^{(i)}$ and $\rho_{\infty}^{(i)}$.
If the QNM frequencies are properly extracted, the same complex numbers should emerge regardless of the choice of $\rho_{H}^{(i)}$ and $\rho_{\infty}^{(i)}$. 
Third, this property may allow us to bypass the estimate of the asymptotic behavior when studying the boundary conditions. 
This simplification could be welcomed when dealing with more sophisticated BHs, for which the estimation of the asymptotic behavior of the solution may be much more difficult. 

\subsection{Other combination of the linearized equations}
\label{sec:other_selections}

\begin{figure*}[htp!]
\centering  
\subfloat{\includegraphics[width=0.47\linewidth]{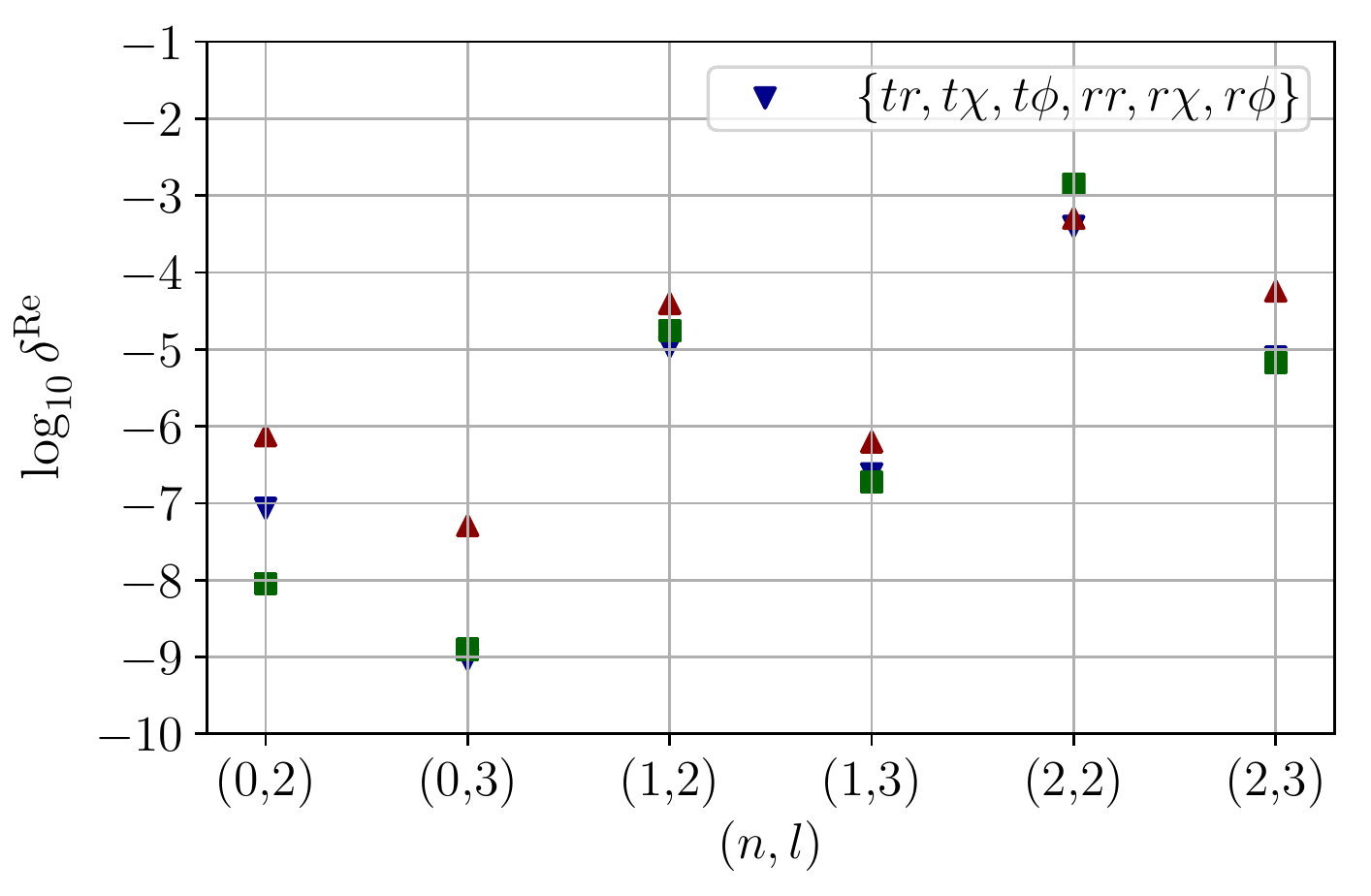}}
\subfloat{\includegraphics[width=0.47\linewidth]{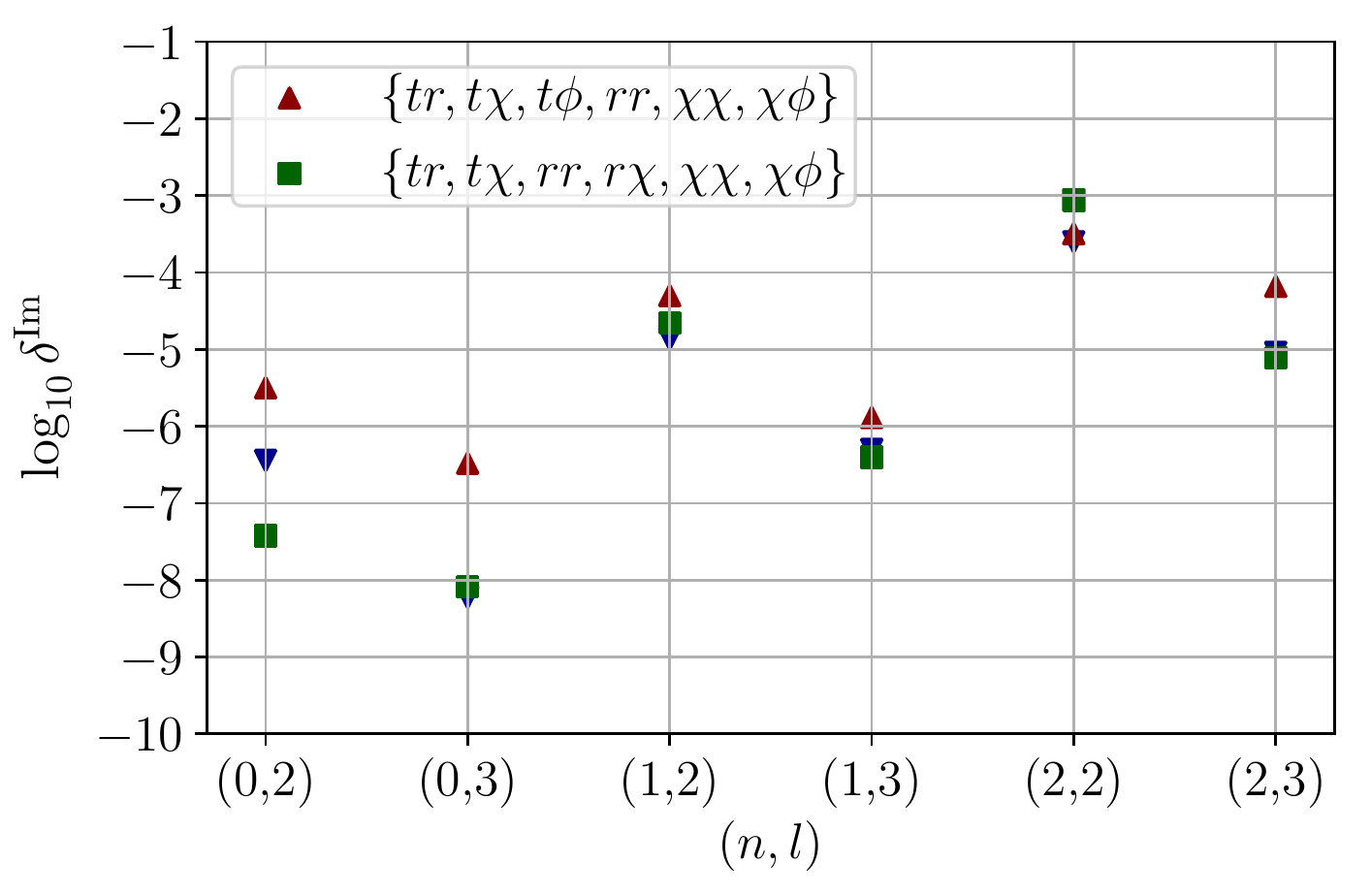}}
\caption{Numerical uncertainty in the real (left) and imaginary parts (right) of different QNM frequencies by spectrally decomposing different sets of components of the linearized Einstein equations.
In all cases, we use at most $25\times25$ spectral functions when computed the QNM frequencies. 
Observe that the accuracy of the QNM frequencies calculated is approximately independent of the choice of components of the linearized Einstein equations that we choose to solve.
}
\label{fig:numerical_uncertainty_other_sets}
\end{figure*}

We have thus focused on the $\{ tr, t \chi, t \phi, rr, r\chi , r \phi \}$ set of linearized Einstein equations, but what if we had chosen a different set? We find that if we select a different set of linearized equations, we can still accurately estimate the QNM frequencies.  
Figure~\ref{fig:numerical_uncertainty_other_sets} compares the $\delta^{\rm Re/Im}$ of the QNM frequencies computed by solving the following systems\footnote{This list of linearized Einstein equations is by no means exhaustive. 
We also calculated various Schwarzschild QNM frequencies by solving other sets, but we found that a larger number of basis functions would then be required to obtain an accurate result. }: 
\begin{itemize}
\item $\{ tr, t \chi, rr, r \chi, \chi \chi, \chi \phi \} $ (red triangles), 
\item $\{ tr, t \chi, t \phi, rr, \chi \chi, \chi \phi \} $ (green circles), 
\end{itemize}
and $\{ tr, t \chi, t \phi, rr, r \chi, \chi \phi \} $ (inverted blue triangles, the system we have been focusing on) solved using at most $25 \times 25$ spectral functions. 
Observe that the choice of the components of the linearized Einstein equations one works with does not affect our ability to solve for the QNM frequencies.  This flexibility allows us to cross-check our results by computing the QNM frequencies using different sets of linearized Einstein equations.  

This flexibility is also an interesting result in its own right. 
Previous calculations of Schwarzschild QNM frequencies relied on solving certain master equations, which are computed by simplifying and eliminating various components of the Einstein tensor~\cite{Regge:PhysRev.108.1063, Zerilli_even, Berti_01, Berti_02}. 
To keep the calculations tractable, those derivations naturally make use of the simplest linearized Einstein tensor components. 
Here, we show that different choices of the components of the linearized Einstein equations that one solves also lead to the accurate computation of Schwarzschild QNM frequencies. 

\section{Concluding remarks}
\label{sec:conclusion}

In this paper, we have developed a spectral method to systematically study gravitational perturbations of a nonrotating BH. 
We first apply spectral decompositions to study the asymptotic behavior of gravitational perturbations at spatial infinity and at the BH event horizon. 
Using this asymptotic behavior, we then construct an \textit{Ansatz} for the metric perturbations. 
The \textit{Ansatz} allows us to spectrally decompose the linearized Einstein field equations along both the radial and polar coordinates, thereby transforming the linearized field equations into a linear eigenvalue problem. 
By solving the matrix equation for the generalized eigenvalues, and through the development of a procedure to identify the QNMs these eigenvalues correspond to, we can calculate the frequency of many QNMs with excellent accuracy. 
For example, using our numerical scheme, we can simultaneously compute six QNM frequencies of the Schwarzschild BH with a relative error always better than (and sometimes much better than) $\leq 10^{-4}$. 
We thoroughly investigate the computational uncertainty of our spectral method, concluding that our calculations are highly accurate and the QNM identification procedures are robust. 
Our approach allows us to verify that, as expected, the QNM frequencies of a perturbed Schwarzschild BH are independent of the set of components of the linearized field equations that one uses to calculate them. 

The spectral method contains several advantages over the existing approaches to studying gravitational perturbations of a BH. 
First, the spectral method can, in principle, be applied to any BH spacetimes irrespective of their classification under the Petrov scheme~\cite{Petrov:2000bs}. 
Unlike the derivation of the Teukolsky equation, our method does not require the background spacetime to be vacuum (i.e., no matter) and Petrov-type D~\cite{Teukolsky:1973ha}. 
This advantage enables us to apply the spectral method to other more complicated and generic BH spacetimes that cannot be easily studied through the Newman-Penrose formalism. 

Second, the spectral method does not require simplifications of the linearized field equations into master equations through special master functions. 
The derivation of the Regee-Wheeler, the Zerilli-Moncrief, or the Teukolsky equation requires the simplification of the perturbed (metric or curvature) equations into several decoupled master equations, obtained through various transformations or redefinitions of perturbation variables. 
These transformations and redefinitions usually need to be modified for non-Schwarzschild or Kerr BHs, and precisely how to do so can be quite difficult~\cite{PratikRingdown, Li:2022pcy}.
By applying the spectral method, we have a unified framework to accurately estimate the QNM frequencies without such simplifications or decouplings, bypassing the difficulties of deriving the necessary transformations or redefinition. 

Third, the spectral method is computationally straightforward. 
When computing the QNM frequencies by solving the Teukolsky equation, one also needs to solve for the angular separation constants. 
The spectral method focuses on calculations of only the QNM frequencies, avoiding the need to compute these separation constants. 
Moreover, previous work had found that more than 100 spectral functions in the radial and angular coordinates are needed to compute higher-mode frequencies by spectrally decomposing the Teukolsky equation, even for the case of the Schwarzschild BH ($a=0$)~\cite{Ripley:2022ypi} \footnote{Though this number can be reduced by using a new-sparse spectral method \cite{ApproxFun.jl-2014}. }. 
In contrast, the spectral method presented here requires a much smaller set of basis functions ($\sim 25$) for the accurate estimation of six QNM frequencies.
These features aid in making the numerical computations more straightforward and convenient. 

Finally, the spectral method does not involve the calculation of the Weyl scalars, making the studies of gravitational perturbations more direct, and perhaps, more physically intuitive. 
The Teukolsky equation expresses all gravitational perturbations in terms of curvature perturbations that are encoded in perturbed Weyl scalars. 
Therefore, if one wishes to find the gravitational metric perturbations using solutions to the Teukolsky equation, one needs to reconstruct the metric from the Weyl scalars through a lengthy procedure~\cite{Chrzanowski:1975wv,Kegeles_Cohen_1979,Yunes_Gonzalez_2006}. 
The spectral method we presented here avoids all of these complications because it works directly with metric perturbations. 

To fully realize the potential of the spectral method we presented here, we need to further develop it so that it can be applied to more sophisticated BH spacetimes. 
Our immediate next step is to apply the spectral method to spinning BH backgrounds, and more concretely to the Kerr background metric.
When doing so, it may be beneficial to consider other basis functions for the spectral decomposition, instead of the associated Legendre polynomials for the angular sector and the Chebyshev polynomials for the radial sector that we used here. 
One option would be to use spheroidal harmonics or spin-weighted spherical harmonics for the angular sector, while one could use a rational polynomial basis for the radial sector. We have started this exploration already and have found some encouraging results, but their detailed presentation will be shown elsewhere. 
Moreover, thus far we have focused on the Regge-Wheeler gauge, which should be applicable to a wide range of modified BHs. 
But to make the spectral method more generally applicable, we also need to explore different gauges.
One could also further investigate how exactly the sources of numerical inaccuracies, mentioned in Sec.~\ref{sec:error_sources}, affect the quasinormal frequencies, and how to improve their precision.

Other than rotating BHs, one still needs to explore the application of our spectral method to beyond-GR BHs whose metric is irrational (e.g. \cite{Supertranslational_hair_01}) or numerical (e.g. \cite{Sullivan:2019vyi, Sullivan:2020zpf, Fernandes:2022gde, Dias:2023nbj, Dias:2022str, Biggs:2022lvi}). 
For irrational BH solutions, a change of variables may rationalize the metric, which allows straightforward applications of our spectral method. 
Numerical BH solutions are commonly expressed in terms of spectral functions when the solutions are being calculated, and thus, our spectral method directly applies. 
Alternatively, we can also fit numerical BH solutions using spectral functions or by numerically evaluating their derivatives to derive the linearized field equations.
Once the linearized field equations are obtained, even via numerical means, our spectral method still applies. 
In the future, we plan to explore various modifications to adapt our spectral method to irrational or numerical BHs.

Once the spectral method has been generalized and developed further, it could be applied to a plethora of problems. The most obvious one is perhaps the calculation of QNM frequencies in modified gravity theories, such as in dynamical Chern-Simons gravity~\cite{QNM_dCS_01, QNM_dCS_02, QNM_dCS_03, QNM_dCS_04} or scalar-Gauss-Bonnet gravity~\cite{QNM_EdGB_01, QNM_EdGB_02, QNM_EdGB_03}. In such theories, and in almost all other theories known to date, QNM frequencies are only known in the slow-rotation limit, a limitation that could be lifted with the spectral method.   
Another possible application of our spectral method is the study of BH spectral instabilities.
Typically, the criterion of spectral instability is characterized by modifications to an effective potential~\cite{spectral_instability_01, spectral_instability_02, spectral_instability_03, spectral_instability_04}. 
In the spectral method, however, the notion of the effective potential is not manifest, as the method does not need master equations governing the gravitational perturbations. 
To apply the spectral method to study spectral instabilities, we would need to reconcile it with the notion of an effective potential. 

\section*{Acknowledgement}

The authors acknowledge the support from the Simons Foundation through Award No. 896696 and the NSF through award PHY-2207650. 
The authors would like to acknowledge Emanuele Berti, Mark H.Y. Cheung, Pedro Ferreira, Thomas Helfer, Tjonnie Li, Lionel London, and Justin Ripley for insightful discussion, and Vitor Cardoso, Jose Santos and Leo Stein for comments on the initial manuscript.
A.K.W.C would like to thank Alan Tsz Lok Lam and Lap Ming Lin for useful advice offered at the beginning of this work. 
The calculations and results reported in this paper were jointly produced using the computational resources of the department of physics at King's College London and the Illinois Campus Cluster, a computing resource that is operated by the Illinois Campus Cluster Program (ICCP) in conjunction with NCSA, and is supported by funds from the University of Illinois at Urbana-Champaign. 

\appendix

\section{Symbols}
\label{sec:Appendix_A}

The calculations presented in this paper involved numerous symbols. 
For convenience of the reader, we provide a list of the symbols and their definitions in this Appendix. 

\begin{itemize}
    \item $ A^{\ell}_i (r) $ is the asymptotic prefactor of the $i$ th perturbation variable, first defined in Equation~\eqref{eq:asym_prefactor}. 
    \item $d_{r}$ is the degree of $r$ of the coefficient of the partial derivative of the linearized Einstein equations, first defined in Equation~\eqref{eq:pertFE-1}
    \item $d_{\chi}$ is the degree of $\chi$ of the coefficient of the partial derivative of the linearized Einstein equations, first defined in Equation~\eqref{eq:pertFE-1}. 
    \item $d_{z}$ is the degree of $z$ of the coefficient of the partial derivative of the compactified linearized Einstein equations, first defined in Equation~\eqref{eq:system_3}. 
    \item $\mathcal{D}(N)$ is the modulus difference of the optimally truncated quasinormal-mode frequency over successive iterations, first defined in Equation~\eqref{eq:diff_over_iterations}. 
    \item $\mathbb{D}(\omega) $ is the coefficient matrix of spectral decomposition, from one particular basis to another, first defined in Equation~\eqref{eq:pertFE-2}. 
    \item $\tilde{\mathbb{D}} (\omega)$ is the augmented matrix of the coefficients of spectral decomposition, first defined in Equation~\eqref{eq:augmented_matrix_02}. 
    \item $\delta^{\rm Re/Im}$ is the numerical uncertainty of the real and imaginary parts of the QNM frequencies computed using the spectral method, first defined in Equation~\eqref{eq:computational_uncertainty}. 
    \item $\Delta^{\rm Re/Im}$ is the relative fractional error in the real and imaginary parts of the QNM frequencies computed using the spectral method and the Teukolsky equations, first defined in Equation~\eqref{eq:relative_error}. 
    \item $\mathcal{E}(N)$ is the absolute error between the QNM frequencies computed using the spectral method, $\omega(\text{spectral})$, and Leaver's method to solve for the QNM modes $\omega (\text{L})$, first defined in Equation~\eqref{eq:absolute_error}. 
    \item $\mathcal{G}_{i, \gamma, \delta, \sigma, \alpha, \beta, j} $ is the coefficient of $\omega^\gamma r^{\delta} \chi^{\sigma} \partial_{r}^{\alpha} \partial_{\chi}^{\beta} h_j$ of the linearized Einstein equations of $h_j$, first defined in Eq.~\eqref{eq:pertFE-1}. 
    \item $h_i(r, \chi)$ is the functions of metric perturbations, first defined in Eqs.~\eqref{eq:odd} and \eqref{eq:even}. 
    \item $i$ the subscript is the component of the metric perturbation functions and $i= 1, ..., 6$, first defined in Eqs.~\eqref{eq:odd} and \eqref{eq:even}. 
    \item $\mathcal{K}_{i, \alpha, \beta, \gamma, \delta, \sigma, j} $ is the coefficient of $\omega^\gamma z^{\delta} \chi^{\sigma} \partial_{z}^{\alpha} \partial_{\chi}^{\beta}(...)$ of the linearized Einstein equations in $z$ and $\chi$, first defined in Equation~\eqref{eq:system_3}. 
    \item $l$ is the azimuthal mode number of the gravitational QNMs, first defined in Equation~\eqref{sec:intro}. 
    \item $\ell$ is the degree of associate Legendre polynomial used in spectral decomposition, first defined in Equation~\eqref{eq:Thetai-def}.
    \item $\lambda (N) $ is the generalized eigenvalue of the linear matrix equation Equation~\eqref{eq:matrix_equation} obtained using $N$ Chebyshev and associated Legendre polynomials, first defined in Equation~\eqref{eq:evalue_convergence}. 
    \item $M$ is the BH mass, which is taken to be $M=1$ throughout this work, first defined in Equation~\eqref{eq:metric}. 
    \item $\mathbb{M}(r)$ is the coefficient matrix of the system of ordinary differential equations , first defined in Equation~\eqref{eq:ODE_system_03}.
    \item $\mathbb{M}_{k}$ is the coefficient matrix of $r^k$ term of the asymptotic expansion of $\mathbb{M}(r)$, first defined in Equation~\eqref{eq:M_r}.
    \item $m$ is the azimuthal number of the metric perturbations, first defined in Eqs.~\eqref{eq:odd} and \eqref{eq:even}. 
    \item $N$ is the number of the Chebyshev and associated Legendre polynomials used in the full spectral decomposition, first defined in Sec.~\ref{sec:setup}. 
    \item $N_{\rm opt}$ is the optimal truncation order for the frequency computation, first defined in Equation~\eqref{eq:w_optimally_truncated}. 
    \item $\mathcal{N}$ is the normalization factor of spectral decomposition, first defined in Equation~\eqref{eq:normalization_const}. 
    \item $\mathcal{N}_{\chi}$ is the number of the associated Legendre polynomials included in the spectral decomposition, first defined in Equation~\eqref{eq:spectral_decoposition_factorized_finite}. 
    \item $\mathcal{N}_{z}$ is the number of the Chebyshev polynomials included in the spectral decomposition, first defined in Equation~\eqref{eq:spectral_decoposition_factorized_finite}. 
    \item $r_\h = 2M$ is the radial coordinate of the position of the event horizon of the Schwarzschild BH, first defined below Equation~\eqref{eq:Schwarzschild_fn}. 
    \item $r_*$ is the tortoise coordinate, first defined in Equation~\eqref{eq:tortoise_coordinate}. 
    \item $p_{H}$ is the Poincar\'e rank of $ - \epsilon^{-2} \mathbb{M}(\epsilon)$ at $r = r_\h $, first defined in Equation~\eqref{eq:ODE_system_H}. 
    \item $p_{\infty}$ is the Poincar\'e rank of $\mathbb{M}(r)$ at $r = \infty $, first defined in Equation~\eqref{eq:ODE_system_03}. 
    \item $\mathbb{Q}$ is the coefficient matrix of $d \textbf{y} / dr $ of the system of ordinary differential equations , first defined in Equation~\eqref{eq:ODE_system_01}. 
    \item $\mathbb{\tilde{Q}}$ is the coefficient matrix of $d \textbf{y} / dr $ of the system of ordinary differential equations, after algebraic variables have been removed, first defined in Equation~\eqref{eq:ODE_system_AV_removed}. 
    \item $\mathbb{R}$ is the coefficient matrix of $ \textbf{y} $ of the system of ordinary differential equations , first defined in Equation~\eqref{eq:ODE_system_01}.
    \item $\tilde{\mathbb{R}}$ is the coefficient matrix of $ \textbf{y} $ of the system of ordinary differential equations , after algebraic variables have been removed, first defined in Equation~\eqref{eq:ODE_system_AV_removed}.
    \item $\rho_{\infty}^{(i)}$ and $\rho_H^{(i)}$ are the parameters that characterize the boundary conditions of $h_i$ in spatial infinity and at the horizon, first defined in Eqs.~\eqref{eq:asymptotic_limits} and \eqref{eq:asymptotic_limits2}. 
    \item $\omega_{q}(\rm L)$ is the frequency of the QNM $q$ computed using the Leaver method, first defined in Equation~\eqref{eq:Leaver_QNMFs}. 
    \item $\omega_{q}^{\rm opt}$ is the optimally truncated frequency of the QNM $q$, first defined in Equation~\eqref{eq:w_optimally_truncated}. 
    \item $y_i^{\ell}$ is the component of $h_i (r, \chi)$ projected along $P_{\ell}^{|m|}$, first defined in Equation~\eqref{eq:separation-variable-2}.
    \item $z = \frac{2 r_\h}{r} - 1$ is the variable that maps $r$ into a finite domain, first defined in Equation~\eqref{eq:z}. 
\end{itemize}

\section{An explicit example of the asymptotic behavior at the event horizon and spatial infinity}
\label{sec:Diagonalization_example}



In this Appendix, we explicitly apply the procedures described in Sec.~\ref{sec:asymptotic_behaviour} to obtain the asymptotic behavior of the metric variables for a Schwarzschild BH, setting its mass $M=1$ and $m=2$ for simplicity.

To estimate the asymptotic behavior, we need to specifically study six equations out of the ten linearized Einstein equations. 
In this example, we focus on $\{ tr, t \chi, t \phi, rr, r \chi, r \phi \} $ because these six equations contain the second-order $r$ derivative of only one perturbation function, $h_5 (r, \chi)$. 
Thus, we have the $Y_5$ element but no other $Y_{i\neq 5}$ elements in $\textbf{y}$. 
Other choices of six equations contain the second-order $r$ derivatives of more perturbation functions, making the calculations less convenient. 
To limit the length of this example, we only include two associate Legendre polynomials ($\mathcal{N}_{\chi} = 2$), 
\begin{equation}
h_i (r, \chi) = \sum_{\ell = 2, 3} y_i^{\ell}(r) P_{\ell}^{|m|}(\chi). 
\end{equation}

The resulting system of ordinary differential equations  contains
\begin{equation*}
\frac{d^2 y_5^2}{d r^2}, ~~~ \text{and} ~~~ \frac{d^2 y_5^3}{d r^2}. 
\end{equation*}
To keep the system of ordinary differential equations  first order, we write 
\begin{equation*}
\begin{split}
\frac{d y_5^2}{d r} = Y_5^2 ~~ \text{and} ~~ \frac{d y_5^3}{d r} = Y_5^3. 
\end{split}
\end{equation*}
Hence, $\textbf{y}$ is a 14-vector [$14 = 2 \times (6+1)$], 
\begin{equation}
\begin{split}
\textbf{y} = (& y_1^2, y_1^3, y_2^2, y_2^3, y_3^2, y_3^3, y_4^2, \\
& y_4^3, y_5^2, y_5^3, y_6^3, y_6^3, Y_5^2, Y_5^3)^{\rm T}. 
\end{split}
\end{equation}
For the sake of clarity, we define the following expressions which are recurring in the coefficient matrices
\begin{equation}
\begin{split}
A & = (r-2) r^2 \\
B & = (r-2)^2 \\
C & = r^2-3 r+2\\
D & = r^3 \omega \\
\end{split}
\end{equation}
In terms of $A, B, C$ and $D$, 
\begin{widetext}
\begin{equation}
\mathbb{Q}(r) = 
{
\begin{pmatrix}
 0 & 0 & 0 & 0 & 0 & 0 & \frac{12}{7} i A \omega  & 0 & 0 & 0 & 0 & 0 & 0 & 0 \\
 0 & 0 & 0 & -\frac{20}{7} A & 0 & 0 & 0 & 0 & 0 & 0 & -\frac{12}{7} A \omega  & 0 & \frac{12}{7} i A & 0 \\
 0 & 0 & \frac{12}{7} i A & 0 & 0 & 0 & 0 & 0 & 0 & 0 & 0 & \frac{20}{7} i A \omega  & 0 & \frac{20}{7} A \\
 -\frac{12}{7} B & 0 & 0 & 0 & 0 & 0 & -\frac{12}{7} C & 0 & 0 & 0 & 0 & 0 & 0 & 0 \\
 0 & \frac{20}{7} A & 0 & 0 & 0 & 0 & 0 & \frac{20}{7} A & -\frac{12}{7} D & 0 & 0 & 0 & 0 & 0 \\
 -\frac{12}{7} i A & 0 & 0 & 0 & 0 & 0 & -\frac{12}{7} i A & 0 & 0 & \frac{20}{7} i D  & 0 & 0 & 0 & 0 \\
 0 & 0 & 0 & 0 & 0 & 0 & 0 & 0 & 1 & 0 & 0 & 0 & 0 & 0 \\
 0 & 0 & 0 & 0 & 0 & 0 & 0 & \frac{4}{3} i A \omega  & 0 & 0 & 0 & 0 & 0 & 0 \\
 0 & 0 & \frac{4}{15} A & 0 & 0 & 0 & 0 & 0 & 0 & 0 & 0 & -\frac{4}{3} A \omega  & 0 & \frac{4}{3} i A \\
 0 & 0 & 0 & \frac{4}{3} i A & 0 & 0 & 0 & 0 & 0 & 0 & -\frac{4}{15} i A \omega  & 0 & -\frac{4}{15} A & 0 \\
 0 & -\frac{4}{3} B & 0 & 0 & 0 & 0 & 0 & -\frac{4}{3} C & 0 & 0 & 0 & 0 & 0 & 0 \\
 -\frac{4}{15} A & 0 & 0 & 0 & 0 & 0 & -\frac{4}{15} A & 0 & 0 & -\frac{4}{3} D & 0 & 0 & 0 & 0 \\
 0 & -\frac{4}{3} i A & 0 & 0 & 0 & 0 & 0 & -\frac{4}{3} i A & -\frac{4}{15} i D  & 0 & 0 & 0 & 0 & 0 \\
 0 & 0 & 0 & 0 & 0 & 0 & 0 & 0 & 0 & 1 & 0 & 0 & 0 & 0 \\
\end{pmatrix}. 
}
\end{equation}

We now perform the procedures described in Sec.~\ref{sec:asymptotic_behaviour} to the system of ordinary differential equations. 
The quantity $\mathbb{Q}(r)$ is a $14 \times 14 $ matrix but has a rank of $10$.  
    Thus, we should have $14 - \text{rank}(\mathbb{Q}) = 4 $ algebraic equations. 
    After a few elementary row operations, we simplify $\mathbb{Q}(r)$ into 
    \begin{equation}
    \mathbb{Q}(r) = 
    {
    \begin{pmatrix}
     0 & 0 & 0 & 0 & 0 & 0 & \frac{12}{7} i A \omega  & 0 & 0 & 0 & 0 & 0 & 0 & 0 \\
     0 & 0 & 0 & -\frac{20}{7} A & 0 & 0 & 0 & 0 & 0 & 0 & -\frac{12}{7} A \omega  & 0 & \frac{12}{7} i A & 0 \\
     0 & 0 & \frac{12}{7} i A & 0 & 0 & 0 & 0 & 0 & 0 & 0 & 0 & \frac{20}{7} i A \omega  & 0 & \frac{20}{7} A \\
     -\frac{12}{7} B & 0 & 0 & 0 & 0 & 0 & -\frac{12}{7} C & 0 & 0 & 0 & 0 & 0 & 0 & 0 \\
     0 & \frac{20}{7} A & 0 & 0 & 0 & 0 & 0 & \frac{20}{7} A & -\frac{12}{7} D & 0 & 0 & 0 & 0 & 0 \\
     0 & 0 & 0 & 0 & 0 & 0 & 0 & 0 & 0 & \frac{20}{7} i r^3 \omega  & 0 & 0 & 0 & 0 \\
     0 & 0 & 0 & 0 & 0 & 0 & 0 & 0 & 1 & 0 & 0 & 0 & 0 & 0 \\
     0 & 0 & 0 & 0 & 0 & 0 & 0 & \frac{4}{3} i A \omega  & 0 & 0 & 0 & 0 & 0 & 0 \\
     0 & 0 & 0 & 0 & 0 & 0 & 0 & 0 & 0 & 0 & 0 & -\frac{16}{9} A \omega  & 0 & \frac{16}{9} i A \\
     0 & 0 & 0 & 0 & 0 & 0 & 0 & 0 & 0 & 0 & -\frac{16}{15} i A \omega  & 0 & -\frac{16}{15} A & 0 \\
     0 & 0 & 0 & 0 & 0 & 0 & 0 & 0 & 0 & 0 & 0 & 0 & 0 & 0 \\
     0 & 0 & 0 & 0 & 0 & 0 & 0 & 0 & 0 & 0 & 0 & 0 & 0 & 0 \\
     0 & 0 & 0 & 0 & 0 & 0 & 0 & 0 & 0 & 0 & 0 & 0 & 0 & 0 \\
     0 & 0 & 0 & 0 & 0 & 0 & 0 & 0 & 0 & 0 & 0 & 0 & 0 & 0 \\
    \end{pmatrix}. 
    }
    \end{equation}
    The nonzero elements of the corresponding $\mathbb{R}(r)$ after the elementary row operations are
    \end{widetext}
    \begin{align*}
    \mathbb{R}_{13}(r) & = \frac{36 (r-2)}{7}, \\
    \mathbb{R}_{15}(r) & = \frac{12}{7} i (r-2) r \omega, \\
    \mathbb{R}_{17}(r) & = -\frac{12}{7} i (r-3) r \omega, \\
    \mathbb{R}_{24}(r) & = \frac{40 r}{7}, \\
    \mathbb{R}_{26}(r) & = -\frac{20}{7} i r^3 \omega, \\
    \mathbb{R}_{28}(r) & = -\frac{20}{7} i r^3 \omega, \\
    \mathbb{R}_{29}(r) & = \frac{24}{7} i (3 r-2), \\
    \mathbb{R}_{2~11}(r) & = \frac{24}{7} (r-2) r \omega, \\
    \mathbb{R}_{33}(r) & = -\frac{24}{7} i r, \\
    \mathbb{R}_{35}(r) & = -\frac{12}{7} r^3 \omega , \\
    \mathbb{R}_{37}(r) & = -\frac{12}{7} r^3 \omega , \\
    \mathbb{R}_{3~10}(r) & = \frac{80}{7} (3 r-1), \\ 
    \mathbb{R}_{3~12}(r) & = - \frac{40}{7} i (r-2) r \omega, \\
    \mathbb{R}_{41}(r) & = - \frac{36}{7} (r-2), \\
    \mathbb{R}_{43}(r) & = \frac{24}{7} i (r-2) r \omega, \\
    \mathbb{R}_{45}(r) & = -\frac{12}{7} (r-2), \\
    \mathbb{R}_{47}(r) & = \frac{12}{7} \left(r^3 \omega ^2-2 r+4\right), \\
    \mathbb{R}_{52}(r) & = \frac{20}{7} (r-3) r, \\
    \mathbb{R}_{54}(r) & = -\frac{20}{7} i r^3 \omega, \\
    \mathbb{R}_{56}(r) & = \frac{20}{7} (r-1) r, \\
    \mathbb{R}_{59}(r) & = -\frac{24}{7} r^2 \omega, \\
    \mathbb{R}_{5~11}(r) & = \frac{12}{7} i \left(r^3 \omega ^2-4 r+8\right), \\
    \mathbb{R}_{61}(r) & = \frac{12}{7} i r (2 r+3), \\
    \mathbb{R}_{63}(r) & = \frac{12 \left(r^3 \omega ^2-3\right)}{7 \omega }, \\
    \mathbb{R}_{67}(r) & = -\frac{12 i r \left(r^4 \omega ^2-2 r^2+3 r+3\right)}{7 (r-2)}, \\
    \mathbb{R}_{6~10}(r) & = \frac{40}{7} i r^2 \omega, \\
    \mathbb{R}_{6~12}(r) & = \frac{20}{7} \left(r^3 \omega ^2-10 r+20\right), \\
    \mathbb{R}_{7~13}(r) & = 1, \\
    \mathbb{R}_{84}(r) & = 8 (r-2), \\ 
    \mathbb{R}_{86}(r) & = \frac{4}{3} i (r-2) r \omega, \\
    \mathbb{R}_{88}(r) & = -\frac{4}{3} i (r-3) r \omega, \\
    \mathbb{R}_{9~10}(r) & = \frac{64}{9} i (3 r-1), \\
    \mathbb{R}_{9~11}(r) & = \frac{32}{9} (r-2) r \omega, \\
    \mathbb{R}_{10~9}(r) & = \frac{32}{15} (2-3 r), \\
    \mathbb{R}_{10~11}(r) & = \frac{32}{15} i (r-2) r \omega, \\
    \mathbb{R}_{11~2}(r) & = \frac{4}{3} \left(-5 r+\frac{6}{r}+7\right), \\
    \mathbb{R}_{11~4}(r) & = \frac{4 i (r-2) \left(r^3 \omega ^2-6\right)}{3 r^2 \omega }, \\
    \mathbb{R}_{11~8}(r) & = \frac{4 \left(r^4 \omega ^2-5 r^2+9 r+3\right)}{3 r}, \\
    \mathbb{R}_{11~9}(r) & = - \frac{8}{5} (r-2) \omega, \\
    \mathbb{R}_{11~11}(r) & = \frac{4 i (r-2) \left(r^3 \omega ^2-4 r+8\right)}{5 r^2}, \\
    \mathbb{R}_{11~13}(r) & = \frac{4}{5} (r-2) r \omega, \\
    \mathbb{R}_{12~1}(r) & = \frac{16}{15} r (2 r+3), \\
    \mathbb{R}_{12~3}(r) & = -\frac{16 i \left(r^3 \omega ^2-3\right)}{15 \omega }, \\
    \mathbb{R}_{12~7}(r) & = -\frac{16 r \left(r^4 \omega ^2-2 r^2+3 r+3\right)}{15 (r-2)}, \\
    \mathbb{R}_{13~9}(r) & = - \frac{32}{15} i r^2 \omega, \\
    \mathbb{R}_{13~11}(r) & = -\frac{16}{15} \left(r^3 \omega ^2-4 r+8\right), \\
    \mathbb{R}_{13~13}(r) & = \frac{16}{15} i r^3 \omega, \\
    \mathbb{R}_{14~1}(r) & = -\frac{6 r+9}{5 r^2 \omega }, \\
    \mathbb{R}_{14~3}(r) & = \frac{3 i}{5}-\frac{9 i}{5 r^3 \omega ^2}, \\
    \mathbb{R}_{14~7}(r) & = \frac{3 \left(r^4 \omega ^2-2 r^2+3 r+3\right)}{5 (r-2) r^2 \omega }, \\
    \mathbb{R}_{14~10}(r) & = -\frac{2}{r}, \\
    \mathbb{R}_{14~12}(r) & = \frac{i \left(r^3 \omega ^2-10 r+20\right)}{r^3 \omega }, \\
    \mathbb{R}_{14~14}(r) & = 1. 
    \end{align*}
\begin{widetext}
By reading the fifth and sixth column of $\mathbb{Q}(r)$, we identify two algebraic variables, $y_3^2$ and $y_3^3$. 
    By solving the ordinary differential equations (ODEs) represented by the first and second row of $\mathbb{Q}(r)$ and $\mathbb{R}(r)$ for $y_3^2$ and $y_3^3$, we have 
    \begin{equation}\label{eq:Subh3}
    \begin{split}
    y_3^2 & = \frac{3 i}{r \omega} y^{2}_{2} + \frac{(r-3)}{r-2} y_{4}^{2} + r \frac{d y_{4}^{2}}{d r}, \\ 
    y_3^3 &= -\frac{1}{5 r^3 \omega} \left( 5 r^3 \omega  y_{4}^{3}+3 r^2(r-2) \frac{d Y^{5}_{2}}{d r} + i r \left((r-2) \left(3 r \omega  \frac{d y_{6}^{2}}{d r} +6 \omega  \frac{d y_{6}^{2}}{d r} +5 r \frac{d y_{2}^{3}}{d r} \right)+10 y_2^3\right) + (12 - 18r) y_5^2 \right). 
    \end{split}
    \end{equation}
    As all algebraic variables have been expressed in terms of the differential variables and at most their first-order $r$ derivative, the system of ordinary differential equations  remains first order if we substitute the algebraic variables back into the system. 

We substitute $y_3^2 $ and $y_3^3$ back to the system of ordinary differential equations . Now $\tilde{\textbf{y}}$ is a 12 vector ($2 = 14 -2$), and $\tilde{\mathbb{Q}}(r)$ and $\tilde{\mathbb{R}}(r)$ are $12\times12$ matrices.
After some elementary row operations to simplify  $\tilde{\mathbb{Q}}(r)$, we have 
\begin{equation}
\tilde{\mathbb{Q}}(r) = 
{
\begin{pmatrix}
 0 & 0 & \frac{12}{7} i A & 0 & \frac{12 r^4 \omega }{7} & 0 & 0 & 0 & 0 & \frac{20}{7} i A \omega  & 0 & \frac{20}{7} A \\
 -\frac{12}{7} B & 0 & 0 & 0 & \frac{12 (r-2)}{7} & 0 & 0 & 0 & 0 & 0 & 0 & 0 \\
 0 & \frac{20}{7} A & 0 & \frac{20 i C}{7 \omega } & 0 & \frac{20}{7} A & -\frac{12 D}{7} & 0 & \frac{12}{7} i C & 0 & \frac{12 C}{7 \omega } & 0 \\
 0 & 0 & 0 & 0 & 0 & 0 & 0 & \frac{20}{7} i D  & 0 & 0 & 0 & 0 \\
 0 & 0 & 0 & 0 & 0 & 0 & 1 & 0 & 0 & 0 & 0 & 0 \\
 0 & 0 & 0 & - \frac{4}{3} B & 0 & \frac{4}{3} i A \omega  & 0 & 0 & - \frac{4}{5} B \omega  & 0 & \frac{4}{5} i B & 0 \\
 0 & 0 & 0 & 0 & 0 & 0 & 0 & 0 & 0 & - \frac{16}{9} A \omega  & 0 & \frac{16}{9} i A \\
 0 & 0 & 0 & 0 & 0 & 0 & 0 & 0 & - \frac{16}{15} i A \omega  & 0 & -\frac{16}{15} A & 0 \\
 0 & 0 & 0 & 0 & 0 & 0 & 0 & 0 & 0 & 0 & 0 & 0 \\
 0 & 0 & 0 & 0 & 0 & 0 & 0 & 0 & 0 & 0 & 0 & 0 \\
 0 & 0 & 0 & 0 & 0 & 0 & 0 & 0 & 0 & 0 & 0 & 0 \\
 0 & 0 & 0 & 0 & 0 & 0 & 0 & 0 & 0 & 0 & 0 & 0 \\
\end{pmatrix}, 
}
\end{equation}
and the corresponding $\tilde{\mathbb{R}}$ has the following nonzero elements, 
\end{widetext}
\begin{align*}
\tilde{\mathbb{R}}_{13}(r) & = - \frac{12}{7} i r (3 r+2), \\
\tilde{\mathbb{R}}_{15}(r) & = -\frac{12 r^3 (2 r-5) \omega }{7 (r-2)} , \\
\tilde{\mathbb{R}}_{18}(r) & = \frac{80}{7} (3 r-1) , \\
\tilde{\mathbb{R}}_{1~10}(r) & = \frac{1}{7} (-40) i (r-2) r \omega, \\
\tilde{\mathbb{R}}_{21}(r) & = - \frac{36}{7} (r-2), \\
\tilde{\mathbb{R}}_{23}(r) & = \frac{12 i (r-2) \left(2 r^2 \omega ^2-3\right)}{7 r \omega }, \\
\tilde{\mathbb{R}}_{25}(r) & = \frac{12}{7} \left(r^3 \omega ^2-3 r+7\right), \\
\tilde{\mathbb{R}}_{32}(r) & = \frac{20}{7} (r-3) r, \\
\tilde{\mathbb{R}}_{34}(r) & = -\frac{20 i \left(r^4 \omega ^2+2 r-2\right)}{7 r \omega }, \\
\tilde{\mathbb{R}}_{36}(r) & = - \frac{20}{7} (r-1) r, \\
\tilde{\mathbb{R}}_{37}(r) & = -\frac{24 \left(r^4 \omega ^2-3 r^2+5 r-2\right)}{7 r^2 \omega }, \\
\tilde{\mathbb{R}}_{39}(r) & = \frac{12 i \left(r^4 \omega ^2-6 r^2+14 r-4\right)}{7 r}, \\
\tilde{\mathbb{R}}_{41}(r) & = \frac{12}{7} i r (2 r+3), \\
\tilde{\mathbb{R}}_{43}(r) & = \frac{12 \left(r^3 \omega ^2-3\right)}{7 \omega }, \\
\tilde{\mathbb{R}}_{45}(r) & = -\frac{12 i r \left(r^4 \omega ^2-2 r^2+3 r+3\right)}{7 (r-2)}, \\
\tilde{\mathbb{R}}_{48}(r) & = \frac{40}{7} i r^2 \omega, \\
\tilde{\mathbb{R}}_{4~10}(r) & = \frac{20}{7} \left(r^3 \omega ^2-10 r+20\right), \\
\tilde{\mathbb{R}}_{5~11}(r) & = 1, \\
\tilde{\mathbb{R}}_{64}(r) & = \frac{8}{3} \left(3 r-\frac{2}{r}-5\right), \\
\tilde{\mathbb{R}}_{66}(r) & = -\frac{4}{3} i r (2 r-5) \omega, \\ 
\tilde{\mathbb{R}}_{67}(r) & = \frac{8 i \left(3 r^2-8 r+4\right)}{5 r^2}, \\
\tilde{\mathbb{R}}_{69}(r) & = \frac{8 (r-2)^2 \omega }{5 r}, \\
\tilde{\mathbb{R}}_{78}(r) & = \frac{64}{9} i (3 r-1), \\
\tilde{\mathbb{R}}_{7~10}(r) & = \frac{32}{9} (r-2) r \omega, \\
\tilde{\mathbb{R}}_{87}(r) & = - \frac{32}{15} (3 r-2), \\
\tilde{\mathbb{R}}_{88}(r) & = \frac{32}{15} i (r-2) r \omega, \\
\tilde{\mathbb{R}}_{92}(r) & = \frac{4}{3} \left(-5 r+\frac{6}{r}+7\right), \\
\tilde{\mathbb{R}}_{94}(r) & = \frac{4 i (r-2) \left(r^3 \omega ^2-6\right)}{3 r^2 \omega }, \\
\tilde{\mathbb{R}}_{96}(r) & = \frac{4 \left(r^4 \omega ^2-5 r^2+9 r+3\right)}{3 r}, \\
\tilde{\mathbb{R}}_{97}(r) & = - \frac{8}{5} (r-2) \omega, \\
\tilde{\mathbb{R}}_{99}(r) & = \frac{4 i (r-2) \left(r^3 \omega ^2-4 r+8\right)}{5 r^2}, \\
\tilde{\mathbb{R}}_{9~11}(r) & = \frac{4}{5} (r-2) r \omega, \\
\tilde{\mathbb{R}}_{10~1}(r) & = \frac{16}{15} r (2 r+3), \\
\tilde{\mathbb{R}}_{10~3}(r) & = -\frac{16 i \left(r^3 \omega ^2-3\right)}{15 \omega }, \\
\tilde{\mathbb{R}}_{10~5}(r) & = -\frac{16 r \left(r^4 \omega ^2-2 r^2+3 r+3\right)}{15 (r-2)}, \\
\tilde{\mathbb{R}}_{11~7}(r) & = - \frac{32}{15} i r^2 \omega, \\
\tilde{\mathbb{R}}_{11~9}(r) & = -\frac{16}{15} \left(r^3 \omega ^2-4 r+8\right), \\
\tilde{\mathbb{R}}_{11~11}(r) & = \frac{16}{15} i r^3 \omega, \\
\tilde{\mathbb{R}}_{12~1}(r) & = -\frac{6 r+9}{5 r^2 \omega }, \\
\tilde{\mathbb{R}}_{12~3}(r) & = \frac{3 i}{5}-\frac{9 i}{5 r^3 \omega ^2}, \\
\tilde{\mathbb{R}}_{12~5}(r) & = \frac{3 \left(r^4 \omega ^2-2 r^2+3 r+3\right)}{5 (r-2) r^2 \omega }, \\
\tilde{\mathbb{R}}_{12~8}(r) & = -\frac{2}{r}, \\
\tilde{\mathbb{R}}_{12~10}(r) & = \frac{i \left(r^3 \omega ^2-10 r+20\right)}{r^3 \omega }, \\
\tilde{\mathbb{R}}_{12~12}(r) & = 1 \\
\end{align*}
\begin{widetext}
    Now the ninth to 12th row of $\tilde{\mathbb{Q}}(r)$ are all zeros. 
    By reading the corresponding rows of $\tilde{\mathbb{R}}(r)$, we obtain the following four algebraic equations, 
    \begin{equation}
    \begin{split}
    & 4 \big(5 i (r-2) \left(r^3 \omega ^2-6\right) y_2^3 -5 r \left(5 r^2-7 r-6\right) \omega  y_1^3 +5 r \omega  \left(r^4 \omega ^2-5 r^2+9 r+3\right) y_4^3 \\
    & \quad \quad \quad \quad +3 (r-2) \omega  \left(i \left(r^3 \omega ^2-4 r+8\right) y_6^2 +r^3 \omega  Y_5^2 -2 r^2 \omega  y_5^2 \right)\big) = 0, \\
    & 16 \left(i (r-2) \left(r^3 \omega ^2-3\right) y_2^2 +r \left(-2 r^2+r+6\right) \omega  y_1^2 +r \omega  \left(r^4 \omega ^2-2 r^2+3 r+3\right) y_4^2 \right) = 0, \\
    & \left(r^3 \omega ^2-4 r+8\right) y_6^2 -i r^3 \omega Y_5^2 +2 i r^2 \omega y_5^2 = 0 \\
    & 3 i (r-2) \left(r^3 \omega ^2-3\right) y_2^2 -3 r \left(2 r^2-r-6\right) \omega y_1^2 +3 r \omega  \left(r^4 \omega ^2-2 r^2+3 r+3\right) y_4^2 \\
    & \quad \quad \quad \quad +5 i (r-2) \omega  \left(\left(r^3 \omega ^2-10 r+20\right) y_6^3 -i r^3 \omega Y_5^3 + 2 i r^2 \omega Y_5^3\right) = 0 \,. 
    \end{split}
    \end{equation}
    These algebraic equations allow us to express four differential variables in terms of the remaining eight $(=12-4)$ differential variables in 81 different combinations. 
    Each of these 81 combinations leads to a $\mathbb{M}(r)$ of $0 \leq p_{\infty} \leq 2$. 
    Eventually, we solve the algebraic equations for $y_4^2, y_4^3, y_7^2$ and $y_7^3$, 
    \begin{equation}\label{eq:eliminate_diff_vars}
    \begin{split}
    y_4^2 & = \frac{(r-2) \left(r (2 r+3) \omega  y_1^2-i \left(r^3 \omega ^2-3\right) y_2^2\right)}{r \omega  \left(r \left(r^3 \omega ^2-2 r+3\right)+3\right)}, \\
    y_4^3 & = \frac{(r-2) \left(r (5 r+3) \omega  y_1^3-i \left(r^3 \omega ^2-6\right) y_2^3\right)}{r \omega  \left(r \left(r^3 \omega ^2-5 r+9\right)+3\right)}, \\
    Y_5^2 & = \frac{2 y_5^2}{r}-\frac{i \left(r^3 \omega ^2-4 r+8\right) y_6^2}{r^3 \omega }, \\
    Y_5^3 & = \frac{2 y_5^3}{r}-\frac{i \left(r^3 \omega ^2-10 r+20\right) y_6^3}{r^3 \omega }, 
    \end{split}
    \end{equation}
    for two advantages. 
    First, eliminating these four variables leads to a $\mathbb{M}(r)$ of $p_{\infty} = 0 $, with both $\mathbb{M}_0$ and $\mathbb{M}_{-1} $ diagonalizable. 
    This is a crucial advantage because it drastically reduces the difficulty to diagonalize $\mathbb{M}(r)$ and study the asymptotic behavior of $\textbf{y}$ for larger $p_{\infty}$. 
    Second, this combination eliminates all the differential variables concerning the second-order $r$ derivative of the metric perturbation functions, which are less relevant to our studies as no metric perturbations are expressed as the $r$ derivatives of $h_i$. 
    
We now have a system of ordinary differential equations of the form of Equation~\eqref{eq:ODE_system_03} concerning $\text{rank} (\mathbb{Q}) - N_{\rm alg} = 10 - 2 = 8 $ differential variables, with
\begin{equation}
\begin{split}
\textbf{z} &= (y_1^2, y_1^3, y_2^2, y_2^3, y_5^2, y_5^3, y_6^2. y_6^3)^{\rm T}.
\end{split}
\end{equation} 
The nonzero elements of $\mathbb{M}(r)$ are
\begin{equation}
\begin{split}
\mathbb{M}_{11} (r) & = \frac{r^5 \omega ^2-4 r^4 \omega ^2+4 r^2-12 r+6}{(r-2) r \left(r^4 \omega ^2-2 r^2+3 r+3\right)}, \\
\mathbb{M}_{13} (r) & = -\frac{i \left(r^6 \omega ^4-4 r^4 \omega ^2+4 r^3 \omega ^2+r^2 \left(9 \omega ^2+6\right)-24 r+24\right)}{(r-2) r \omega  \left(r^4 \omega ^2-2 r^2+3 r+3\right)}, \\
\end{split}
\end{equation}
\begin{equation}
\begin{split}
\mathbb{M}_{22} (r) & = \frac{r^5 \omega ^2-4 r^4 \omega ^2+7 r^2-18 r+6}{(r-2) r \left(r^4 \omega ^2-5 r^2+9 r+3\right)}, \\
\mathbb{M}_{24} (r) & = -\frac{i \left(r^6 \omega ^4-10 r^4 \omega ^2+16 r^3 \omega ^2+r^2 \left(9 \omega ^2+30\right)-120 r+120\right)}{(r-2) r \omega  \left(r^4 \omega ^2-5 r^2+9 r+3\right)}, \\
\end{split}
\end{equation}
\end{widetext}
\begin{align*}
\mathbb{M}_{31} (r) & = -\frac{i r \omega  \left(r^4 \omega ^2-4 r^2+4 r+9\right)}{(r-2) \left(r^4 \omega ^2-2 r^2+3 r+3\right)}, \\
\mathbb{M}_{33} (r) & = \frac{r^5 \omega ^2-4 r^4 \omega ^2+r^2-6}{(r-2) r \left(r^4 \omega ^2-2 r^2+3 r+3\right)}, \\
\mathbb{M}_{42} (r) & = -\frac{i r \omega  \left(r^4 \omega ^2-10 r^2+16 r+9\right)}{(r-2) \left(r^4 \omega ^2-5 r^2+9 r+3\right)}, \\
\mathbb{M}_{44} (r) & = \frac{r^5 \omega ^2-4 r^4 \omega ^2+4 r^2-6 r-6}{(r-2) r \left(r^4 \omega ^2-5 r^2+9 r+3\right)}, \\
\mathbb{M}_{55} (r) & = \frac{2}{r}, \\
\mathbb{M}_{57} (r) & = -\frac{i \left(r^3 \omega ^2-4 r+8\right)}{r^3 \omega }, \\
\mathbb{M}_{66} (r) & = \frac{2}{r}, \\
\mathbb{M}_{68} (r) & = -\frac{i \left(r^3 \omega ^2-10 r+20\right)}{r^3 \omega }, \\
\mathbb{M}_{75} (r) & = -\frac{i r^2 \omega }{(r-2)^2}, \\
\mathbb{M}_{77} (r) & = -\frac{2}{(r-2) r}, \\
\mathbb{M}_{86} (r) & = -\frac{i r^2 \omega }{(r-2)^2}, \\
\mathbb{M}_{88} (r) & = -\frac{2}{(r-2) r}. 
\end{align*}
\begin{widetext}
    At $r = \infty$, we express $\mathbb{M}(r)$ as a power series of $r$ and discard terms that drop faster than $\mathcal{O}(r^{-2})$, 
    \begin{equation}
    \begin{split}
    \mathbb{M}(r) & \approx \mathbb{M}_0 + \frac{\mathbb{M}_{-1}}{r}, 
    \end{split}
    \end{equation}
    \begin{equation}
    \begin{split}    
    \mathbb{M}_0 & = 
    \begin{pmatrix}
     0 & 0 & -i \omega  & 0 & 0 & 0 & 0 & 0 \\
     0 & 0 & 0 & -i \omega  & 0 & 0 & 0 & 0 \\
     -i \omega  & 0 & 0 & 0 & 0 & 0 & 0 & 0 \\
     0 & -i \omega  & 0 & 0 & 0 & 0 & 0 & 0 \\
     0 & 0 & 0 & 0 & 0 & 0 & -i \omega  & 0 \\
     0 & 0 & 0 & 0 & 0 & 0 & 0 & -i \omega  \\
     0 & 0 & 0 & 0 & -i \omega  & 0 & 0 & 0 \\
     0 & 0 & 0 & 0 & 0 & -i \omega  & 0 & 0 \\
    \end{pmatrix} ~~\text{and}~~\mathbb{M}_{-1} = 
    \begin{pmatrix}
    1 & 0 & -2 i \omega  & 0 & 0 & 0 & 0 & 0 \\
     0 & 1 & 0 & -2 i \omega  & 0 & 0 & 0 & 0 \\
     -2 i \omega  & 0 & 1 & 0 & 0 & 0 & 0 & 0 \\
     0 & -2 i \omega  & 0 & 1 & 0 & 0 & 0 & 0 \\
     0 & 0 & 0 & 0 & 2 & 0 & 0 & 0 \\
     0 & 0 & 0 & 0 & 0 & 2 & 0 & 0 \\
     0 & 0 & 0 & 0 & -4 i \omega  & 0 & 0 & 0 \\
     0 & 0 & 0 & 0 & 0 & -4 i \omega  & 0 & 0 \\
    \end{pmatrix}. 
    \end{split}
    \end{equation}
    Both $\mathbb{M}_0$ and $\mathbb{M}_{-1}$ are diagonalizable. 
    We first diagonalize $\mathbb{M}_0$ by writing $\mathbb{M}_0 = \mathbb{P}_{1} \mathbb{M}^{(1)}_0 \mathbb{P}^{-1}_{1}$ such that 
    \begin{equation}
    \mathbb{M}^{(1)}_0 = 
    \begin{pmatrix}
     -i \omega  & 0 & 0 & 0 & 0 & 0 & 0 & 0 \\
     0 & -i \omega  & 0 & 0 & 0 & 0 & 0 & 0 \\
     0 & 0 & -i \omega  & 0 & 0 & 0 & 0 & 0 \\
     0 & 0 & 0 & -i \omega  & 0 & 0 & 0 & 0 \\
     0 & 0 & 0 & 0 & i \omega  & 0 & 0 & 0 \\
     0 & 0 & 0 & 0 & 0 & i \omega  & 0 & 0 \\
     0 & 0 & 0 & 0 & 0 & 0 & i \omega  & 0 \\
     0 & 0 & 0 & 0 & 0 & 0 & 0 & i \omega  \\
    \end{pmatrix}
    ~~\text{and}~~
    \mathbb{P}_{1}  = 
    \begin{pmatrix}
     0 & 0 & 0 & 1 & 0 & 0 & 0 & -1 \\
     0 & 0 & 1 & 0 & 0 & 0 & -1 & 0 \\
     0 & 0 & 0 & 1 & 0 & 0 & 0 & 1 \\
     0 & 0 & 1 & 0 & 0 & 0 & 1 & 0 \\
     0 & 1 & 0 & 0 & 0 & -1 & 0 & 0 \\
     1 & 0 & 0 & 0 & -1 & 0 & 0 & 0 \\
     0 & 1 & 0 & 0 & 0 & 1 & 0 & 0 \\
     1 & 0 & 0 & 0 & 1 & 0 & 0 & 0 \\
    \end{pmatrix}. 
    \end{equation}
    We change $\textbf{z}$ into  $\textbf{z}^{(1)} = \mathbb{P}_{1} \textbf{z}$, which satisfies another system of ordinary differential equations , 
    \begin{equation}
    \begin{split}
    \frac{d \textbf{z}^{(1)}}{d r} & = \mathbb{M}^{(1)}(r) \textbf{z}^{(1)}, \\
    \mathbb{M}^{(1)} (r) & = \mathbb{P}^{-1}_{1} \mathbb{M}(r) \mathbb{P}_{1} - \mathbb{P}_{1}^{-1} \frac{d \mathbb{P}_{1}}{d r} = \mathbb{M}^{(1)}_{0} +  \frac{\mathbb{M}^{(1)}_{-1} }{r}, \\
    \mathbb{M}^{(1)}_{-1}  & = 
    \begin{pmatrix}
     1-2 i \omega  & 0 & 0 & 0 & 2 i \omega -1 & 0 & 0 & 0 \\
     0 & 1-2 i \omega  & 0 & 0 & 0 & 2 i \omega -1 & 0 & 0 \\
     0 & 0 & 1-2 i \omega  & 0 & 0 & 0 & 0 & 0 \\
     0 & 0 & 0 & 1-2 i \omega  & 0 & 0 & 0 & 0 \\
     -2 i \omega -1 & 0 & 0 & 0 & 2 i \omega +1 & 0 & 0 & 0 \\
     0 & -2 i \omega -1 & 0 & 0 & 0 & 2 i \omega +1 & 0 & 0 \\
     0 & 0 & 0 & 0 & 0 & 0 & 2 i \omega +1 & 0 \\
     0 & 0 & 0 & 0 & 0 & 0 & 0 & 2 i \omega +1 \\
    \end{pmatrix}. 
    \end{split}
    \end{equation}
    We can diagonalize $\mathbb{M}^{(1)}_{-1}$ while keeping $\mathbb{M}^{(1)}_{0} $ unchanged by further changing $\textbf{z}^{(1)}$ into $ \textbf{z}^{(2)} = \mathbb{P}_{2} \textbf{z}^{(1)}$, where $\mathbb{P}_2 = 1 + \frac{\Sigma}{r}$, provided that $\Sigma$ satisfies the matrix equation 
    \begin{equation}
    D_{-1}=\mathbb{M}_{-1}^{(1)}+\left[D_{0}, \Sigma\right]. 
    \end{equation}
    Here $D_{0}$ and $D_{-1}$ are respectively the diagonal part of $\mathbb{M}^{(1)}_{0}$ and $\mathbb{M}^{(1)}_{-1}$. 
    The matrix equation gives
    \begin{equation}
    \Sigma = \frac{1}{2 \omega}
    \begin{pmatrix}
     0 & 0 & 0 & 0 & -i (2 i \omega -1) & 0 & 0 & 0 \\
     0 & 0 & 0 & 0 & 0 & -i (2 i \omega -1) & 0 & 0 \\
     0 & 0 & 0 & 0 & 0 & 0 & 0 & 0 \\
     0 & 0 & 0 & 0 & 0 & 0 & 0 & 0 \\
     i (-2 i \omega -1) & 0 & 0 & 0 & 0 & 0 & 0 & 0 \\
     0 & i (-2 i \omega -1) & 0 & 0 & 0 & 0 & 0 & 0 \\
     0 & 0 & 0 & 0 & 0 & 0 & 0 & 0 \\
     0 & 0 & 0 & 0 & 0 & 0 & 0 & 0 \\
    \end{pmatrix}. 
    \end{equation}
    With this $\mathbb{P}_2 $, $\textbf{z}^{(2)}$ satisfies the system of ordinary differential equations  whose coefficient matrix $\mathbb{M}^{(2)} (r) $ is given by
    \begin{equation}
    \mathbb{M}^{(2)} (r) = 
    \begin{pmatrix}
     \frac{1-2 i \omega }{r}-i \omega  & 0 & 0 & 0 & 0 & 0 & 0 & 0 \\
     0 & \frac{1-2 i \omega }{r}-i \omega  & 0 & 0 & 0 & 0 & 0 & 0 \\
     0 & 0 & \frac{1-2 i \omega }{r}-i \omega  & 0 & 0 & 0 & 0 & 0 \\
     0 & 0 & 0 & \frac{1-2 i \omega }{r}-i \omega  & 0 & 0 & 0 & 0 \\
     0 & 0 & 0 & 0 & i \omega +\frac{2 i \omega +1}{r} & 0 & 0 & 0 \\
     0 & 0 & 0 & 0 & 0 & i \omega +\frac{2 i \omega +1}{r} & 0 & 0 \\
     0 & 0 & 0 & 0 & 0 & 0 & i \omega +\frac{2 i \omega +1}{r} & 0 \\
     0 & 0 & 0 & 0 & 0 & 0 & 0 & i \omega +\frac{2 i \omega +1}{r} \\
    \end{pmatrix}. 
    \end{equation}
\end{widetext}
Since $\mathbb{M}^{(2)} (r) $ is now diagonal, we can readily solve the system of ordinary differential equations  for $\textbf{z}^{(2)}$, 
\begin{equation} \label{eq:asymptotic_behaviour}
\textbf{z}^{(2)} = 
\begin{pmatrix}
c_1 r^{1-2i \omega} e^{- i \omega r} \\
c_2 r^{1-2i \omega} e^{- i \omega r} \\
c_3 r^{1-2i \omega} e^{- i \omega r} \\
c_4 r^{1-2i \omega} e^{- i \omega r} \\
c_5 r^{1+2i \omega} e^{+ i \omega r} \\
c_6 r^{1+2i \omega} e^{+ i \omega r} \\
c_7 r^{1+2i \omega} e^{+ i \omega r} \\
c_8 r^{1+2i \omega} e^{+ i \omega r} \\
\end{pmatrix}, 
\end{equation}
where $c_1, c_2, ..., c_8 $ are constants. 
The asymptotic behavior of $\textbf{z}$ can be obtained by the inverse transformations 
\begin{equation}
\textbf{z} = \mathbb{P}_1 \mathbb{P}_2 \textbf{z}^{(2)}. 
\end{equation}
As QNMs correspond to GWs that are purely outgoing at spatial infinity, we can just set $c_1 = c_2 = c_3 = c_4 = 0 $. 
By setting these constants to be zero, using the algebraic equations and the relations between the algebraic variables and differential variables, we find 
\begin{equation}\label{eq:y_ell_infinity}
y_{i}^{\ell = 2, 3} (r \rightarrow + \infty) \propto 
\begin{cases}
r^{1+2i \omega} e^{i \omega r} \text{  for } i \neq 4, \\
r^{2i \omega} e^{i \omega r} ~~~\text{ for } i = 4. 
\end{cases}
\end{equation}

Equation~\eqref{eq:asymptotic_behaviour} makes good physical sense. 
First, we simultaneously obtain the ingoing and outgoing asymptotic behavior at spatial infinity. 
This is consistent with the wave nature of the metric perturbations, since these can be ingoing and outgoing at spatial infinity. 
Second, we recognize that $r^{ \pm 2i \omega} e^{\pm i \omega r - i\omega t} \approx e^{\pm i \omega r_* - i\omega t} $ at spatial infinity, which implies that the waves are propagating at the speed of light relative to observers at spatial infinity. 
Finally, we observe that Equation~\eqref{eq:y_ell_infinity} does not depend on $\ell$. 
We confirm this observation by extending our calculations to $\mathcal{N}_{\chi} = 19 $ (thus 20 associated Legendre polynomials are included) and we obtain the same asymptotic behavior.
The independence on $\ell$ is consistent with the existing calculations of the asymptotic behavior of the gravitational perturbations around a Schwarzschild BH

The asymptotic behavior of $y^{\ell}$ at the event horizon can be similarly obtained. 
We shall omit the details of the calculations at the horizon as they are completely analogous to the above, and simply report the asymptotic behavior, which is purely ingoing at the horizon, 
\begin{equation}\label{eq:y_ell_horizon}
y_{i}^{\ell = 2, 3} (r \rightarrow r_\h) \propto 
\begin{cases}
(r-r_\h)^{-1-2i \omega} \text{  for } i \neq 4\text{ and }5, \\
(r-r_\h)^{-2i \omega}  ~~~\text{ for } i = 4\text{ and }5. 
\end{cases}
\end{equation}

We would like to point out the flexibility of our estimates of the asymptotic behavior of the metric perturbation functions. 
We can eliminate the algebraic variables by solving two differential equations, such as those corresponding to the first and the fifth row. 
We can also eliminate different differential variables using the obtained algebraic equations. 
One can show that eliminating other differential variables will not affect the QNM frequencies. 
To see this, we go back to step four and eliminate $y_2^{\ell = 2, 3}$ instead, which leads to another vector, 
\begin{equation}
\bar{\textbf{z}} = (y_1^2, y_1^3, y_4^2, y_4^3, y_5^2, y_5^3, y_6^2. y_6^3)^{\rm T}. 
\end{equation}
According to Equation~\eqref{eq:eliminate_diff_vars}, $\bar{\textbf{z}}$ and $\textbf{z}$ are related by a transformation matrix, 
\begin{equation}
\bar{\textbf{z}} = \bar{\mathbb{P}}(r, \omega) \textbf{z}, 
\end{equation}
where 
\begin{widetext}
\begin{equation}
\bar{\mathbb{P}}(r, \omega) = 
\begin{pmatrix}
1 & 0 & 0 & 0 & 0 & 0 & 0 & 0 \\
 0 & 1 & 0 & 0 & 0 & 0 & 0 & 0 \\
 \frac{(r-2) (r (2 r+3) \omega}{r \omega  \left(r \left(r^3 \omega ^2-2 r+3\right)+3\right)} & 0 & - \frac{(r-2) \left(r^3 \omega ^2-3\right)}{r \omega  \left(r \left(r^3 \omega ^2-2 r+3\right)+3\right)} & 0 & 0 & 0 & 0 & 0 \\
 0 & \frac{(r-2) \left(r (5 r+3) \omega\right)}{r \omega  \left(r \left(r^3 \omega ^2-5 r+9\right)+3\right)} & 0 & -i \frac{(r-2) \left(i \left(r^3 \omega ^2-6\right)\right)}{r \omega  \left(r \left(r^3 \omega ^2-5 r+9\right)+3\right)} & 0 & 0 & 0 & 0 \\
 0 & 0 & 0 & 0 & 1 & 0 & 0 & 0 \\
 0 & 0 & 0 & 0 & 0 & 1 & 0 & 0 \\
 0 & 0 & 0 & 0 & 0 & 0 & 1 & 0 \\
 0 & 0 & 0 & 0 & 0 & 0 & 0 & 1 \\
\end{pmatrix}. 
\end{equation}
\end{widetext}
Thus, the system of ordinary differential equations satisfied by $\bar{\textbf{z}}$ and that by $\textbf{z}$, 
\begin{equation}
\begin{split}
\frac{d \bar{\textbf{z}}}{d r} & = \bar{\mathbb{M}}(r, \omega) \bar{\textbf{z}}, \\
\frac{d \textbf{z}}{d r} & = \mathbb{M}(r, \omega) \textbf{z}, 
\end{split}
\end{equation}
is related by 
\begin{equation}
\mathbb{M} = {\bar{\mathbb{P}}}^{-1} \bar{\mathbb{M}} \bar{\mathbb{P}} - {\bar{\mathbb{P}}}^{-1}  \frac{d \bar{\mathbb{P}}}{d r}. 
\end{equation}
In other words, the two systems of ordinary differential equations are equivalent, even though the $p_{\infty}$ and $p_{H}$ of $\mathbb{M}$ and $ \bar{\mathbb{M}}$ may be different. 
Moreover, both systems of ordinary differential equations admit the same QNM frequencies, even though $\mathbb{M}$ and $ \bar{\mathbb{M}}$ are seemingly different, because $\omega$ is not altered through the transformation of $\bar{\mathbb{P}}$. 
We have checked that all these changes eventually lead to the same asymptotic behavior of the perturbation functions, despite the calculations in the middle being slightly different. 
This flexibility allows us to adjust the details of the calculations to make them the most convenient. 

Finally, using the above calculations, we can derive the ODE satisfied by every $y_{i=1, 2, ..., 6}^{\ell = 2, 3}$. 
The explicit equations are contained in a Mathematica notebook which is available upon request. 
A key feature of these ODEs is that those governing $y_{i=1, 2, ..., 4}^{\ell = 2, 3}$ contain only $y_{i=1, 2, ..., 4}^{\ell = 2, 3}$, and those governing $y_{i= 5, 6}^{\ell = 2, 3}$ contain only $y_{i= 5, 6}^{\ell = 2, 3}$. 
This property is consistent with the fact that, for perturbations of a Schwarzschild BH, the odd- and even-parity modes decouple. 

\section{Comparison with the existing calculations}
\label{sec:Asymptotic_behaviour_check}

In this Appendix, we check the validity of our calculations by comparing their details to those in the existing literature \cite{Regge:PhysRev.108.1063, Zerilli_even, Berti_02, Maggiore_vol_2}. 
We find that the first equation in Eq.~\eqref{eq:Subh3} is equivalent to 
\begin{equation} \label{eq:Zerilli_eqn_1}
\begin{split}
& \frac{d K(r)}{dr}-\frac{1}{r} H_0(r)-\frac{i(\lambda+1)}{\omega r^2} H_1(r)+\frac{1}{r} \frac{2 r-3 r_s}{2\left(r-r_s\right)} K(r)\\
&=0, 
\end{split}
\end{equation}
in the literature, where $\lambda = \ell (\ell+1) - 2$, and the definition of $K$, $H_0$ and $H_1$ is given by \cite{Zerilli:1971wd, Langlois:2021xzq, Maggiore_vol_2, Berti_02}. 
The second equation in Eq.~\eqref{eq:Subh3} is seemingly different from Equation~\eqref{eq:Zerilli_eqn_1}, but substituting the ordinary differential equations into Equation~\eqref{eq:Subh3}, both equations are simplified to 
\begin{equation}
y_{3}^{\ell = 2, 3} = - y_{1}^{\ell = 2, 3}, 
\end{equation}
which is equivalent to 
\begin{equation}\label{eq:Zerilli_eqn_2}
H_0(r) = H_2(r) 
\end{equation}
in the existing literature. 
We note also that the first two lines of Equation~\eqref{eq:eliminate_diff_vars} correspond to 
\begin{equation}\label{eq:Zerilli_eqn_3}
\begin{split}
& \left(\frac{3 r_\h}{r}+2 \lambda\right) H_0(r)+\left(\frac{i r_\h(\lambda+1)}{\omega r^2}-2 i \omega r\right) H_1(r)\\
& +\frac{3 r_\h^2+2 r_\h(2 \lambda-1) r-4 \lambda r^2+4 \omega^2 r^4}{2 r\left(r-r_\h\right)} K(r)=0. 
\end{split}
\end{equation}
This relation is consistent with previous calculations of even-parity perturbations of Schwarzschild BHs \cite{Zerilli_even}. 
We have checked that our calculations are consistent with Eqs.~\eqref{eq:Zerilli_eqn_1}, \eqref{eq:Zerilli_eqn_2}, and \eqref{eq:Zerilli_eqn_3} as we expand our calculations to $\mathcal{N}_{\chi} = 20 $. 
Finally, we compare the asymptotic behavior obtained in this paper with those in the existing literature.
Our calculations of the asymptotic behavior are clearly consistent with that of previous calculations. 

\begin{table*}[htb]
\centering
\begin{tabular}{llllll}
\hline
\multicolumn{3}{c}{In this paper}                                                             & \multicolumn{3}{c}{In the existing literature}                                                  \\ \hline
Symbols & $r \rightarrow \infty$                             & $r \rightarrow r_\h$              & Symbols & $r \rightarrow \infty$                             & $r \rightarrow r_\h$              \\
$h_1$ & $ r^{1+2i \omega} e^{i \omega r} $ & $(r-r_\h)^{-1-2i \omega}$ & $-H_0$  & $ r^{1+2i \omega} e^{i \omega r} $ & $(r-r_\h)^{-1-2i \omega}$ \\
$h_2$ & $ r^{1+2i \omega} e^{i \omega r} $ & $(r-r_\h)^{-1-2i \omega}$ & $H_1$   & $ r^{1+2i \omega} e^{i \omega r} $ & $(r-r_\h)^{-1-2i \omega}$ \\
$h_3$ & $ r^{1+2i \omega} e^{i \omega r} $ & $(r-r_\h)^{-1-2i \omega}$ & $H_2$   & $ r^{1+2i \omega} e^{i \omega r} $ & $(r-r_\h)^{-1-2i \omega}$ \\
$h_4$ & $ r^{2i \omega} e^{i \omega r} $   & $(r-r_\h)^{-2i \omega}$   & $K$     & $ r^{2i \omega} e^{i \omega r} $   & $(r-r_\h)^{-2i \omega}$   \\
$h_5$ & $ r^{1+2i \omega} e^{i \omega r} $ & $(r-r_\h)^{-2i \omega}$   & $h_0$   & $ r^{1+2i \omega} e^{i \omega r} $ & $(r-r_\h)^{-2i \omega}$   \\
$h_6$ & $ r^{1+2i \omega} e^{i \omega r} $ & $(r-r_\h)^{-1-2i \omega}$ & $h_1$   & $ r^{1+2i \omega} e^{i \omega r} $ & $(r-r_\h)^{-1-2i \omega}$ \\ \hline
\end{tabular}\label{tab:asymptotic_behaviour}
\caption{Comparison of the metric perturbation variables and asymptotic behavior derived in this paper and those in the existing literature \cite{Regge:PhysRev.108.1063, Zerilli:1971wd, Moncrief:1974am, Maggiore_vol_2}. }
\end{table*}

\section{Normwise scaling of quadratic eigenvalue problem}
\label{sec:Normwise scaling}

For the completeness of this paper, we briefly summarize the procedures of normwise scaling of the quadratic eigenvalue problem. We refer the reader for the details of this scaling to \cite{doi:10.1137/S0895479803434914, 2022arXiv220407424K}. 

Consider a quadratic eigenvalue problem
\begin{equation}
\left[ \tilde{\mathbb{D}}_0 + \tilde{\mathbb{D}}_1 \omega + \tilde{\mathbb{D}}_2 \omega^2 \right] \textbf{v} = \textbf{0}. 
\end{equation}
This quadratic eigenvalue problem is equivalent to 
\begin{equation}
\left[ \tilde{\mathbb{M}}_0 + \tilde{\mathbb{M}}_1 \omega + \tilde{\mathbb{M}}_2 \omega^2 \right] \textbf{v} = \textbf{0}, 
\end{equation}
where 
\begin{equation}
\begin{split}
\tilde{\mathbb{M}}_2 & = \sqrt{\frac{\|\tilde{\mathbb{D}}_0\|_2}{\|\tilde{\mathbb{D}}_2\|}_2} \tilde{\mathbb{D}}_2 \\
\tilde{\mathbb{M}}_1 & = \frac{1}{\sqrt{\|\tilde{\mathbb{D}}_0\|_2 \|\tilde{\mathbb{D}}_2\|_2}} \tilde{\mathbb{D}}_1 \\
\tilde{\mathbb{M}}_0 & = \frac{\tilde{\mathbb{D}}_0}{\| \tilde{\mathbb{D}}_0 \|_2}, 
\end{split}
\end{equation}
where $\| \mathbb{A} \|_2 $ is the two-norm of the matrix $\mathbb{A}$, defined as 
\begin{equation}
\|\mathbb{A}\|_2=\sup _{\textbf{x} \neq 0} \frac{\|\mathbb{A} \textbf{x}\|_2}{\|\textbf{x}\|_2}, 
\end{equation}
and $\|\textbf{x}\|_2$ is the two-norm of the vector $\textbf{x}$. 
It is shown that this definition is equivalent to \cite{meyer2000matrix}
\begin{equation}\label{eq:matrix_2_norm_eigenvalue}
\| \mathbb{A} \|_2 = \sqrt{\max |\lambda (\mathbb{A}^{\dagger} \mathbb{A})|}, 
\end{equation}
where $\max |\lambda (\mathbb{A}^{\dagger} \mathbb{A})|$ stands for the maximum modulus of the eigenvalue of the Hermitian matrix $\mathbb{A}^{\dagger} \mathbb{A}$. 
Equation~\eqref{eq:matrix_2_norm_eigenvalue} is also how we computed the two-norm of $\tilde{\mathbb{D}}_0$, $\tilde{\mathbb{D}}_1$ and $\tilde{\mathbb{D}}_2$ for the scaling before computing the generalized eigenvalues. 

\bibliography{ref}

\end{document}